\documentclass[aps,pre,floats,onecolumn,showpacs,superscriptaddress,floatfix]{revtex4-1}

\usepackage{graphicx,epsfig}
\usepackage{rotate}
\usepackage{amsmath}
\usepackage{amssymb}
\usepackage{amsthm}
\usepackage{comment,color}
\usepackage{bbold}
\usepackage{bbm}

\newcommand{\ie}{\emph{i.e.}}

\newcommand{\ER}{Erd\H{o}s-R\'{e}nyi}

\newcommand{\X}{ |C|} 
\newcommand{\Y}{P} 
\global\long\def\Rv{\mathcal{R}_v}

\begin{document}
\title{Explosive phenomena in complex networks}

\author{Raissa M. D'Souza}
\affiliation{Department of Computer Science,
Department of Mechanical and Aerospace Engineering, Complexity Sciences Center, University of California, Davis 95616, USA}
\affiliation{Santa Fe Institute, 1399 Hyde Park Rd Santa Fe New Mexico 87501, USA}

\author{Jesus G\'omez-Garde\~nes}
\affiliation{Department of Condensed Physics, University of Zaragoza, Zaragoza 50009, Spain}
\affiliation{GOTHAM Lab, Institute for Biocomputation and Physics of Complex Systems (BIFI), University of Zaragoza, Zaragoza 50018, Spain}

\author{Jan Nagler}
\affiliation{Deep Dynamics Group \& Centre for Human and Machine Intelligence, Frankfurt School of Finance \& Management, Frankfurt, Germany}

\author{Alex Arenas}
\affiliation{Departament d'Enginyeria Inform\`{a}tica i Matem\`{a}tiques, Universitat Rovira i Virgili, 43007 Tarragona, Spain}

\date{\today}

\begin{abstract}

\ \ \  \\

\vspace{0.1in}

The emergence of large-scale connectivity and synchronization are crucial to the structure, function and failure of many complex socio-technical networks. Thus, there is great interest in analyzing phase transitions to large-scale connectivity and to global synchronization, including how to enhance or delay the onset. These phenomena are traditionally studied as second-order phase transitions where, at the critical threshold, the order parameter increases rapidly but continuously. In 2009, an extremely abrupt transition was found for a network growth process where links compete for addition in attempt to delay percolation. This observation of ``explosive percolation" was ultimately revealed to be a continuous transition in the thermodynamic limit, yet with very atypical finite-size scaling, and it started a surge of work on explosive phenomena and their consequences. Many related models are now shown to yield discontinuous percolation transitions and even hybrid transitions. Explosive percolation enables many other features such as multiple giant components, modular structures, discrete scale invariance and non-self-averaging, relating to properties found in many real 
phenomena such as  explosive epidemics, electric breakdowns and the emergence of molecular life. 
Models of explosive synchronization provide an analytic framework for 
the dynamics of abrupt transitions and reveal the interplay between the distribution in natural frequencies and the network structure, 
with applications ranging from epileptic seizures to waking from anesthesia. 
Here we review the vast literature on explosive phenomena in networked systems and synthesize the fundamental connections between models and survey the application areas. We attempt to classify explosive phenomena based on underlying mechanisms and to provide a coherent overview and perspective for future research to address the many vital questions that remained unanswered.
\end{abstract}

\newcommand{\bc}{\begin{center}}
\newcommand{\ec}{\end{center}}
\newcommand{\be}{\begin{equation}}
\newcommand{\ee}{\end{equation}}
\newcommand{\bea}{\begin{eqnarray}}
\newcommand{\eea}{\end{eqnarray}}
\newcommand{\la}{\langle}
\newcommand{\ra}{\rangle}
\maketitle
\tableofcontents

\section{Introduction}
This review of explosive phenomena in complex networks is an attempt to provide a coherent overview of an area with vigorous activity and extensive literature dating back to 2009.  We discuss the paradigms of Explosive Percolation (EP) and Explosive Synchronization (ES) in complex networks, surveying the broad range of phenomena observed in models that display these types of abrupt transitions. We identify that these models share a common underlying mechanism with a microscopic evolution dynamics that delays the formation of a macroscopic (connected or synchronized) component. We show that similar underlying cluster dynamics can be associated with many real systems displaying abrupt transitions.
We also discuss examples of explosive phenomena in a variety of scenarios such the emergence of globalization from local individual decisions, epileptic seizures, explosive epidemics, and the emergence of molecular life.

\subsection{Percolation, synchronization, and phase transitions}
\label{intro_transitions}
Percolation theory is used to analyze the extent of connectivity in an underlying random media 
such as a lattice or a random graph. It has been used extensively for many decades to model phenomena ranging from flow through porous media to connectivity and spreading on random graphs~\cite{StaufferPercBook,SahimiBook,dgm07}. It was first formulated as a lattice model
where each of the ``bonds" between adjacent neighbors could be occupied with probability $p$ or unoccupied with probability $(1-p)$~\cite{BroadHamms1957}. A cluster of adjacent occupied bonds is considered to be connected, allowing a fluid or an epidemic to spread throughout that cluster. Of great interest is the percolation phase transition that describes the sudden, but typically smooth, onset of large scale connectivity at a critical value $p_c$.

While percolation focuses on the structural properties of a system, synchronization constitutes a paradigm for analyzing its dynamical behavior.  Synchronization is a fundamental phenomenon which is pervasive in systems ranging from coupled mechanical clocks to neuronal firing patterns in the brain. Although percolation and synchronization consider different aspects of a system, both phenomena are traditionally known to emerge via a second-order phase transition.  In other words, starting from a disconnected system, long-range connectivity (i.e., percolation) emerges smoothly at a critical point when increasing internal system density. In turn, starting from a system of uncoupled identical oscillators, global synchronization emerges smoothly at a critical point when increasing the
coupling between oscillators. In both cases, the global behavior (percolation or synchronization) arises as a consequence of increasing the strength of local interactions and are, in general, second-order phase transitions.  

Thermal equilibrium phase transitions are typically classified as first-order or second-order based on how the order parameter changes at the critical point, with a discrete jump in the value for first-order, and a sharp but smooth change for second-order.  Second-order transitions are additionally accompanied by a range of interrelated scaling behaviors, whereas first-order transitions typically display instead other phenomena like phase coexistence and hysteresis loops indicative of irreversibility.  See, for instance,~\cite{stanley1971}, and the discussion therein. 

In the scope of percolation, several examples of first-order transitions were known 
before the introduction of EP, such as $k$-core percolation~\cite{kcore83,kcore84}. In the scope of synchronization, several works pointed out that the character of the phase transition may be first-order in specific 
setups 
that are unrelated to the underlying topology of connections, but  instead are related to the particular distribution of natural frequencies, to random fields, or to inertial terms in the dynamical equations~\cite{Kuramoto84,Bonilla1992,vw93,tno97,p05}. Here our focus is on the more recently discovered phenomena of explosive transitions. These are abrupt transitions that arise as a consequence of  
the use of microscopic rules that, supported by the networked architecture, aim to hinder and postpone the formation of macroscopic components. In such a way, once the emergence of collective states becomes unavoidable, they emerge abruptly. 

\subsection{Percolation and synchronization on complex networks}

With the increasing prevalence and reliance of modern society on a collection of networks, including social, transportation, biological, ecological, and communication networks, a coherent scientific study of the principles of networks started emerging in the late 1990s~\cite{WS1998,BarabasiAlbert99,KKRRT}.
The paradigms of percolation and synchronization have provided theoretical underpinnings for analyzing the structure, function, resilience and robustness of network systems (see for instance~\cite{Barrat2008,Newman2010}).  
The classic starting points are the \ER~model of a random network discussed in Sec.~\ref{subsec:ER} and the Kuramoto model of synchronization discussed in Sec.~\ref{sec:KM_original}.

In any network system, the extent of connectivity is a fundamental property that shapes what functions that network can support. For instance, ensuring large-scale connectivity is essential for transportation networks such as the world-wide airline network or for communication systems like the internet. Yet, when an infectious disease is spreading over a network, large-scale connectivity becomes a liability; a virus spreading on a well-connected social or computer network can reach enough nodes to cause an epidemic. Thus, in some contexts large-scale connectivity is highly desirable, yet in other contexts large-scale connectivity is quite hazardous. 

The phenomena of ``explosive percolation" was first discovered in an attempt to use small  
interventions to  
delay the onset of large-scale connectivity~\cite{EPScience}.  
The quest to understand the true 
nature of the resulting abrupt transition and the associated underlying mechanisms 
sparked a flurry of activity into alternative models,  e.g.,~\cite{daCosta2010,naglerPRX,riordanPRE2012}, 
thermodynamics of cluster growth, e.g.,~\cite{ChoPRL09}, and explosive phenomena as a modeling paradigm, leading to the discovery of additional phenomena. A number of connections of EP to real-world systems have begun to emerge, e.g.,~\cite{clusella2016}.

In parallel, explosive synchronization in networks was first reported by~\cite{ggam11} when coupling the intrinsic frequencies of a dynamical system of phase oscillators with a topological characteristic of individual units, namely their number of local connections. An abrupt transition was observed in the order parameter that 
measures the global level of synchronization, for certain values of the coupling between units. 
This abrupt change is accompanied by an hysteresis cycle characteristic of first-order phase transitions,~\cite{sdkkbs93}, and has been proven to exist as a bistability region in the phase space. 
These results have stimulated a large number of follow-on studies, 
e.g.,~\cite{boccaletti2016}, including  
establishing connections between ES and  
the phenomenology of several real systems, e.g.~\cite{kmmvt16, pgnl16}. Similar to EP,  in ES the emergence of large-scale collective behavior is delayed until it emerges dramatically. 

\subsection{Overview of Review} 
Our attempt here is to provide a coherent overview organizing the literature and identifying the most fruitful open directions for future work. The review is organized as follows.  First in Sec.~\ref{sec:EP} we review Explosive Percolation phenomena. In Sec.~\ref{sec:ES} we review Explosive Synchronization phenomena. In Sec.~\ref{sec:EPES} we review the connections between EP and ES. In Sec.~\ref{sec:Other} we discuss other classes of dynamical processes on complex networks that show explosive transitions, including some recent works on multilayer and interdependent networks. Finally, in Sec.~\ref{sec:Conclusions} we provide conclusions and open challenges for the field.  

Note that the scope of this review goes beyond being 
a summary and overview of the literature, and also provides a comprehensive  
analysis of the roots behind explosive behavior and the connections between the apparently different methods and models that display the associated phenomena. 
To this aim, in those parts devoted to EP and ES the reader will find brief introductions to the basics of percolation and synchronization phenomena on networks (due to the variety of phenomena covered in Sec.~\ref{sec:Other}, we do not devote such attention there). In addition, while the EP section describes in detail the statistical-physics based models and analysis  of network growth leading to EP, 
the section devoted to ES is more focused on theoretical analysis of the 
equations underlying  the nonlinear models leading to ES. The difference between both approaches is rooted in the differing natures of EP and ES and offers a complementary view of explosive transitions that we believe will pave the way for a deeper understanding of explosive phenomena.

\section{Explosive percolation (EP)}\label{sec:EP}
Explosive percolation (EP) describes the critical behavior of a general class of graph evolution models with microscopic dynamics that delay the growth of large components until large-scale connectivity inevitably emerges in a dramatic and abrupt manner, hence the term ``explosive". 
EP transitions were first introduced in~\cite{EPScience}, 
by a variant on standard percolation where a simple selection criteria is used to determine how edges (i.e., links) are added between a collection of $N$ originally isolated nodes (i.e., vertices). 
Much more is now understood about 
the broad range of models and behaviors that fit the EP paradigm. 
EP transitions have now been shown to exhibit an array of novel universality classes, and anomalous critical and supercritical behaviors. For instance, in the supercritical regime the order parameter is not necessarily a function of the control parameter; multiple discontinuous transitions including a ``Devil's staircase'' of supercritical discontinuous transitions arbitrarily close to the initial percolation transition are possible; multiple giant components can coexist, which is not possible to classic percolation, and this feature also gives rise to modular structure. 

Many underlying mechanisms that give rise to EP have now been identified including growth by overtaking, directly suppressing the largest component, preferentially keeping all clusters similar in size to the average, and using correlated percolation processes.
It has been shown that EP processes can exhibit a discrete scale invariance predicting the location of the percolation transition. Many connections to real-world networks have been identified and EP is an emerging modeling paradigm. In this section we aim to synthesize these existing results 
and 
also the fundamental concepts and mechanisms underlying EP. 

\subsection{Percolation on random graphs}\label{subsec:ER}
Here we briefly review the foundational models of percolation on random networks, including the paradigm of cluster aggregation and the mechanism of ``multiplicative coalescence" that underlies the continuous emergence of a unique giant component. We refer to these models as classic random graph models of percolation. 

\subsubsection{Static formulation}
Several related models defining the concept of a random graph~\cite{Solomonoff1951,gilbert1959,ER1959,ER} were introduced in the 1950's. They consider a collection of $N$ vertices with edges 
connecting vertices uniformly at random under two slightly different formulations.  The first formulation considers that every possible edge has probability $p$ of being present (i.e., ``occupied"), which is referred to as the $G(N,p)$ model~\cite{Solomonoff1951,gilbert1959}. The second formulation considers that exactly $m$ edges are chosen to be occupied uniformly at random between the $N$ vertices, which is referred to as the $G(N,m)$ model~\cite{ER1959,ER}.  The two formulations are coincident in expectation, with the expected number of edges $\left<T\right> = m = p N (N-1)/2 \approx p N^2/2$ (with the final approximation becoming exact as $N \rightarrow \infty$). Thus, in the large $N$ limit, the ratio of the expected number of edges to nodes is $t = m/N = p N/2$, which provides the control parameter to parameterize the process. 

Nodes that are connected to one another following a path of occupied edges are considered to be in the same component (i.e., cluster) and we are interested in the distribution of component sizes for a given value of $t$.  
Of particular interest is the size of the largest connected component, $|C|$. This is an order parameter that displays a second order phase transition at $t_c=1/2$. For $t < t_c$, $|C|$ is of order logarithmic in $N$ and, for $t > t_c$, there is a unique largest component with size linear in $N$, with the value of $|C|$ changing rapidly but smoothly close to $t_c$~\cite{bolloRGbook}. In fact, for $t$ slightly greater than $t_c$ the size of the largest component is described by the function $|C|\sim (4t - 2)N$ ~\cite{bolloRGbook}.

One can further characterize the nature of the random graph phase transition by analyzing associated response functions, such as what is referred to in the literature as the ``susceptibility", $\chi$, defined as the second moment of the component sizes, see e.g.,~\cite{bolloRGbook,Grimmett2010,Newman2010}. In more detail, this can be written
\be
\chi = \frac{1}{N}  \sum_j j^2 N_j
\label{eqn:chi}
\ee
where $N_j$ is the number of clusters that are of size $j$.   Note $j N_j/N$ is the fraction of vertices in components of size $j$, so this second moment, $\chi$, is the expected size of the component to which an arbitrary vertex belongs. In the critical regime we see divergence of $\chi$ as expected for a standard second order phase transtion
\be
\chi \sim | t - t_c |^{-\gamma}. 
\ee
Classic random graph percolation obeys the mean-field exponents, with $ \gamma =1$. For an explicit treatment, see for instance ~\cite{bolloRGbook,SaberiPhysReports2015}.  It should be noted that although this definition of susceptibility does show a divergence at the critical point, it does not obey the fluctuation-dissipation theorem that gives the connection between susceptibility and variances in statistical thermodynamics~\cite{bizhani2012}.

\subsubsection{Kinetic formulation} \label{subsubsec:ClusterAgg}
The model above can be recast as a kinetic process~\cite{ZiffStell1980,BenNaimPRE05} using the framework of cluster aggregation and the celebrated Smoluchowski equation~\cite{smoluchowski}.
Beginning from a collection of $N$ isolated nodes, edges are added, one at a time, uniformly at random to the graph in a discrete time process, with $T$ denoting the total number of edges added.  The control parameter is once again the relative number of introduced edges $t = T /N$. 

Since the process begins with a collection of $N$ isolated nodes, the first edge necessarily joins together two previously distinct components, forming a component of size two. Similarly, while the component sizes are sufficiently small, the likelihood that an edge chosen at random is internal to a component is negligible and with high probability
each added edge joins together two previously disjoint components.
With this assumption, one can then describe the random graph evolution process 
as a Smoluchowsky rate equation of cluster aggregation~\cite{smoluchowski}. Here the likelihood that a random edge merges together a component of size $i$ with a component of size $j$ is described by a collision kernel $K_{ij}$.  
The evolution of the corresponding cluster density is described by the Smoluchowski rate equation as follows: 
\begin{equation}
\frac{dn_k}{dt} = \sum_{i+j=k} K_{ij} n_i n_j- 2 n_k \sum_j K_{kj} n_j, 
\label{eq:SmEqu}
\end{equation}
where $n_k$ is the density of $k$-size clusters 
(the number of clusters of size $k$, divided by the total number of clusters).
The particular collision kernel, $K_{ij}$, captures the adhesion properties and clusters' diffusivity.
The rate $K_{ij}n_i n_j$ reflects the probability that two clusters of sizes $i$ and $j$ merge per unit volume and unit time thus producing a cluster of size $k=i+j$. 
The negative term accounts for the case that a $k$-size cluster merges with any of the remaining clusters. 

Cluster aggregation analysis is used in many studies discussed throughout this review. Modern perspective and rigorous details on the approach can be found in the recent 
textbook~\cite{krapivsky2010}. The approach ignores intra-cluster link structure of the underlying network, but from the computational perspective, this allows for much faster simulations as only the distribution of cluster sizes needs to be propagated. In particular the celebrated Newman-Ziff algorithm based on union-find~\cite{NewmanZiff01} is pervasively used 
for efficient computation of percolation. 

For the random graph model discussed thus far, two vertices are chosen uniformly at random and connected by an edge. The likelihood of choosing a vertex in a particular component of size $i$ is proportional to $i$ (the number of vertices in that component). Thus, the likelihood of merging two components is proportional to the product of their sizes and $K_{ij} \propto i j$, 
which is referred to as ``multiplicative coalescence"~\cite{Aldous99}.  This 
leads to a phase transition in the size of the largest component (called gelation) that is mathematically equivalent to static formulations of percolation on a random network discussed above. This mechanism of multiplicative coalescence, whereby large clusters grow proportionately more rapidly than small clusters, leads to the emergence of one unique giant component as discussed in more depth in~\cite{Grimmett2010,SpencerAMS2010}. For more discussion on kernels, the gelation transition and connections to ``mean-field" probability theory of random graphs see~\cite{Aldous99,Ziff1980}. Also see~\cite{stockmayer1943} for the kinetic theory approach to branched polymers which is equivalent to percolation on the Bethe lattice.  

Note that the cluster aggregation approach, Eq.~\ref{eq:SmEqu}, assumes that each added edge connects two previously disjoint components, thus a cluster aggregation process necessarily ends at $t = (N-1)/N $ when only one component remains. 
If a giant component only appears at this extreme limit of $t=1$, in the limit $N\rightarrow \infty$, then this is considered lack of gelation, or trivial gelation, as there is only one cluster remaining. In contrast, alternative models of network growth beyond cluster aggregation can allow for the addition of edges that are internal to a cluster, in which case, the maximum edge density attainable would be $t =(N-1)/2$ on an undirected network.
Also note that due to the equivalence of the static definition and the kinetic approach, the random graph percolation model is reversible, an issue we will return to in Sec.~\ref{subsec:insights} in the context of explosive transitions.

\begin{table*}[tb]
\begin{center}
\begin{tabular}{|l|l|l|} \hline
\ Model name & \ Process considered & \ First reference\\ \hline

\ Bootstrap percolation & \ Pruning of sites with too few neighbors & \ \cite{kcore-mixed1}\\

\ $k$-core percolation & \ Pruning of insufficiently connected nodes & \ \cite{kcore83,kcore84}\\

\ Rigidity percolation & \ Dynamics of rigid and floppy regions in glassy structures \ \ & \ \cite{Thorpe1985} \\

\ Jamming percolation & \ Jamming transition in sphere packing & \  \cite{liunagel} \\

\ Generalized contagion & \ SIR threshold dynamics & \  \cite{Dodds2004} \\

\ Tricritical dynamic percolation \ \ & \ 4-species spreading dynamics & \ \cite{Janssen2004} \\

\hline 

\end{tabular}
\end{center}
\caption{Seminal models of discontinuous percolation transitions which pre-date the work on EP.}
\label{fig:historyDPT}
\end{table*}%

\subsubsection{Discontinuous percolation transitions}

Before the notion of Explosive Percolation (EP) was introduced in~\cite{EPScience}, 
there were many well-known studies of discontinuous percolation transitions.
Several of these seminal models are summarized in Table~\ref{fig:historyDPT}. 
These models belong to three distinct classes.
The first class is based on site or link {\em removal}, or culling, with the most prominent examples being bootstrap percolation~\cite{kcore-mixed1} and $k$-core percolation~\cite{kcore83,kcore84}.  
An in-depth discussion of k-core percolation on complex networks can be found in~\cite{dorogovtsev2006,dgm07}. The second class is based on glassy dynamics~\cite{Thorpe1985}. The third class is based on multiple-species models, in particular infected-susceptible type of dynamics~\cite{Dodds2004},
or the 4-species models~\cite{Janssen2004}. 

Also worth noting is the jamming transition in sphere packings as introduced in~\cite{liunagel}.  The nature of that transition was revealed by mapping the process onto analogous models related to $k$-core percolation~\cite{schwarz2006onset,PhysRevLett.96.035702}.
These models of ``jamming percolation'' on low-dimensional lattices incorporate spatial correlations intended to capture glassy dynamics.  They exhibit hybrid phase transitions with a discontinuous jump in an order parameter, but diverging length scales characteristic of second-order transitions~\cite{schwarz2006onset,PhysRevLett.96.035702,jengPRE2010,cao2012correlated}.

\subsection{EP from competitive percolation}

Many types of graph evolution processes are now known to lead to EP. These processes delay the formation of large components and break the ``multiplicative coalescence" rule of traditional models of percolation discussed in Sec.~\ref{subsubsec:ClusterAgg}. This allows for the many unique 
phenomena manifested during the emergence of large-scale connectivity, as surveyed herein. EP was first demonstrated for graph evolution processes with edge competition, which is the focus of this current Section. 

\begin{figure*}[tb]
  \begin{center}
\includegraphics[width=.95\textwidth]{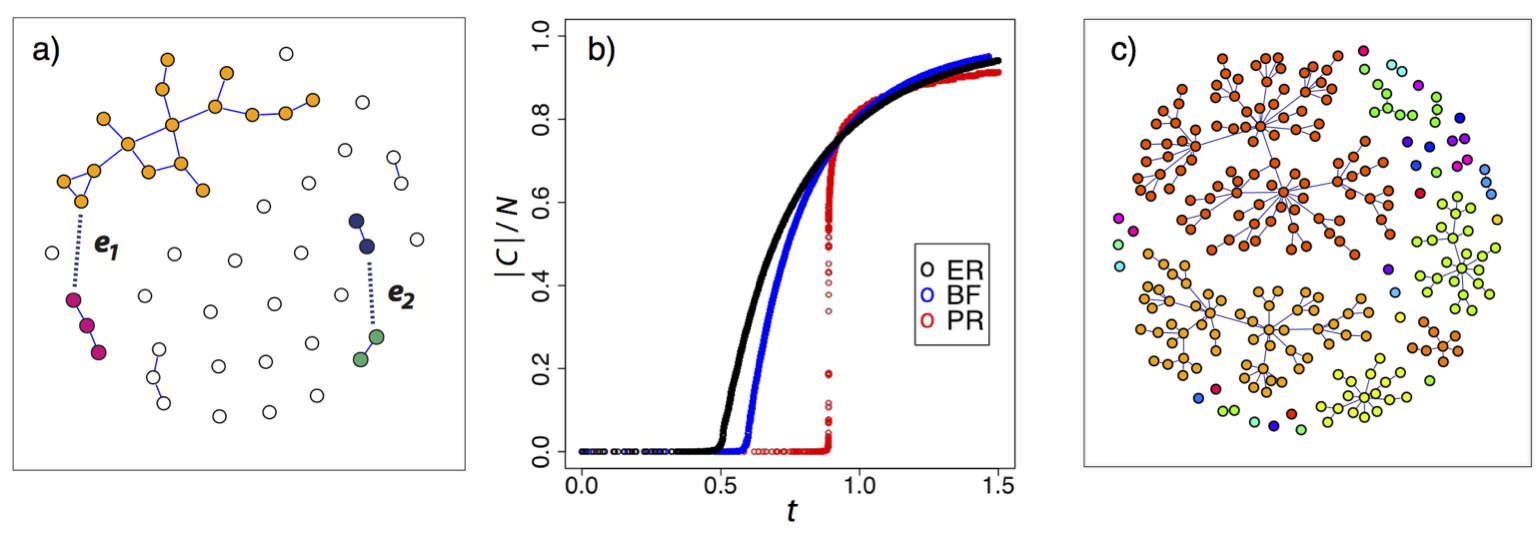}
  \end{center}
  \vspace{-0.2in}
  \caption{
  Schematic of the Product Rule.
  (a)  At each time step of the Product Rule (PR) process, two edges, $e_1$ and $e_2$, compete for addition.
  Here the product of the components merged by $e_1$ is 
  $3\times 16=48$ and by $e_2$ is $2\times 2=4$,
  so $e_2$ is accepted in and $e_1$ rejected.
   (b) Typical evolution of an \ER~(ER), Bohman Frieze (BF), and PR process on a system of size $N=10^6$.
    Plotted is the fractional size of the largest component, $|C|/N$, as a function of edge density $t$.
  (c) A sample network of $N=250$ grown via the Product Rule, with the nodes in each distinct component rendered in the same color. The largest component, of size $|C|$, is indicated in red.  [Figs (a) and (b) reproduced from~\cite{RDJN2015}.]
  }
\label{figBasics}
\end{figure*}

\subsubsection{Achlioptas Process (AP)}

Random graphs are long-studied topics in combinatorics and probability theory. At a workshop entitled ``Probabilistic graph theory" held at the Fields Institute in February of 2000, Dimitris Achlioptas proposed that ``the power of two choices" paradigm, a recent innovation in randomized algorithms, could be used in the \ER~graph evolution process to enhance or delay the percolation transition. 
This impact of two choices was first introduced in 1994 in the context of hash tables, showing that if there are two choices for the potential hash function it greatly reduces the needed storage~\cite{AzarBroder94,AzarBroder99,Adler98}. This paradigm was shown to be particularly effective for problems such as load balancing, where giving arriving jobs a choice between the shorter of two randomly selected queues leads to exponential improvement in performance~\cite{mitzenmach01}.  
The ``power of two choices" was also introduced  
into networking protocols 
based on random walks and shown to enhance performance~\cite{avin2008power}. This latter model is similar to the ``true" self-avoiding walk of~\cite{amit1983asymptotic}, where they find surprising results, such as an upper critical dimension of $d=2$.

Introducing the element of design to percolation on a lattice  can also be attributed to studies of Highly Optimized Tolerance (HOT)~\cite{CD1999HOT}. The intent of HOT is to use optimization, via natural selection or engineering design, to create power law distributions of cluster sizes with the clusters strategically situated to minimize damage to particular classes of random fluctuations. Specific implementations of HOT include forest fire models on a lattice. The lattice configurations achieved by the HOT process are extreme configurations that form a set of measure zero, having compact clusters that avoid a spanning cluster even at the highest densities studied. Thus, HOT models use the element of design to avoid percolation throughout the 
process. Occasional large failures are seen and the highly optimized states are fragile to the nature of the fluctuations. 

The power of two choices for a random graph evolution process delays, but does not avoid, percolation. It can be formulated as a kinetic process as
discussed in Sec.~\ref{subsubsec:ClusterAgg}, where edges were added uniformly at random in a discrete time manner~\cite{BenNaimPRE05}.
But now, instead of adding one randomly chosen edge at each discrete time step, $T$, first choose two candidate edges, $\{e_1,e_2\}$, and examine what would be the consequence of adding each one to the graph (an example is shown in Fig.~\ref{figBasics}(a)). Add whichever edge is more favorable by your selection criteria and discard the remaining candidate edge making it available to be a future candidate, and move on to the next discrete time step, $T+1$. Thus edges are added via a competition process that has become known as an ``Achlioptas Process". The selection criteria can vary. For instance, to enhance the onset of percolation, choose the edge that leads to the larger resulting component, and to delay the onset, choose the edge that leads to the smaller.  

In 2001 Bohman and Frieze formally introduced the terminology ``Achlioptas Process" (AP) and analyzed such an AP~\cite{BohmanFrieze}. Their goal was to use this approach to delay the formation of a giant component in an \ER-like graph evolution process. In order to make the model amenable to rigorous analysis, they are limited to the context of ``bounded-size'' rules where all components of size $K$ or greater are treated as if they were equivalent in size. 
For the Bohman and Frieze process (BF), $e_1$ is accepted if it joins two isolated nodes (and $e_2$ rejected), otherwise $e_2$ is accepted (and $e_1$ rejected). Thus, only components of size one (isolated nodes) are distinguished, and all components of size $K \ge 2$ are treated equivalently. A rigorous proof shows that BF delays the percolation transition when compared to ER~\cite{BohmanFrieze}, but this does not address the nature of the transition.

BF can be analyzed using rate equations similar to the cluster aggregation approach. 
Early in the evolution, well before the emergence of the giant component, all components are of size at most $O(\log N)$, and edges that are added to the graph join previously disjoint components with high probability.
But in the critical regime, where components can grow sufficiently large, internal component edges become more likely to be chosen to be added to the graph.  Rigorous analysis of the error introduced from violating the spanning component assumption of cluster aggregation models leads to the conjecture that all bounded-size rules lead to a continuous percolation phase transition~\cite{SpencWorm}.

\subsubsection{AP with $m$-edge rules on random graphs}\label{subsubsec:APedge}

Achlioptas Processes with unbounded-size selection rules (where components of different sizes are treated uniquely) are not amenable to the analysis of the approaches described above~\cite{BohmanFrieze,SpencWorm}. Thus unbounded-size rules were first studied via numerical simulation of a random graph evolution process using an AP called the Product Rule (PR)~\cite{EPScience}.
\\
\paragraph{The Product Rule and related AP's}
\ \\

The Product Rule is described as follows and illustrated in Fig.~\ref{figBasics}(a). Starting from $N$ isolated nodes, two candidate edges $\{e_1, e_2\}$ are chosen uniformly at random at each discrete time step, $T$. Let the vertices linked by $e_1$ be denoted $a$ and $b$, and the vertices linked by $e_2$ be denoted $c$ and $d$, and let $|C_i|$ denote the size of the component that contains vertex $i$.  
If $| C_a|  | C_b| < | C_c| | C_d|$, then $e_1$ is added to the graph. Otherwise, $e_2$ is added. In other words, we retain the edge that minimizes the product of the two components that would be joined by that edge. 
Note the likelihood of choosing an edge internal to a component of size $|C_i|$ is proportional to $|C_i|^2/N^2$, which approaches zero as $N \rightarrow \infty$ in the subcritical regime when components are sub-linear in system size $N$.
Thus the PR selection criteria can be used if vertices $a=b$ and/or $c=d$ without impact on the percolation transition.

Figure~\ref{figBasics}(b) shows a typical realization of a PR process, 
an \ER~(ER) process~\cite{ER1959,ER,gilbert1959} and a Bohman-Frieze (BF) process~\cite{BohmanFrieze} on a system of size $N = 10^6$. Note that the onset of large-scale connectivity is considerably delayed for the PR process and that it emerges drastically, going from sublinear to a level approximately equal to the corresponding \ER~and BF processes during an almost imperceptible change in edge density. Here the numerical simulations make use of the commonly used Newman-Ziff algorithm 
~\cite{NewmanZiff01}.

To quantify the abruptness of the transition, the scaling window as a function of system size $N$, denoted $\Delta_N(\gamma,A)$, was analyzed. This measures the number of edges required for $|C|$ to transition from being smaller than $N^\gamma$ to being larger than $AN$. With the specific parameters choices used in~\cite{EPScience}, $\gamma=A=1/2$, this measures the number of edges that need to be added for the order parameter to go from $|C| \le \sqrt{N}$ to $|C| \ge 0.5 N$.  
Systems up to size $N\sim 6\times 10^7$ were studied and the results indicated a sublinear scaling window,
\be
\Delta_N(0.5,0.5) \propto N^{2/3}.
\ee
The associated change in edge density
\begin{equation}
\Delta_N(0.5,0.5)/N \propto N^{-1/3} \rightarrow 0 {\rm \ \ as \ \ } N\rightarrow \infty
\end{equation}
converging on $t_c \approx 0.888$~\cite{EPScience}. 
This provides evidence of a discontinuous phase transition, but as discussed below, the Product Rule on a random graph will ultimately be shown to lead to a continuous transition in the thermodynamic limit, albeit one with an unusual universality class. 

The Product Rule was then analyzed in settings beyond the \ER~random graph, including percolation on a 2D lattice~\cite{ZiffPRL09} and on scale-free networks~\cite{ChoPRL09,RadFortPRL09}. 
Similar sub-linear scaling windows were observed. The scale-free networks showed additional interesting behavior. Here the degree distribution follows a power law distribution with exponent $\lambda$ and evidence suggested that there was a critical value $\lambda_c$ above which the system displayed a discontinuous transition, with $\lambda_c$ being a tricritical point~\cite{ChoPRL09,RadFortPRL09}. Yet, in addition, it was shown that all the models display scaling behaviors characteristic of continuous, second-order phase transitions such as power-law cluster-size distribution with an exponent close to two and scaling of susceptibility $\chi \sim | t - t_c |^{-\gamma}$~\cite{ChoPRL09,ZiffPart2,RadFortPart2}, but with very unusual values of scaling exponents; for a table see~\cite{RadFortPart2}.    

Many generalizations of the Product Rule and variants followed, in particular the introduction of APs with $m$-edge selection rules.  Here, rather than two candidate edges,  $m > 2$ candidates edges are considered at each discrete time step, where $m$ is a fixed constant.  For instance, 
in~\cite{powderkeg09} they introduced the ``triangle rule"  
where the $m=3$ edges that are possible between three randomly chosen vertices are considered as candidate edges. This process was later analyzed in~\cite{RDMMclustagg}. 
Likewise, 
the Sum Rule (which minimizes the sum of the components sizes that would be joined) was analyzed~\cite{PRL.107.275703,riordanPRE2012}. 
These all show similar results of sublinear scaling windows accompanied by critical scaling behaviors.
We refer the reader to the comprehensive review of these models in~\cite{bastas2014review}, including detailed tables of critical exponents.
\\

\paragraph{The impact of a single edge}
\ \\

Complementary to a scaling window analysis, 
the impact of a single edge was studied 
in Refs.~\cite{naglerNP,Manna}. 
This is the maximum change in the relative size of the largest component from the addition of a single edge, defined as 
\begin{equation}
\Delta C_{\rm max} \equiv {\rm max_{\{T\}}}\left[\ |C(T+1)| - |C(T)|\ \right] /N.
\end{equation}
For an $m$-edge AP, it is found that $\Delta C_{\rm max}$ decays as a power law with system size, $\Delta C_{\rm max} \sim N^{-\beta}$~\cite{naglerNP,Manna}. 
Thus, the size of this single-edge addition gap decays to zero in the thermodynamic limit.
The rate of decay is typically quite small ($\beta = 0.065$ for the Product Rule~\cite{Manna}), leading to large discrete jumps in system sizes orders of magnitude larger than real-world networks. 
Note that the gap decaying to zero typically indicates a continuous transition, yet this also requires verifying that the growth of the order parameter displays a finite slope at the critical point as discussed formally in Sec.~\ref{subsec:Anomalous}.

A rigorous advance was made in~\cite{naglerNP} with a proof that if the largest component is allowed to grow directly in any way (i.e., by merging with another component), denoted $P_{Gr}>0$, this can lead to a continuous percolation transition. In contrast, if $P_{Gr}=0$ this necessary leads to a discontinuous percolation transition during the process.  $P_{Gr}=0$ implies that growth in the size of the largest component can only happen via overtaking, when two smaller components merge together to become the new largest component. 
This allows for analytic calculation of the strict lower bound on the relative size of the discrete jump $\Delta C_{\rm max}  \ge 1/3$~\cite{naglerNP}.   
The mechanism of growth by overtaking is now seen in several processes that lead to EP.\\

\paragraph{The powder keg}
\ \\

In~\cite{powderkeg09} they establish that for any percolation transition to be discontinuous there must exist a ``powder keg".  This is essentially a collection of a sub-extensive number of components that together contain a total of $cN$ nodes for some constant $c > 0$. Thus the number of nodes in these components diverges to infinity as $N\rightarrow \infty$. 
Formally the size of a powder keg, $W$, can be expressed as
\be
W = \frac{1}{N} F\left( t(N^\alpha), N^{1-\beta} \right)
\label{eqn:powderKeg}
\ee
where $F(\tau,a)$ is the number of nodes in clusters of size greater than or equal to $a$ after the addition of the {$\tau$-th} edge. 
If $W > 0$ this means there is a non-zero fraction of nodes in components ranging in size from $N^{1-\beta}$ to $N^\alpha$. 
Merging the components of such a powder keg requires only a sub-linear number of edges, leading to a discontinuous percolation transition. In fact, 
as shown in~\cite{powderkeg09}, if a system is initialized with a powder keg, then even a random edge addition rule causes a discontinuous transition. 
For a more detailed treatment of the powder keg see~\cite{Hooyberghs2011}. \\

\begin{figure}[tb]
  \begin{center}
  \includegraphics[width=.4\textwidth]{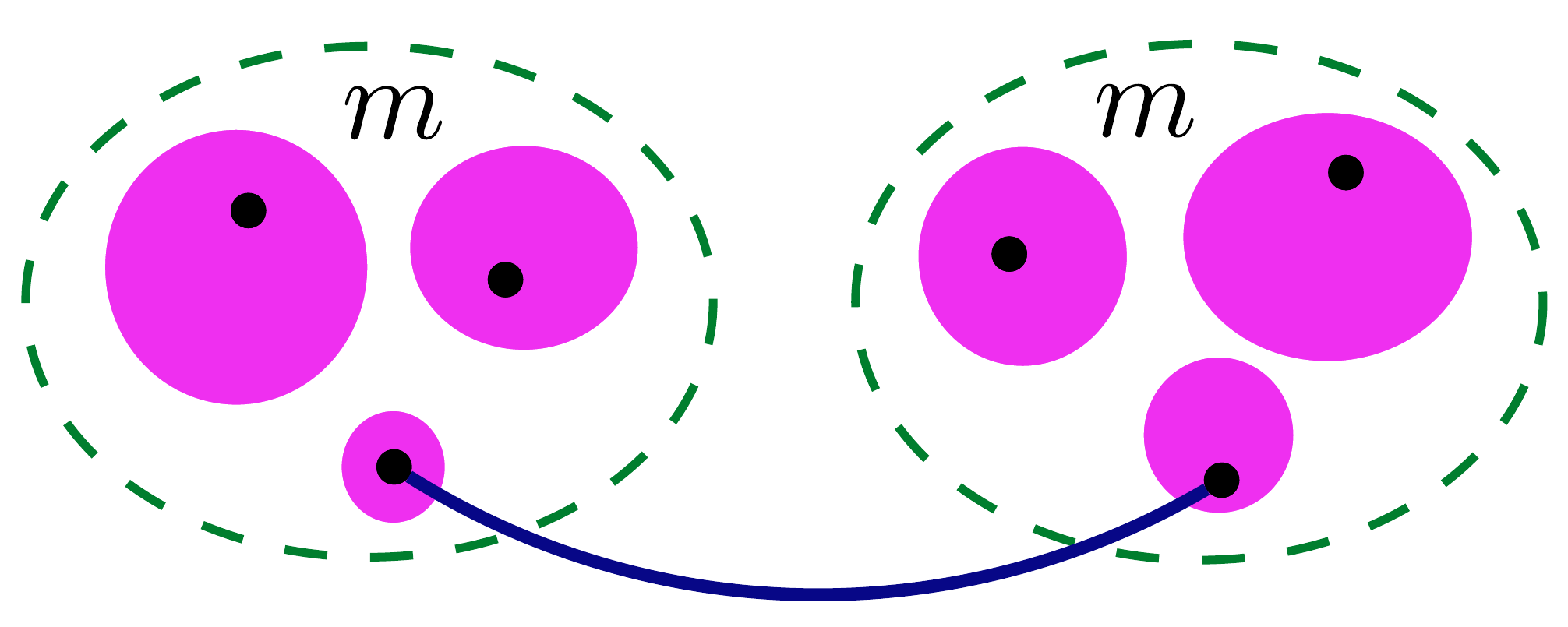}
  \end{center}
  \vspace{-0.2in}
  \caption{The definition of the dCDGM model. At each discrete time step, two sets of $m$ nodes are chosen uniformly at random, and the node in the smallest component of the first set is linked by an edge to the node in the smallest component of the second set. [Reprinted from~\cite{daCostaPRE2014}.]}
\label{fig:dCDGMmodel}
\end{figure}

\paragraph{Continuous transition with unusual finite size scaling}
\ \\ 

A series of seminal works 
soon followed showing via 
numerical evidence and rate equation analysis that an AP leads, in fact, to a continuous phase transition~\cite{daCosta2010,grassberger2011,lee2011continuity,tian2012nature},
but with many new distinct universality classes~\cite{grassberger2011,tian2012nature}.  

In particular in~\cite{daCosta2010} they analyze a representative model of a competitive percolation process, referred to as the dCDGM model and shown in Fig.~\ref{fig:dCDGMmodel}. Here two sets of $m$ nodes are chosen uniformly at random. The node in the smallest component of the first set is identified and then linked by an edge to the node in the smallest component of the second set. They denote the relative size of a giant component as $S$ (which is $|C|/N$ in the notation used thus far).  From direct numerical simulation they show that  
at the critical point $S \sim (t-t_c)^{\beta}$, indicating a continuous transition.  They also develop analytic models using rate equations for the evolution in time $t$ of the size distribution $P(s,t)$ for a finite cluster of $s$ nodes to which a randomly chosen node belongs. Specifically, 
\be
P(s,t) = \frac{s n(s)} {\left< s\right> },
\ee
where $n(s)$ is the fraction of finite components of size $s$ nodes, and $\left< s\right>$ is the mean cluster size defined as the number of nodes $N$ divided by the total number of clusters that exist at time $t$.  
Thus, they can write the order parameter as the fraction of nodes that are not in finite components, $S(t) = 1 - \sum_{s}P(s,t)$.

\begin{figure}[b]
  \begin{center}
  \includegraphics[width=.48\textwidth]{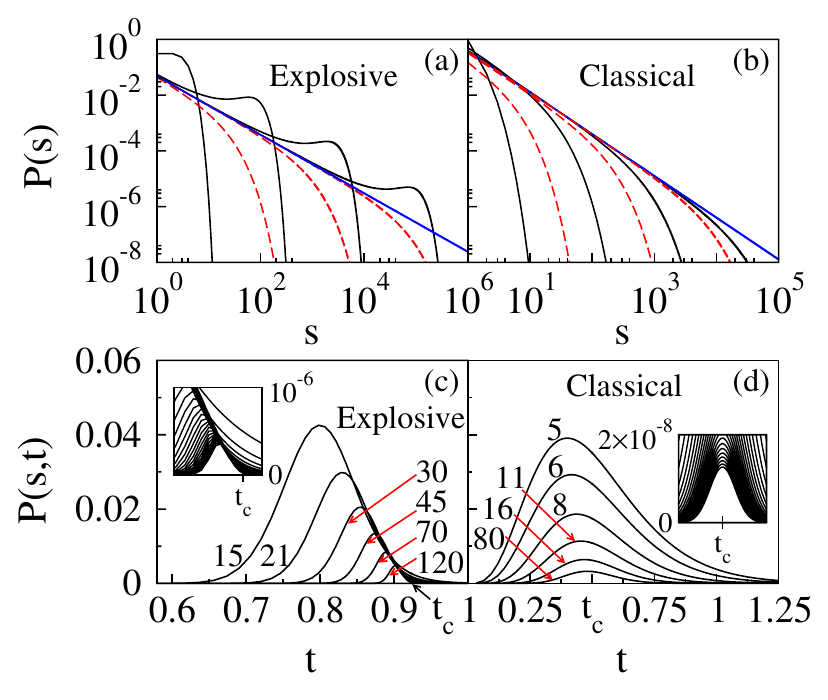}
  \end{center}
  \vspace{-0.2in}
  \caption{Comparing the dCDGM model with classic random graph percolation.  
  (a)-(b) The distribution of finite clusters of size $s$, denoted $P(s$), shown in the subcritical region (solid black line), at $t_c$ (solid blue line), and in the supercritical region (dashed red line).  (c)-(d) The evolution of this size distribution, $P(s,t)$. 
  [Reprinted from~\cite{daCosta2010}.]}
\label{fig:dCDGMscaling}
\end{figure}

Figure~\ref{fig:dCDGMscaling} shows the evolution of $P(s,t)$ for both the dCDGM model and classical random graph percolation. As shown in Fig.~\ref{fig:dCDGMscaling}(a), although the early evolution of $P(s,t)$ for the dCDGM model (the solid black lines) suggest the build up of a powder keg, we see that at $t_c$ (the solid blue lines) this distribution obeys a power law, with no powder keg. 
Numerical solution of the rate equation for $S \approx 1 - \sum_{s=1}^{10^6} P(s)$ leads to the value of the order parameter critical exponent $\beta=0.0555(1) \approx 1/18.$
 
In~\cite{grassberger2011} they also establish the continuity of APs, but using finite size scaling approaches. They denote the order parameter by $s_{\rm max}/N$ and the control parameter by $p$ (correspondingly $|C|/N$ and $t$ in this review). In particular, they focus on $P_{p,N}(s_{\rm max}/N)$ which is the distribution of the order parameter in finite systems for a given value of the control parameter. From finite size scaling theory we know that at the critical point this distribution behaves as 
\be
P_{p=p_c,N}\left(\frac{s_{\rm max}}{N} \right) \sim N^{\eta} f\left(\frac{s_{\rm max}}{N} \cdot N^{\eta}\right).
\ee
(For details see for instance~\cite{BrucePRL1992,BinderBook2010}.)  For a first-order (i.e., discontinuous) transition we expect $P_{p=p_c,N}(s_{\rm max}/N)$ to be a bi-modal distribution with two distinct peaks, each one corresponding to one of the two distinct co-existing phases. We further expect that the typical distance between the peaks is the expected jump in the size of the order parameter at the critical point. In contrast, for second-order (i.e., continuous) transitions, even if the system shows the bi-modal distribution for finite $N$, we expect that as $N\rightarrow \infty$ the distance between the peaks shrinks to zero. 

\begin{figure}[tb]
  \begin{center}
  (a)\includegraphics[width=.53\textwidth]{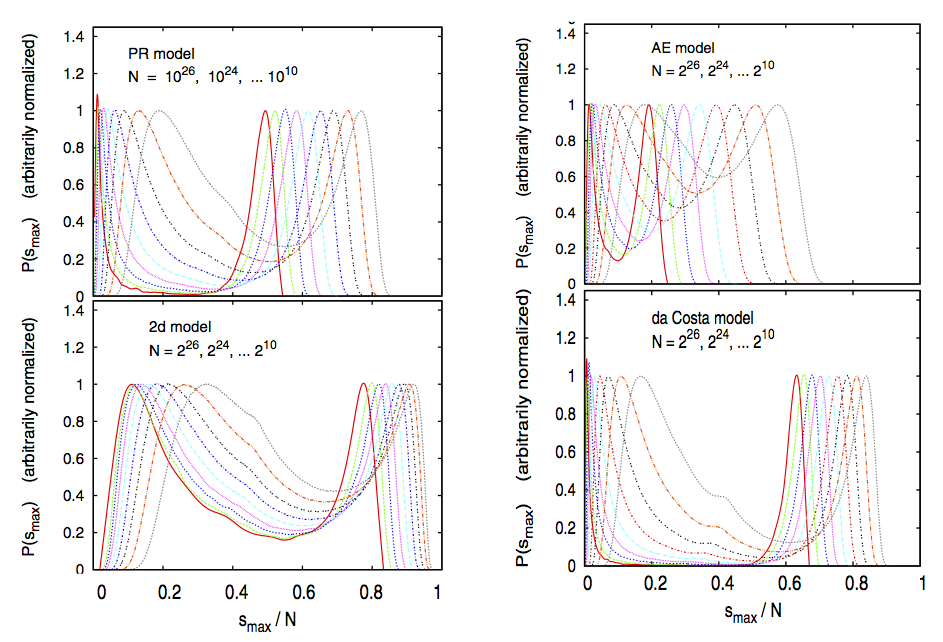}\hfill
    (b)\includegraphics[width=.42\textwidth]{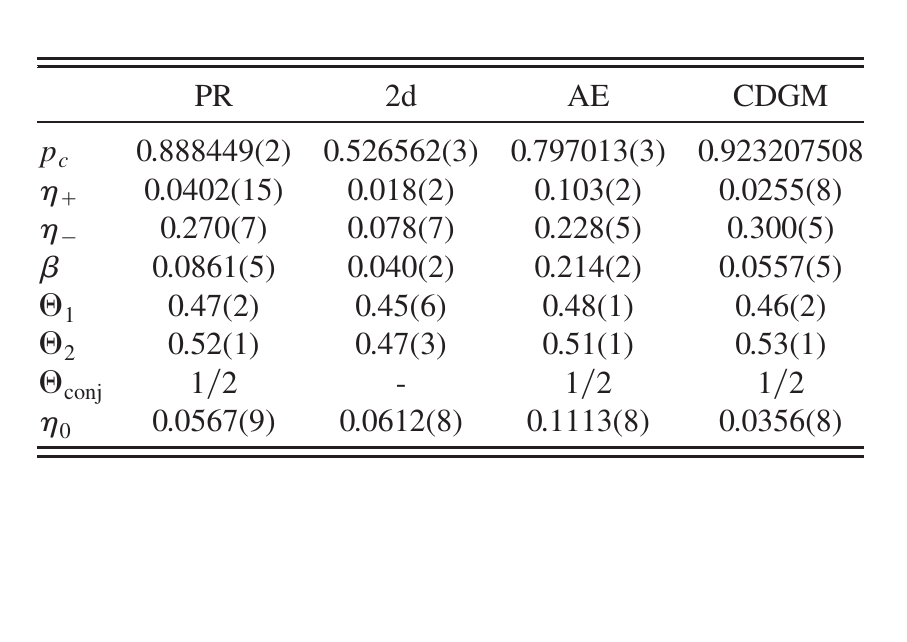}
  \end{center}
  \vspace{-0.2in}
  \caption{(a) The distribution of $P_{p,N}(s_{\rm max}/N)$ with varying $N$ for four different competitive percolation processes. (b) Table showing the differing critical points and numerical estimates of the critical exponents for the four different models. [Adapted and reprinted from~\cite{grassberger2011}.]}
\label{fig:GrassPRL2011}
\end{figure}

Shown in Fig.~\ref{fig:GrassPRL2011}(a) is a main result from~\cite{grassberger2011}, of $P_{p,N}(s_{\rm max}/N)$ for four different models of competitive percolation (the original Product Rule, the Product Rule on a 2D lattice~\cite{ZiffPRL09}, the dCDGM model~\cite{daCosta2010}, and the ``Adjacent edge" rule~\cite{RDMMclustagg}). The value of the control parameter $p$ is set such that the heights of both peaks are the same. They study varying $N$, extrapolating to the $N \rightarrow \infty$ limit.  Although they show that all four models have continuous transitions in the thermodynamic limit, each one displays a distinct universality class, in particular values of $\beta$ and $\eta$ (see the table from~\cite{grassberger2011} reprinted in Fig~\ref{fig:GrassPRL2011}(b)).  Furthermore, all four models show double-peaked order parameter distributions, with the sharpness of the peaks increasing initially with system size. Each has different scaling laws for the width of the scaling region and for the shift of the effective $p_c(N)$.   They conjecture that the specific non-locality of the APs studied give rise to these unusual features.

The important work of~\cite{tian2012nature} also quantifies this coexistence of a strongly double peaked distribution in the histogram of the order parameter at the percolation threshold. They are able to further show that although the peaks can become more defined and separated with increasing system size, nevertheless the distance between the two peaks does shrink to zero, in a power law manner, as $N \rightarrow \infty$. They also find novel scaling exponents and universality classes.  
Extensive numerical studies in~\cite{lee2011continuity} further show through a finite size scaling collapse that an AP has a well defined convergence to a continuous percolation threshold. 
\\

\paragraph{A rigorous proof of continuity}
\ \\

A rigorous mathematical proof was finally achieved in 2011~\cite{RWscience2011}, providing analytic resolution of the $N\rightarrow \infty$ limit. In~\cite{RWscience2011} they prove that any AP on a random graph (with a fixed number of choices) 
leads to a continuous percolation transition in the thermodynamic limit.
In essence they show that the number of subcritical components that join together to form the emergent 
macroscopic-sized component 
is not sub-extensive in system size. There is no ``powder keg". 
Yet, Riordan and Warnke also showed that for an AP on a random graph, if the number of random candidate edges, $m$, is allowed to increase in any way with system size $N$,  so that $m\rightarrow \infty$ as $N\rightarrow \infty$
(for instance, even as slowly as $m \sim \log(\log N))$, 
then this is sufficient to allow for a discontinuous percolation transition.  
In~\cite{Waagen2014}, the authors show that once $m\rightarrow \infty$ as $N\rightarrow \infty$ simple selection rules based on node degree, rather than component size, can lead to EP.  

In summary, rigorous analytic arguments now show that an AP on a random graph with a fixed number $m$ of candidate edges  leads to a globally continuous percolation transition in the thermodynamic limit.  
We use the term ``globally continuous" as introduced in~\cite{RWscience2011} to
mean that the order parameter is a continuous function in both the sub- and super-critical regimes. 
In contrast, an AP with vertex based edge selection rules (rather then edge based rules) can lack global continuity and display instead non-self averaging behaviors and discontinuous percolation transitions as discussed in 
Secs.~\ref{subsubsec:APvertex} and~\ref{sec:type6}. 
For additional details on finite-size scaling, scaling functions and critical exponents for 
EP and percolation more broadly, we refer the reader to a number of recent reviews~\cite{bastas2014review, AraujoPercReview2014, SaberiPhysReports2015, boccaletti2016} 
\\

\paragraph{AP's to enhance percolation}
\ \\

As a final discussion of a standard AP on a random graph we consider {\it enhancing} (rather then delaying) the onset of percolation.  
For instance, consider the Product Rule but where the edge selected is the one that leads to the {\it largest} product of the sizes of the two components. 
This indeed leads to the earlier emergence of a giant component. But the percolation phase transition appears to have the same qualitative properties as the \ER~model.
Enhancing the onset of percolation  
was studied more formally in a variant that inverts the Achlioptas rule~\cite{daCosta2015}.
The authors found a transition similar to the ordinary percolation one, though occurring in less connected systems ($p_c \ll 0.5$).

\subsubsection{AP with $m$-edge rules on lattices}\label{subsubsec:APlattice}
\begin{figure}[tb]
  \begin{center}
  \includegraphics[width=.5\textwidth]{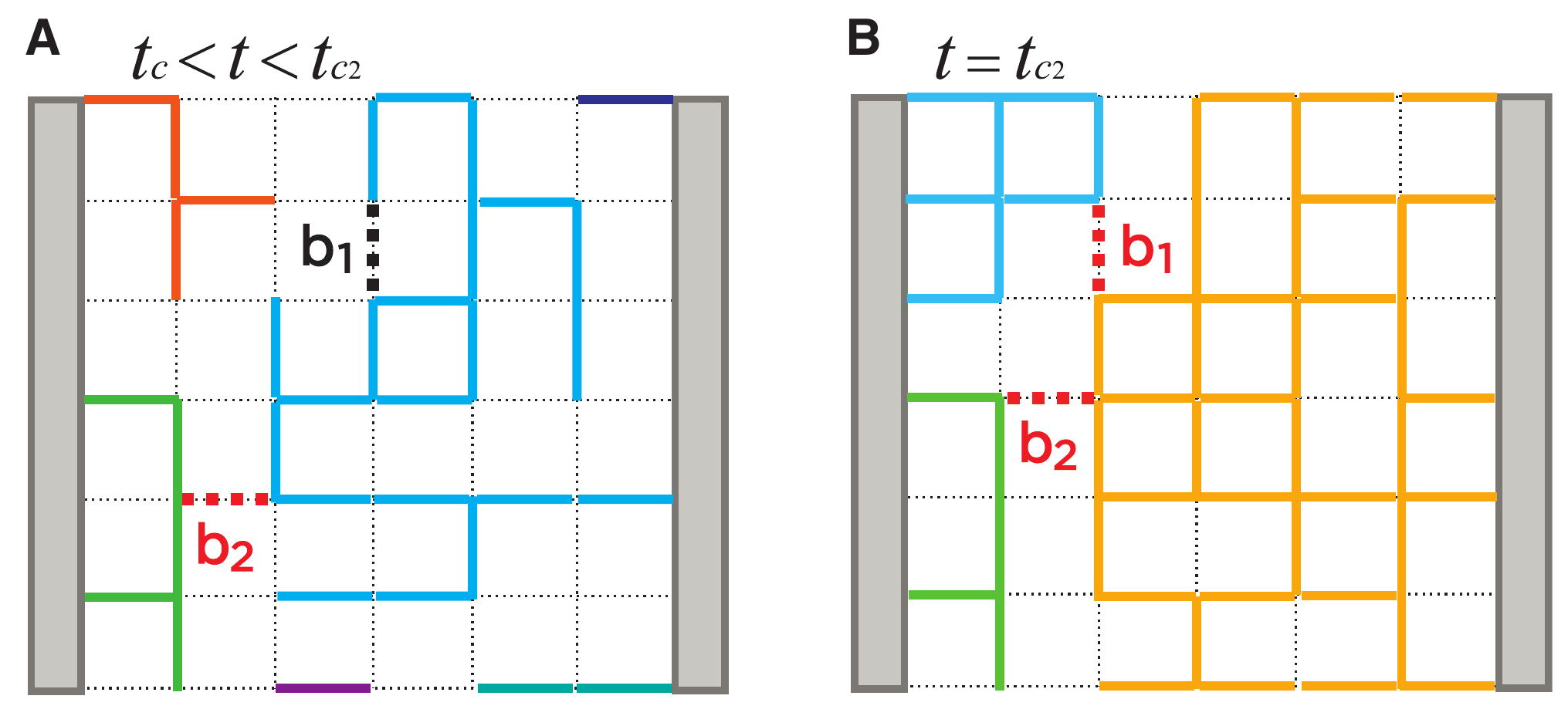}
  \end{center}
  \vspace{-0.2in}
  \caption{The spanning cluster avoidance (SCA) model shown for $m=2$. (a) At a discrete time step, bonds $b_1$ and $b_2$ are randomly selected. If one is a ``bridge bond" that would form a spanning cluster (e.g., $b_2$ here), the other bond is accepted.  (b) When ultimately two bridge bonds are selected, one of them will be chosen at random and added to the lattice forming the spanning cluster.   [Reprinted from~\cite{cho2013avoiding}.]}
\label{fig:choScience}
\end{figure}

On a lattice, an AP with a fixed number $m$ of candidate edges can yield a truly discontinuous EP transition. 
Percolation on a lattice can be quantified by the emergence of a cluster of macroscopic size or by the emergence of a spanning cluster, the latter being a path of occupied bonds or sites that connect sites from one side of the lattice to another. Both the works by~\cite{ZiffPRL09} and~\cite{cho2013avoiding} study bond percolation. But, in~\cite{ZiffPRL09} (as discussed above) an AP is used to delay the growth of a macroscopic size cluster on a 2D square lattice. And, in~\cite{cho2013avoiding} an AP is used to delay the emergence of a spanning cluster, referred to as the spanning cluster avoidance (SCA) model, and is studied on square lattices of various dimensions $d$. The schematic of the SCA model is given in Fig.~\ref{fig:choScience}. 

The SCA model is shown to lead to a discontinuous percolation transition for a lattice with dimension $d < d_c = 6$ as long as the number of candidate edges for the AP is $m \ge m_c$, where $m_c$ is a constant.  Here the critical value $m_c$ is related to $d_{BB}$, the fractal dimension of the ``backbone'' cluster structure which emerges close to the critical point. This is  calculated analytically and measured numerically and it is shown that $m_c = d/(d- d_{BB})$, where $d$ is the dimension of the lattice~\cite{cho2013avoiding} . When $d = 2$, $m_c \approx 2.554$.

An interesting distinction occurs for $m = m_c$ versus $m > m_c$. For $m = m_c$, the discontinuous percolation transition occurs at some intermediate $t_c$ during the process. 
In contrast, for $m > m_c$ the process acts globally, so when the spanning cluster emerges, it encompasses the entire system. 
Such {\it global} percolation also happens for an $m$-edge AP on a random graph in the limit $m\rightarrow \infty$. 
As first shown in in~\cite{RozenEPJB2010} a giant percolating component only emerges in the final step of the process when only one component remains. This process is illustrated in Fig.~\ref{fig:Rozen}(a).
Instead of metric or geometrical confinements, the rule has unrestricted access to the entire collection of components.

\subsubsection{AP with $k$-vertex rules on random graphs}\label{subsubsec:APvertex}

\begin{figure}[tb]
  \begin{center}
  \fbox{\includegraphics[width=.3\textwidth]{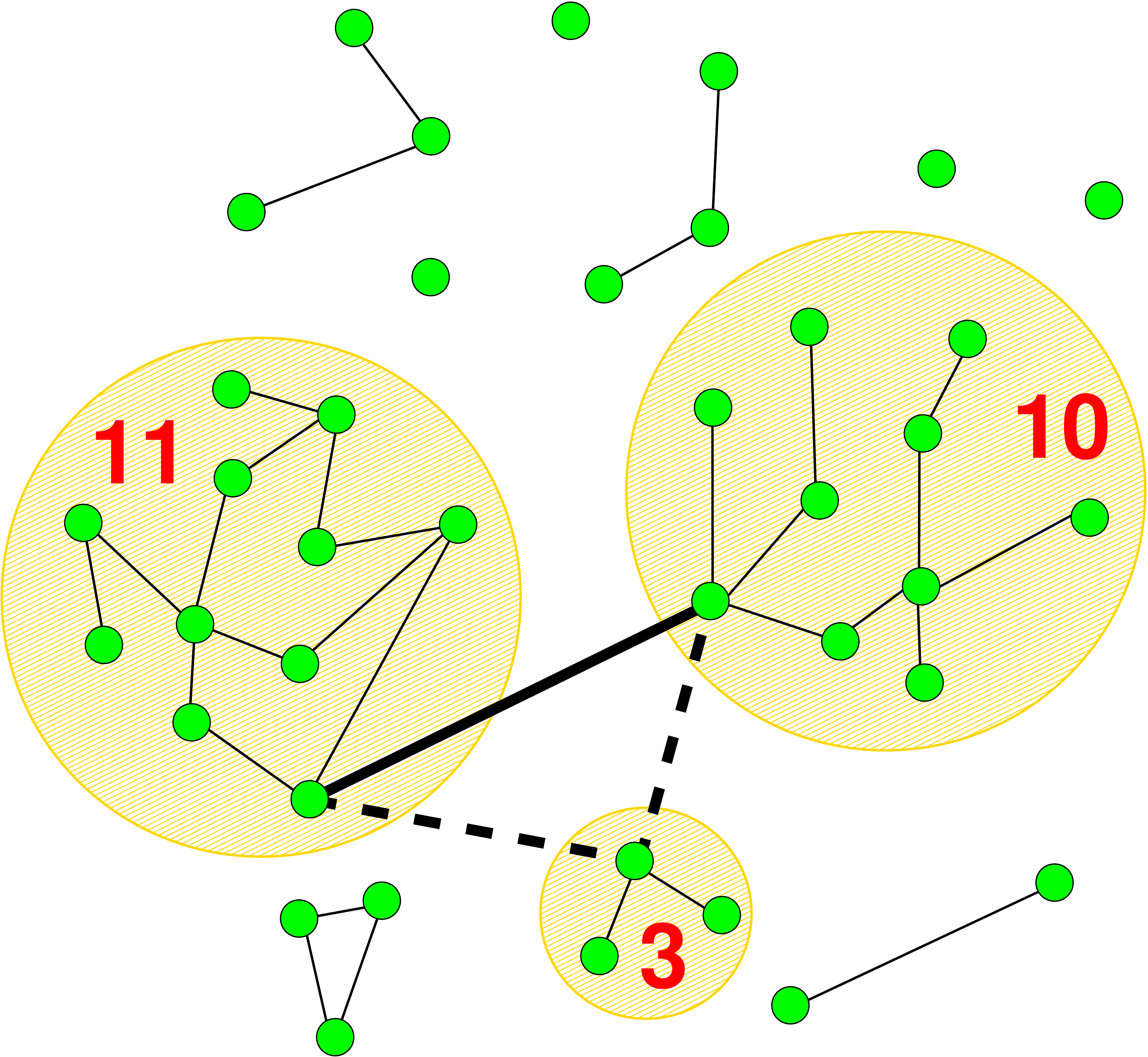}}
  \end{center}
  \vspace{-0.2in}
  \caption{The ``Devil's staircase" (DS) rule, a $k$-vertex rule with $k=3$. Three vertices are chosen uniformly at random and all three possible edges between them are examined. The edge which merges the two components that are most similar in size is added. In the example, the bold edge would be accepted. [Reprinted from~\cite{naglerPRX}.]}
\label{fig:DS}
\end{figure}

The more general class of ``$k$-vertex rules'' (which consider a fixed number of candidate vertices rather than edges) allows for new possibilities~\cite{naglerPRX,schroder2013crackling}. Here, rather than choosing $m$ candidates edges, $k$ vertices are chosen uniformly at random and all of the $k(k-1)/2$ edges possible between them are the candidate edges.   Edge selection criteria that achieve EP from $k$-vertex rules can include preference for keeping components similar in size, rather than aiming to minimize the largest component.  
An example of a $k$-vertex rule for $k=3$ is show in Fig.~\ref{fig:DS}.

As discussed in~\cite{RDJN2015}, this change from random-edge selection to random-vertex based edge selection ensures the ability to prohibit the direct growth of the largest component. For instance, in a $m$-edge rule, so long as there are at least two components in the system, there is a non-zero probability that all $m$ candidate edges have exactly one vertex in the largest component. It this scenario, the largest component would grow directly, thus $P_{Gr}>0$, which is compatible with  
a continuous transition~\cite{naglerNP}. In contrast, for a $k$-vertex rule, if two or more of the $k$ vertices are in the same component, an internal component edge can be added avoiding direct growth and maintaining $P_{Gr}=0$ throughout the percolation process, which necessarily induces a discontinuous transition during the process. This process occurs in the sparse regime, where the number of edges in the graph $T \sim O(N)$, so that the probability of sampling the same edge twice decays as $1/N$. 

For $k$-vertex rules the order parameter necessarily exhibits a continuous initial transition to large-scale connectivity, but a series of secondary discontinuous jumps in the order parameter are possible, with the first discontinuous jump arbitrarily close to the initial continuous transition~\cite{naglerPRX,schroder2013crackling}.  This series of stochastic jumps and ``devil's staircase" transitions demonstrate an array of behaviors worthy of in-depth discussion as found in Sec.~\ref{sec:type6}.

\subsubsection{Thermodynamic formulations of EP}\label{subsubsec:thermo}

In a classic work from 1972, Fortulyn and Kesten (FK) showed how to unify the framework of classic percolation on a lattice and the $q$-state Potts model~\cite{FK1972}. For an explicit treatment see, for instance,~\cite{Grimmett2006}. In~\cite{bradde2009}, the authors generalize those original results of FK to extend the theory to standard percolation on complex networks.  
But this connection is laking for more intricate percolation problems.  
As discussed above, there has been intense study of the scaling exponents characterizing the critical behavior of EP processes.  We refer the reader to~\cite{bastas2014review} for a detailed table of critical exponents. Likewise, a Hamiltonian formulation of EP~\cite{Moreira} is discussed in Sec.~\ref{subsec:EPothers}, which gives insights into the underlying mechanisms of EP.  
In~\cite{daCostaPRE2014} the authors develop a complete scaling theory of the transition for the dCDGM model show in Fig.~\ref{fig:dCDGMmodel}. They indicate the order parameter and the generalized susceptibility, find the full set of scaling relations and functions, including the relations between the critical exponents
and the upper critical dimension.

Most recently, there have been two promising developments.
First, in~\cite{Bianconi2017}, the author develops a large deviation theory of percolation characterizing the response of a sparse network to rare events. The theory reveals that rare configurations of initial damage, for which the size of the giant component is suppressed, leads to a discontinuous percolation transition. A corresponding partition function is developed that allows the formulation of percolation in terms of thermodynamic quantities.  

Second, the recent work of~\cite{Hassan2017} makes a significant advance in formulating $m$-edge Achlioptas Processes as thermal continuous phase transitions. Typically, in past work on random graphs the ``susceptibility" of the random graph percolation process, denoted $\chi$, has been defined as the second moment of the component sizes as shown in Eq.~(\ref{eqn:chi}). 
Instead, in~\cite{Hassan2017} they define susceptibility $\chi$ as a function of the ratio of successive jumps observed in the size of the largest component from addition of single edges.  
Although it is unexpected that such a definition would obey the fluctuation-dissipation theorem of equilibrium statistical physics, with this in place, they define thermodynamic quantities and connect explosive percolation transitions with thermal continuous phase transitions. They derive the associated scaling relationships, show that they obey the Rushbrooke inequality, and that both the Product Rule and the Sum Rule belong to the same universality class. This raises the possibility to analyze other $m$-edge AP models as thermodynamic phase transitions and may point towards the possible existence of larger universality classes in EP.

It should be noted that the results obtained in~\cite{Hassan2017} require that there exists a traditional scaling behavior connecting sub- and supercritical regimes.  Indeed, one of the major achievements of the Wilson-Kadanoff Renormalization Group theory was to prove that such a scaling function provides a smooth connection, see for instance~\cite{wilson1983}. But, as stressed in~\cite{grassberger2011}, one of the most unusual features of EP is that it seems to show completely different scalings in the sub- and supercritical regimes. So still an open question is how to resolve the results in~\cite{grassberger2011} and ~\cite{Hassan2017}. 

An important advance connecting discontinuous percolation transitions and equilibrium thermodynamics is presented in~\cite{bizhani2012}. There they provide a thermodynamic formulation for a discontinuous percolation transition expressed in terms of ``Hamiltonian" graphs (i.e. exponential random graph models) and, more specifically, a partition function. This allows for precise definition of thermodynamic quantities.  
Here chemical potentials control the number of edges, triangles and two-stars present in the graph at thermal equilibrium. They show that, for ranges of the chemical potential, the percolation transition can coincide with a first-order phase transition in the density of links, and also exhibits hysteresis loops of mixed order. Although it is a discontinuous percolation transition, it does not appear to be in the class of EP transitions which require microscopic evolution rules that delay the formation of macroscopic components. Yet the framework in~\cite{bizhani2012} may provide a promising direction for future research into the thermodynamics of explosive transitions.  

\subsection{EP from alternate processes}\label{subsec:EPothers}

There are now several graph evolution processes known to give rise to EP transitions that do not involve direct competition between edges.  These are reviewed here and reveal common mechanisms of delaying macroscopic connectivity 
and the common theme of the interplay of local versus global information, which is yet to be fully understood.

 \begin{figure}[b]
  \begin{center}
  \includegraphics[width=.43\textwidth]{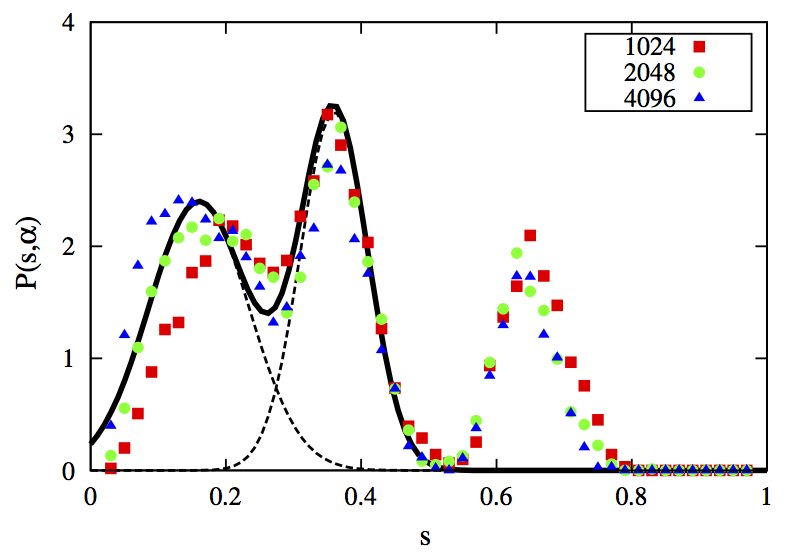}
  \end{center}
  \vspace{-0.2in}
  \caption{Cluster size distribution for the Gaussian model showing coexisting peaks. [Reprinted from~\cite{Araujo}.]}
\label{fig:Aruajo1}
\end{figure}

\subsubsection{Suppressing the largest component} 

Several approaches achieve EP with a truly discontinuous onset of the giant component by mechanisms that directly suppress the growth of the largest component, without need for edge competition.  In the ``Gaussian model" of~\cite{Araujo}, a regular lattice is considered for the the underlying substrate and a single edge is examined at a time (in other words $m$ = 1). If the randomly chosen edge would not increase the current size of the largest component then it is accepted. Otherwise it is rejected with a probability function that decays as a Gaussian distribution centered on the average cluster size, written as 
\be
\min \left\{1, \exp \left[-\alpha \left( \frac{s - \la s \ra}{\la s \ra} \right)^2 \right] \right\},
\ee
where $\la s \ra$ is the mean cluster size.
Thus, forming components that are similar in size to the average is favored. Clear signatures of a first-order transition are observed, such as bimodal peaks for the cluster size distribution, indicating the coexistence of percolating and non-percolating regions in finite systems at $t_c$ as shown in Fig.~\ref{fig:Aruajo1}. 

\begin{figure}[tb]
  \begin{center}
  \includegraphics[width=.43\textwidth]{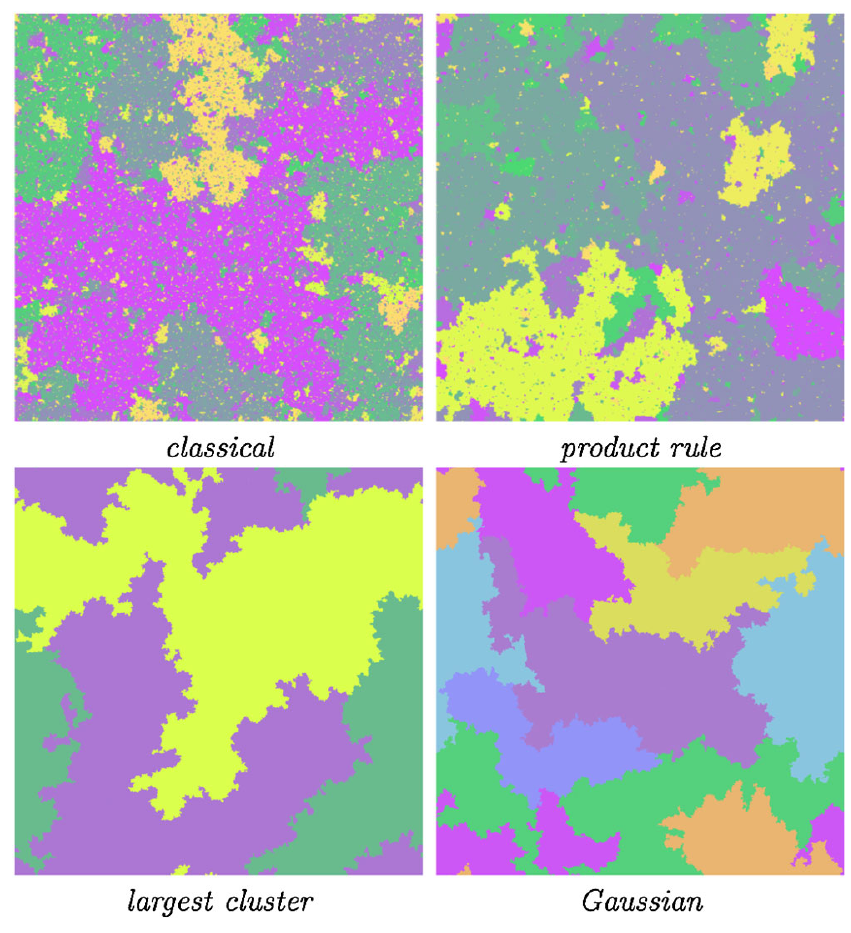}
  \end{center}
  \vspace{-0.2in}
  \caption{Examples of the clusters that form on a square lattice with $1024^{2}$ sites under various processes. [Reprinted from~\cite{Araujo}.]}
\label{fig:Aruajo2}
\end{figure}

Shown in Fig.~\ref{fig:Aruajo2} are examples of the clusters that form on a square lattice with $1024^{2}$ sites.  In contrast to the lattice model, the equivalent random graph version of this Gaussian model exhibits a discontinuous transition at the end of the process when the system condenses into a single component~\cite{Araujo}.

In~\cite{Moreira} a Hamiltonian formalism is developed, providing a connection between equilibrium statistical
mechanics and EP. They have shown that the key for obtaining a discontinuous percolation transition is that the size of the
growing clusters should be kept approximately uniform. More surprisingly, they show that for random graphs, adding edges which merge together previously distinct components should dominate over adding edges internal to a component.

For more discussion of EP processes achieved by control of only the largest component,  in particular by suppressing the growth of a cluster differing significantly in size from the average one, see~\cite{Herrmann2011}

\subsubsection{Tuning edge rejection rate (The BFW model)}\label{subsubsec:BFW}

A model proposed by Bohman, Frieze and Wormald (BFW) also achieves a discontinuous percolation transition by suppressing growth of the largest component and similarly considering only a single edge at a time, $m$=1~\cite{BohFriezWormRSA2004}. Yet, the BFW model displays a broad array of phenomena worthy of discussing in its own right.  

In the BFW model the system is initialized with $N$ isolated vertices, a cap on the maximum allowed component size set to $k=2$, and a budget for rejecting undesired edges where the number of edges that can be rejected  
increases initially with increasing $k$.  In detail, a random edge is sampled from the graph, if it would not lead to an increase in $|C|$ the edge is accepted.  Otherwise the edge is rejected so long as a stringent lower bound on the number of edges that must be accepted, denoted $g(k)$, is maintained. If the lower bound would be violated, instead the cap $k$ is repeatedly increased incrementally to $k+1$ until either the cap is large enough that the edge can be accepted, or $g(k)$ has decayed sufficiently that the edge can be rejected.  

\begin{table}[t]
\caption{{\bf The BFW algorithm}. At each discrete time step $u$, the randomly selected edge $e_u$ is examined via this algorithm, where $u$ denotes the total number of edges sampled, $A$ the set of edges which have been accepted thus far (initially $A =\emptyset$), and $T=|A|$ the number of accepted edges. }
\begin{center}
\begin{tabular}{|l|}
\hline
{Set $l=$ maximum size component in $A \cup \{e_{u}\}$}\\

if $\left(l\leq k\right) \{$\\
 
 \ \ \ $A \leftarrow A \cup \{e_{u}\}$\\
 
 \ \ \ $u\leftarrow u+1.$ (Get next edge.)\}\\
  
   else if $\left(T  / u \ge g(k)\right) \{u \leftarrow u+1 $.  (Get next  edge.)\}\\
 
  else \{\ $k\leftarrow k+1$. Then repeat this block.\}\\ \hline

\end{tabular}
\label{BFWalg}
\end{center}
\end{table}%

The formal BFW process is defined in Table~\ref{BFWalg}, along with the standard notation used in the literature. 
So long as the fraction of accepted edges $T/u \ge g(k)$, then any candidate edge can be rejected. Here $T$ represents the number of edges accepted, and $u$ the number of discrete time steps into the process. 

The lower bound $g(k)$ can be written in the most general terms as:
\be
g(k) = \alpha + (2k)^{-\beta}
\label{eqn:BFWalpha}
\ee
where $\alpha$ and $\beta$ are parameters.  In the original BFW model, these are not parameters, but simply constants $\alpha = 1/2$ and $\beta=1/2$~\cite{BohFriezWormRSA2004}. 
The system is initialized with the cap $k=2$, thus $g(k=2)=1$ and initially all edges must be accepted.  The lower bound decays with increasing $k$ reaching the asymptotic limiting value of $\lim_{k \rightarrow \infty} g(k)=1/2$, 
meaning that asymptotically greater than or equal to one-half of all sampled edges must be accepted.

Note the tradeoff utilized by the BFW model. Early in the process when all components are small almost all edges must be accepted.  But as the components grow in size, and rejection becomes more crucial for avoiding the formation of a giant component, we can reject more edges up until reaching the limiting value $g(k \rightarrow \infty)=1/2$.  

\begin{figure}[b]
  \begin{center}
  \includegraphics[width=.35\textwidth]{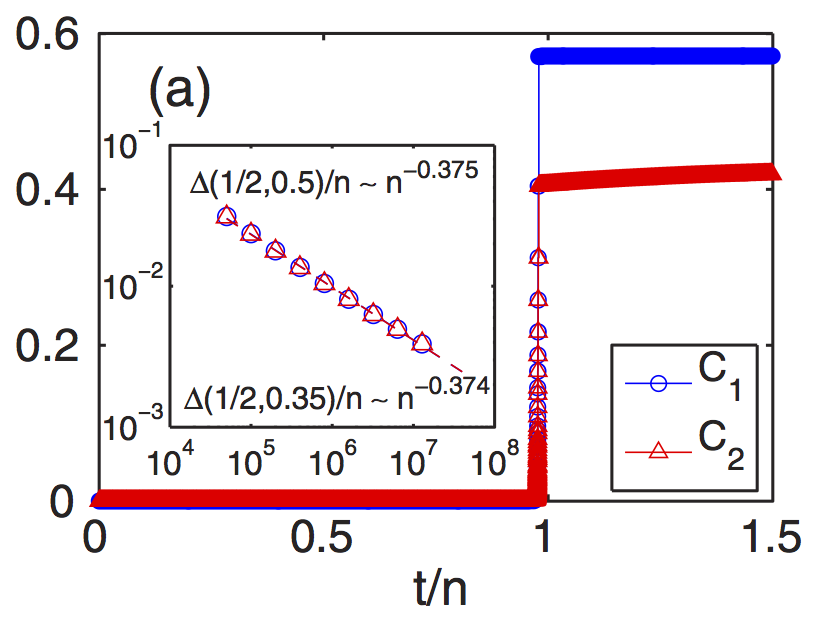}
  \includegraphics[width=.35\textwidth]{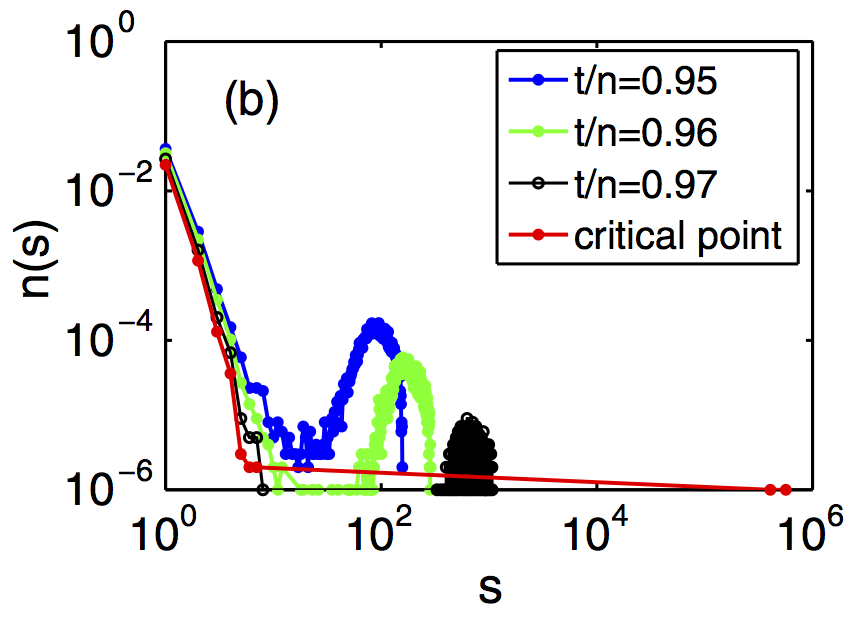}
  \end{center}
 \vspace{-0.3in}
  \caption{(a) Simultaneous emergence of two stable giant components in the BFW model. (b) The evolution of the powder keg in the BFW model, showing the merger at $t_c$. Here $t_c=0.976$.  [Reprinted from ~\cite{ChenPRL2011}.]}
\label{fig:BFW1}
\end{figure}

The critical behavior of the BFW model was analyzed in~\cite{ChenPRL2011}, where the authors show that the BFW model leads to the simultaneous emergence of two co-existing giant components in a truly discontinuous percolation transition. Details are shown in Fig.~\ref{fig:BFW1}. They also show that if edges are sampled uniformly at random from the complete graph, the co-existing giants are asymptotically stable. (Any edge that would lead to merging the two co-existing giants can always be rejected by a small sub-linear increase in the cap $k$.) If edges that are sampled
are restricted to only those that join previously disjoint components, the same sub-critical behavior is observed 
but eventually in the supercritical regime the two components necessarily merge together causing a discontinuous jump in $|C|$.  

\begin{figure}[t]
  \begin{center}
  \includegraphics[width=0.65\textwidth]{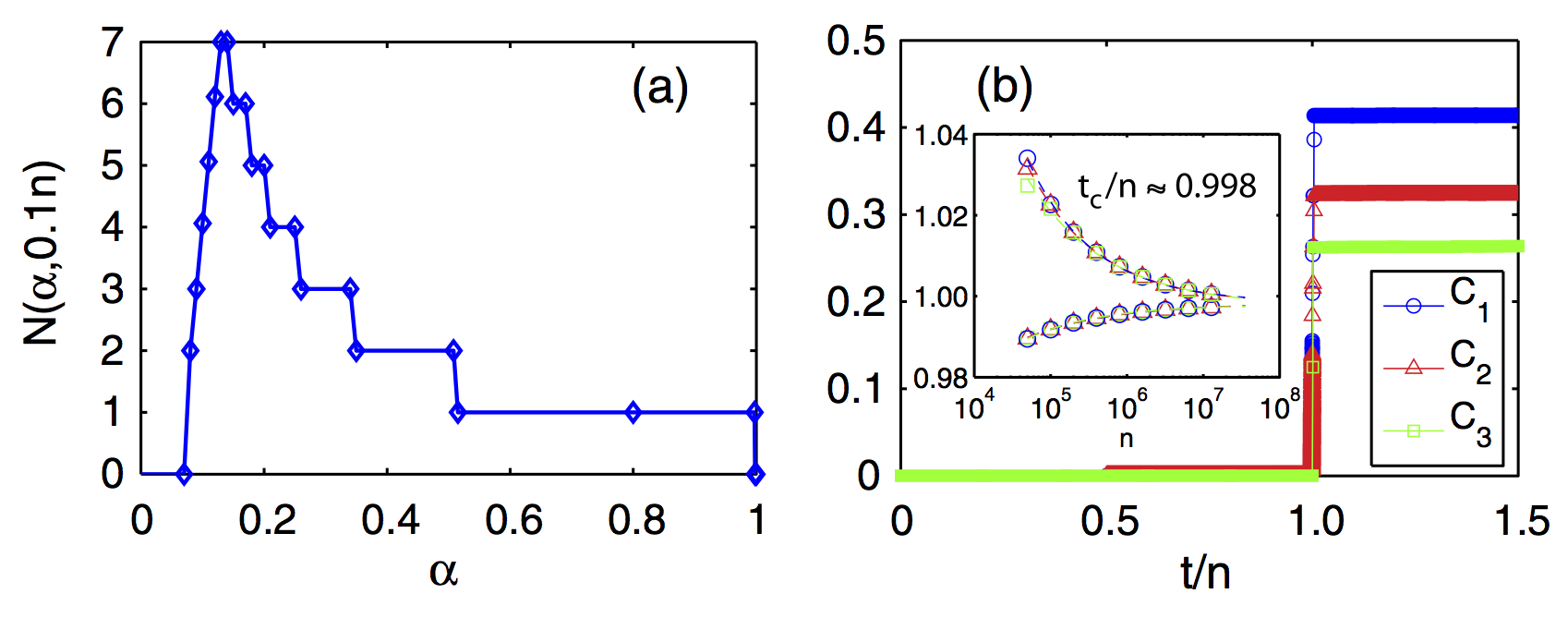}
  \end{center}
  \vspace{-0.3in}
  \caption{(a) For the generalized BFW model shown in Eq.~(\ref{eqn:BFWalpha}), the asymptotic fraction of accepted edges $\alpha$ determines the number of stable giant components. (b) Simultaneous emergence of three stable giant components for $\alpha=0.3$. [Reprinted from ~\cite{ChenPRL2011}.]}
\label{fig:BFW2}
\end{figure}

In~\cite{ChenPRL2011} the authors introduce a generalized version of the original BFW model, with the $\alpha$ parameter shown in Eq.~(\ref{eqn:BFWalpha}). 
Thus the limiting fraction of accepted edges is bounded from below, not by a constant, but by a parameter $\alpha$.  They show that the 
value of $\alpha$ determines the
number of coexisting giant components that simultaneously emerge in a discontinuous transition, 
as shown in Fig.~\ref{fig:BFW2}.

In~\cite{ChenEPL2012} the authors further generalize the BFW model 
by introducing the $\beta$ parameter shown in Eq.~(\ref{eqn:BFWalpha}).
They show that the value of $\beta$ governs 
the extent of direct growth allowed for the largest component. 
For $\beta < 1$ it is always possible to reject a sampled edge that would lead to significant direct growth. Instead the  growth of the size of the largest component is dominated by overtaking, when two smaller components merge together to become the new largest, giving rise to a discontinuous percolation transition.  In contrast, for $\beta > 1$ once the cap size $k\ge N^{1/\beta}$ there are situations when a sampled edge cannot be rejected and the largest component experiences significant direct growth, leading to a continuous percolation transition.  

The BFW model displays a rich set of phenomena. Of particular interest is that $\alpha$ controls the number of co-existing giant components, and that $\beta$ controls the extent of direct growth of $|C|$ allowed.  The BFW model has now been analyzed on 2D and 3D lattices, showing strong evidence of a discontinuous transition including compact clusters with fractal surfaces~\cite{SchrenkPRE2012}.  The supercritical properties and additional phase transitions of the BFW model on various substrates have also now been studied in for instance~\cite{ChenPRE2013a,ChenPRE2013b}.

\subsubsection{Cluster aggregation}\label{sec:cluster}

A great number of models that display EP phenomena have been built on cluster aggregation processes, thereby ignoring the intra-cluster link structure
of the underlying network.
Sec.~\ref{subsubsec:ClusterAgg} contains a discussion of the approach and the foundational Smoluchowski rate equation, Eq.~(\ref{eq:SmEqu}), along with the definition of the collision kernel $K_{ij}$ which is proportional to the probability that components of size $i$ and $j$ merge per unit time. 

The solution of the rate equation is still unknown for the great majority of collision kernels $K_{ij}$, and analytical insights are difficult. 
In linear polymerization, two clusters are merged by the molecules at two reactive ends, where the kernel $K_{ij}$ is a constant. 
When two clusters have a compact shape and merge, the kernel is given by $K_{ij}\sim (ij)^{1-1/d}$, where $d$ is the spatial dimension. 
A series of detailed 
papers based on this approach have had a deep impact on our understanding of EP systems~\cite{ChoPRL09,ChoClusterAgg,ChoKhangPRE2011,daCosta2010,KahngArxiv2014,nagler2016a}.

Cho {\it et al.}\ considered the important and more general case  of a power law  kernel,
 $K_{ij} \sim (ij)^{\omega}$, where $\omega$ is a positive exponent~\cite{ChoClusterAgg}. 
Models with $0.5< \omega \le1$ account for steric hindrance and intramolecular bonding~\cite{leyvraz1982, ziff1982, ziff1983, leyvraz2003}.
Below the critical value $\omega_{\text{c}}=0.5$ 
aggregation based on Eq.~(\ref{eq:SmEqu}) exhibits a violation of mass conservation.
In addition, for $\omega <0.5$ gelation does not take place in finite time~\cite{leod1962}.
For the modified Smoluchoswki equation with
a (normalized) power law kernel 
\begin{equation}\label{eq:powerlawkernel}
{K}_{ij} =(ij)^{\omega}/(\sum_{s}s^{\omega}n_s)^2
\end{equation}
total mass is conserved  and gelation is guaranteed to occur in finite time~\cite{ChoClusterAgg}.
The authors showed that for $\omega > 0.5$ the transition is continuous but discontinuous for $\omega<0.5$. %

As discussed earlier, controlling the growth of the largest cluster constitutes a key mechanism to delay percolation~\cite{Araujo}.
This motivated Cho and coworkers to  study cluster aggregation where the collision rate of the {\em largest} cluster is controlled. 
Specifically, 
the exponent is taken as $\omega=\alpha$ for all reactions that do not involve the largest cluster,
and otherwise $\omega=\beta$, for the general case $\alpha \neq \beta$.
This model, as well as generalizations of it, show an extremely rich
 phenomenology of four distinct phase transition types, depending on the combination of $\alpha$ and $\beta$ (Ref.\ ~\cite{nagler2016a}), namely
continuous percolation, discontinuous percolation at the very end of the process, ultra-slow convergent discontinuous transition, 
and a non-self-averaging staircase behavior. 
We will come back to this phenomenology in Sec.~\ref{sec:type6}. 
More generally, these anomalous behaviors are expected for aggregation processes
with two (or more) coalescence time scales, as artificially introduced by the two exponents~\cite{nagler2016a}.

In~\cite{KahngArxiv2014} they derive the necessary conditions for the occurrence of a genuinely discontinuous phase transition for a broad class of processes that use cluster aggregation approaches. 
Their theory not only predicts discontinuous transitions, but can also tell the type of the transition.
They showed that the key characteristic of whether or not the cluster kinetic rule is homogeneous with 
respect to the cluster sizes determines the phase transition type. 
This only requires examination of the cluster size distribution immediately before the transition. 
The theory is based on a mean-field two-species cluster aggregation model 
but may hold for a much wider range of models. As a limitation, however, the theory is expected to
be non-exact for percolation processes on networks with non-random locally non-loop-free topologies, on lattices,
or for other correlated or clustered underlying structures~\cite{son2011,ccgg15,grassberger2015,grassberger2016}.

A related model to cluster aggregation is called ``agglomerative percolation"~\cite{christensen2012}. Here, clusters  
exist on a two-dimensional lattice, but rather than diffusive motion, a discrete time process is considered. Starting from a collection of isolated clusters of size one, at each time step a random cluster is picked and bonds are added to the entire surface of the cluster in order to link it to all adjacent clusters. This is motivated in part to mimic the non-locality seen in an Achlioptas Process: If a cluster of length scale $l$ is picked, links are added simultaneously at distances $\mathcal{O}(l)$ apart. The process proceeds until the system is reduced to one cluster.  Similar to multiplicative coalescence, if clusters are picked with probability proportional to their mass, this leads to a single ``runaway" compact cluster.  If instead all clusters are equally likely to be chosen, these leads to a continuous transition in a new universality class for the square lattice, while the transition on the triangular lattice has the same critical exponents as ordinary percolation showing intriguing violations of standard universality classes.

\subsubsection{Correlated processes}\label{subsubsec:correlated}

The role that correlations can play in creating discontinuous percolation on a lattice has been known for some time for models of jamming, e.g.,~\cite{PhysRevLett.96.035702,Toninelli2008,jengPRE2010}.  But with the growing interest in EP phenomena, in~\cite{cao2012correlated} they explore the connections between such known models and EP.  In particular they analyze models that are a mixture of the correlated models and more traditional models of percolation, and they search for tricritical points separating the region of discontinuous from continuous transitions. For random graph models, they analyze a $k$-core percolation  for a mixture of $k$=2-core and $k$=3-core vertices, and find a tricritical point. But for two-dimensional lattice models the behavior is not  so clear, showing crossover behaviors but no tricritical point.  Yet the work suggests interesting connections between EP, glassy dynamics, and jamming. 

Cooperative interactions and the existence of tricritical points is also explored in~\cite{bizhani2012}. In particular they analyze the tricritical point of a generalized epidemic process which maps onto a form of complex social contagion.  
Note several models of explosive phenomena in cooperative epidemics have also been introduced and analyzed, as discussed in-depth in Sec.~\ref{subsec:explosiveEpidemics}.  

\begin{figure}[tb]
  \begin{center}
\includegraphics[width=.21\textwidth]{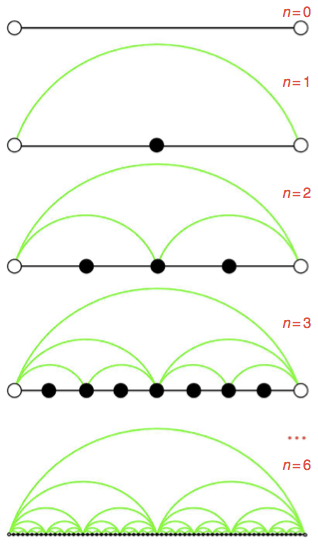} \ \ \ \ \ \ \ 
\includegraphics[width=.23\textwidth]{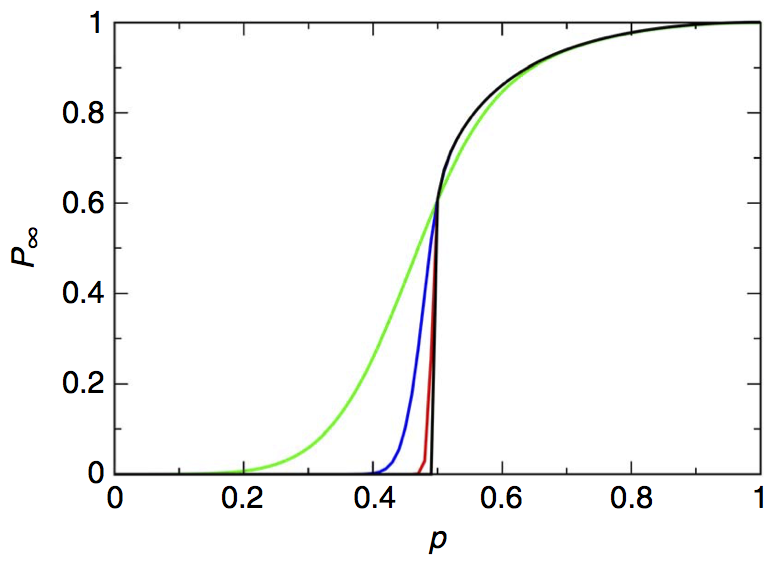}
  \end{center}

  \vspace{-0.1in}
\hfill{\small (a)} \hspace{1.6in} \  {\small (b)} \hfill\hfill\hfill
  \caption{(a) Recursive generation of the hierarchical network studied in~\cite{boettcher2012}. (b) Even ordinary percolation leads to a discontinuous percolation transition on this network structure. Plotted is the order parameter with increasing recursive iteration of the hierarchical structure.  [Both (a) and (b) are reprinted from~\cite{boettcher2012}.]}
\label{fig:hierarchical}
\end{figure}

\subsubsection{Hierarchical lattices}

In~\cite{boettcher2012} the authors further illustrate the interplay between local and global information by examining simple percolation on a hierarchical lattice. In particular they analyze a one-dimensional lattice dressed up with a hierarchy of long-range bonds as shown in Fig.~\ref{fig:hierarchical}.  It is well known that percolation on a one-dimensional lattice is trivial process with percolation only in the limit that the bond occupation probability $p=1$.  Boettcher {\it et al.}\ recursively build up a hierarchical structure and remark that it is similar to a hyperbolic geometry where most nodes, as in a tree, are close to the periphery.  They show an extensive percolation cluster arises for $p < 1$ and emerges in an instantaneous jump to a finite value. 
Although it is not yet clear whether the transition follows the paradigm of EP phenomena, which requires microscopic dynamics that delay formation of macroscopic components, this model by Boettcher {\it et al.}\ provides a systematic and rigorous model to explore the interplay of local and global information that is crucial to EP but is, as of yet, still 
not fully explored.

\subsection{Categories of EP transitions}\label{subsec:Anomalous}

EP transitions have been shown to display a variety of anomalous behaviors. 
Here we attempt to categorize the types of behaviors observed.
Our motivation is to provide an umbrella that covers the main types of the critical and supercritical phenomena that have emerged from the extensive body of work on EP.
It is important to note that this current classification, into only six classes, is limited and could be further expanded to incorporate more fine-grained details or  
as new phenomena are discovered. 

As we are concerned with general universality classes, throughout this section we use the generic notation $\Y$ to denote the order parameter, which is the fractional size of the largest component, and $p$ to denote the control parameter.

\subsubsection{Type I: 
Anomalous continuous
 transition \label{sec:type1}}
 
\begin{figure*}[t]
\includegraphics[width=0.68\linewidth]{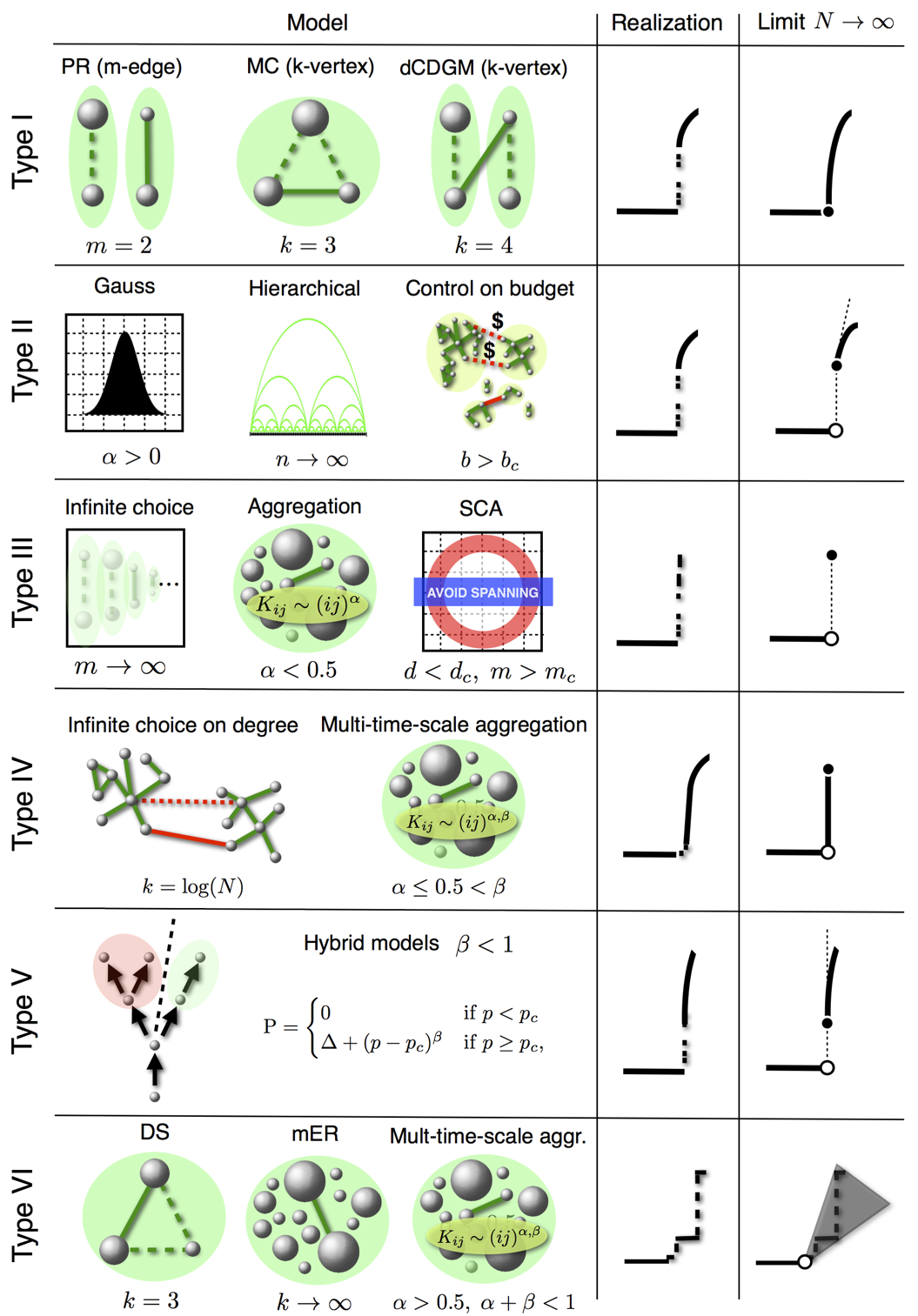}
\caption{
Classes of Explosive Percolation.
{\bf Type I}, the product rule (PR)~\cite{EPScience}, minimal cluster rule (MC)~\cite{powderkeg09} and the dCDGM model~\cite{daCosta2010}
are examples of EP
processes that are continuous in the thermodynamic limit but nevertheless exhibit substantial jumps in $|C| /N$  for any finite system. 
For $m$-edge rules, $m$ links compete for addition. For $k$-vertex rules, all possible $k(k-1)/2$ node pairs compete. 
{\bf Type II}, models that exhibit a single genuine jump in the order parameter $\Y$ well in advance of the end of the process. 
The hierarchical model is based on  $n$ generations of long-range bonds, in the limit of $n\rightarrow\infty$. 
If percolation is controlled on a budged, for a sufficiently large budget $b>b_c$,
a genuine discontinuous transition is observed~\cite{nagler2017}.
{\bf Type III}, models (infinite choice~\cite{RozenEPJB2010}, cluster aggregation~\cite{ChoClusterAgg}, SCA~\cite{cho2013avoiding})
 that exhibit a single discontinuous jump in the order parameter $|C| /N$ at the end of the process, 
resulting in a jump encompassing the full system. 
{\bf Type IV}, same as in Type III, except that single step gaps vanish and the single discontinuous jump in the order parameter 
 at the end of the process results from an infinite slope, rather from the addition of a single link.
This jump is usually approached very slowly, as shown for the infinite choice on degree model~\cite{Waagen2014}
and multi-time-scale aggregation~\cite{nagler2016a}. 
{\bf Type V}, hybrid transitions were found in models as diverse as half-restricted percolation~\cite{panagiotou2011explosive},
$k$-core percolation~\cite{SchwarzLiuChayes06} and cascade failure model in multiplex networks~\cite{clzgb13}.
A critical and a supercritical branching process underly hybrid transitions~\cite{lee2017},
with a specific behavior of the order parameter $P$.
{\bf Type VI},  non-convergent, non-self-averaging models that exhibit a staircase with genuinely discontinuous steps, 
including the devil's staircase (DS)~\cite{naglerPRX}, 
 modified \ER\ (mER) model~\cite{riordanPRE2012}, and multiple-time-scale aggregation~\cite{nagler2016a}. 
In the thermodynamic limit the staircases are stochastic (both in size of steps and location). 
}
\label{fig:EP_classes}
\end{figure*}

We call a continuous transition that 
exhibits a significant 
macroscopic jump in size of the order parameter for any large but finite system  
by the term anomalous continuous,
or for shorthand, we say it is of type I (see Fig.~\ref{fig:EP_classes}).
A type I transition is typically characterized by the scaling of the largest gap in the order parameter as
$\Delta C_{\text{max}} \sim N^{-\beta}$, for $\beta \ll 1$.  
This property has been used as a definition for a so-called weakly discontinuous transition~\cite{RDJN2015}.
Here we further require that for a type I transition the order parameter $P$ must  have a finite slope at the critical point. This is because a gap in scaling can also be compatible with a diverging slope of $P$ (which is a combination that we call instead a type IV transition in Sec.~\ref{sec:type5}).
As a more specific example of type I, consider 
some competitive rule that leads to EP and shows a weak decay of the largest gap scaling, say,
$\Delta C_{\text{max}}\sim  N^{-0.01}$ and no apparent divergence of slope of $P$.
As discussed in~\cite{RDJN2015} this means that even for system sizes of the order of Avogadro's number, $N\approx 10^{23}$, well beyond the size of real-world networks,
the EP transition would still be effectively indistinguishable from a process exhibiting a genuine discontinuous transition.

Anomalous continuous EP transitions possess atypical finite-size scaling behaviors, as seen in~\cite{grassberger2011,tian2012nature},
which are very different from classic continuous percolation.
Once again we refer the reader to a number of recent reviews for extensive details regarding finite-size scaling, scaling functions and critical exponents for processes that lead to EP~\cite{bastas2014review, AraujoPercReview2014, SaberiPhysReports2015, boccaletti2016} and the literature therein, e.g.,~\cite{daCostaPRE2014}.

\subsubsection{Type II: Discontinuous phase transition \label{sec:type2}} 

A genuine
discontinuous phase transition is given by a finite jump $\Delta$ of the order parameter $\Y$ at the percolation threshold $p_c<1$ as follows:  

 \begin{equation}\label{eq:typeI}
 \Y = 
\begin{cases}
0 &\text{if } p <  p_c\\
\Delta + f(p)  &\text{if } p \ge p_c,
\end{cases}
  \end{equation}
where $f(p)$ is a positive function of $p$ and describes the growth in the supercritical region.
Note that Eq.~(\ref{eq:typeI}) (and the  ones that will follow for the other types) hold in the thermodynamic limit.
For $f(p)=(p-p_c)^{\beta}$ and $0 < \beta < 1$, the discontinuous transition coincides with a second order transition
 (see Type IV, hybrid transitions). 

Typically, however, the growth in the supercritical region, $f(p)$, is not a power law function, or
 if it is one, it is characterized by $\beta \ge 1$, leading to a finite slope of $P$ for $p\rightarrow p_c^+$.
The required finite slope of $P$ as approached from below the critical point 
is indicated by the dashed line in Fig.~\ref{fig:EP_classes}.

For type II, the percolation threshold $p_c$ is required to be not at full link (or occupation) density,  which implies that
there exists an extended supercritical region.
From the perspective of the kinetic approach this means that the discontinuity is required {\em not} to be at the end of the process, or in other words, gelation does occur.

Worth noting is that the majority of models of EP are not of this type II, but are either continuous, lack a supercritical region, or are of other types.
Nevertheless, there are a number of models that show evidence for a type II transition on the lattice, in particular using finite size scaling~\cite{ZiffPRL09}.
However, given the challenges arising from  small exponents, for the majority of models it remains open if
there is a discontinuity in the thermodynamic limit, or just a weak discontinuity~\cite{ZiffPRL09}.

For the Gaussian model~\cite{Araujo} and the
spanning cluster avoidance (SCA) model, for $d<d_c$ and $m=m_c$~\cite{cho2013avoiding},
numerical evidence indicates a type II transition.
As shown in~\cite{boettcher2012}, ordinary percolation on a hierarchical network can be of type II, where small-world bonds are grafted onto a one-dimensional lattice.

Schr\"oder and coworkers 
studied the emergence of EP 
from a dynamical process with link rejection on a budget specified as $b$.
An efficient  link rejection protocol is used to maximally delay percolation and 
it is assumed that  rejections are costly.
If the budget is small, percolation is delayed but, in contrast to other EP models, it remains in the same universality class as continuous percolation. 
For sufficiently large budgets, however, the percolation is maximally delayed 
and the transition is genuinely discontinuous~\cite{nagler2017} at $p_c<1$
 (see Fig.~\ref{fig:EP_classes}). 
 
The BFW model discussed in~\ref{subsubsec:BFW} also implements EP by use of a ``budget". Such mechanisms  
break the multiplicative coalescence of classic percolation and allow for the stable coexistence of multiple giant components. As shown in detail for the BFW model, a tunable parameter $\alpha$ (the asymptotic edge rejection rate) controls the number of stable coexisting giants. 
Coexisting giants in the supercritical region is a novel feature of EP transitions which provides connections with
modular organization and community structure in networks~\cite{RozenEPJB2010,PanPRE2011}, with more details in Sec.~\ref{subsec:realworld}. 


\subsubsection{Type III: Discontinuity at the end of the process \label{sec:type3}}

In many models of EP, percolation is delayed to such an extent  that, in the thermodynamic limit,
it takes place just at the end of the process. 
We assume this to coincide with full link density $p_c=1$, 
as it is on lattices and for many cluster aggregation processes
(otherwise replace $p_c=1$ by some $p_c^{\text{full}}$).
Thus,
 \begin{equation}\label{eq:type3}
 \Y = 
\begin{cases}
0 &\text{if } p <  1\\
1 &\text{if } p = 1.
\end{cases}
\end{equation}
In the kinetic approach, full link density indicates the end of the process (as discussed in Sec.~\ref{subsubsec:ClusterAgg}),
or equivalently that a giant component only emerges at the very end of the process and gelation does not occur. 

Examples for a type III transition have been discussed in the literature~\cite{boettcher2012, RDJN2015} and include 
1d percolation, $m$-edge rules with infinite choice (as shown in Fig.~\ref{fig:Rozen}(a)), cluster aggregation for power law kernels $K_{ij}\sim (ij)^{\alpha}$,
 for $\alpha<0.5$ (section~\ref{sec:cluster}), and
the SCA model (discussed in Sec.~\ref{subsubsec:APlattice}) below the critical dimension $d<d_c$ but for sufficiently large choice $m>m_c$.
A type III transition shows an order parameter jump at the end of the growth process (which encompasses the entire system) and  
lacks a supercritical region, see Fig.~\ref{fig:EP_classes}.

\subsubsection{Type IV: Single-step-continuous discontinuity \label{sec:type4}} 

In the random network model introduced by Waagen and D'Souza~\cite{Waagen2014}, $k$ nodes are sampled at random and their respective degrees are considered. The model preferentially favors connecting nodes of small degree. 
If the choice parameter $k$ is an increasing function of system size, 
 then a discontinuous transition occurs at the end of the process. 
Hence, the model reproduces the prediction of Riordan and Warnke regarding $k$-vertex rules 
 for increasing $k$~\cite{RWscience2011}.
Perhaps the most remarkable particularity of the model lies in the finite size gap behavior. 
For increasing system size, 
single step gaps in the order parameter vanish in the thermodynamic limit.
Yet, a genuine discontinuity emerges at the end of the process,
which results from an diverging slope of $P$, see Fig.~\ref{fig:EP_classes}.
Thus, the order parameter is described by Eq.~(\ref{eq:type3}), yet 
the largest gap in the relative size of the largest component $\Delta C_{\text{max}}$ decreases as a function of $N$. 

\begin{figure}[tb]
\includegraphics[width=0.45\textwidth]{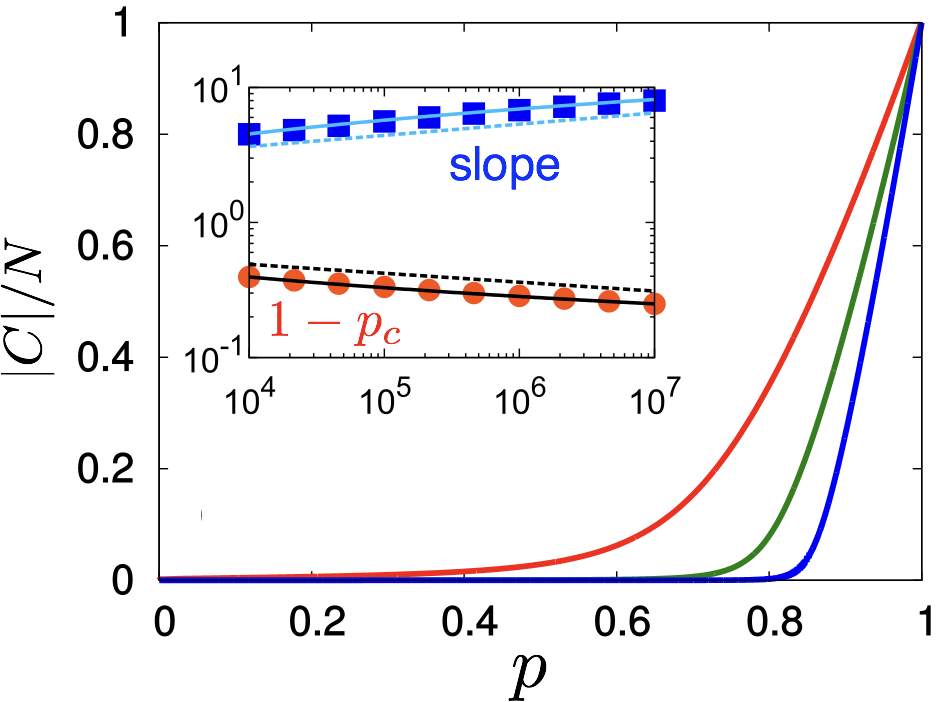}
\caption{\label{fig:type4stat}
Single realizations of the slope and gap scaling for type IV.
Adapted from Ref.~\cite{nagler2016a}.  
Shown are results for the model parameters $\alpha=0$, $\beta=1$.
$\Y$ vs $p$  for  system sizes $N = 10^3$, $10^5$, and $10^7$.
Inset: 
Single-step gaps scale away but the slope of the order parameter is increasing with $N$.
The transition point convergence is extremely slow.
Plot of $1-p_{c}$ vs $N$ ($\bullet$). Dashed line is power law with exponent $-0.07$.
Solid line is a logarithmic dependence for $\sim{\log(x)}^{-0.82}$ for comparison.
Plot of slope $dP/dp|_{\text{max}}$ (Ref.\ 
\cite{nagler2016a}) vs $N$ $(\blacksquare)$.
Dashed line has slope $0.08$. 
Here, $p_c(N)$ is taken as $\text{argmin}_p (|C_1|\ge N^{1/2})$. 
Solid line a the logarithmic dependence for $\sim{\log(x)}^{1.05}$ for comparison.
}
\end{figure}

A similar behavior was found for a model where multiple 
time scales determine irreversible cluster aggregation~\cite{nagler2016a}.
Fig.~\ref{fig:type4stat} shows that the transition point is approached ultra slowly
whereas the slope is an increasing function of $N$.

\subsubsection{Type V:  Hybrid phase transition\label{sec:type5}}

Random graph models displaying EP have been designed with the objective to delay percolation and typically avoid mergers of large components.
The competitive percolation models discussed thus far require sampling of at least two edges, or at least three nodes in a time step.
Yet, in Basatas {\it et al.}~\cite{bastas2011} and Panagiotou {\it et al.}~\cite{panagiotou2011explosive} the authors consider 
``half-restricted processes" and 
demonstrate that sampling of only {\em two} nodes can lead to genuinely discontinuous percolation.
This requires, however,  one node to be sampled from a restricted set of small components,
where the other node is sampled at random.
Soon, a number of models based on half-restricted processes
analyzed the mechanism leading to the discontinuous transition
and the nature of the transition~\cite{Panagiotou2013, herrmann2016a, lee2017}.

They found that the genuinely  discontinuous transition coincides with a second-order transition.
This combination is called a hybrid phase transition.
Hybrid phase transitions exhibit both the typical critical divergence of a second-order phase transition and a genuine discontinuity,
 \begin{equation}\label{eq:hybrid}
 \Y = 
\begin{cases}
0 &\text{if } p <  p_c\\
\Delta + (p-p_c)^{\beta}  &\text{if } p \ge p_c,
\end{cases}
  \end{equation}
where $\Delta$ denotes the gap in the order parameter which coincides 
with a typical critical behavior of a power law divergence. We wish to emphasize, that a
hybrid transition is 
a discontinuous transition with a specific class of critical behavior.  Hybrid transitions require a supercritical behavior with diverging slope of the order parameter at $p_c$, hence $\beta<1$.
Note that an exponent $\beta\ge 1$, or other supercritical behaviors with a {\em finite} slope of $\Y$ 
for $p\rightarrow p_c^+$ 
do {\em not} constitute a hybrid transition. 

Hybrid transitions are known 
for models of $k$-core percolation~\cite{SchwarzLiuChayes06} and for cascade failure models in multiplex networks~\cite{clzgb13}.

Cai {\it et al.}\ first reported universal mechanisms leading to hybrid transitions~\cite{ccgg15}.
This work was followed by a study by
Lee and coworkers~\cite{lee2017}, 
where they show that
a critical branching process underlies the continuous component of the transition, whereas the
discontinuous component stems from supercritical branching~\cite{lee2017} with a very specific form as shown in 
Fig.~\ref{fig:EP_classes}.
More generally, Cai and coworkers and Grassberger~\cite{ccgg15, grassberger2015, grassberger2016}
provided compelling numerical evidence
that hybrid transitions may exhibit a much larger variety of first and second order features,
depending on the underlying structure or network, in particular emphasizing the role of local loops and the limitations 
of mean-field approximations for cascading processes~\cite{grassberger2015}.

\subsubsection{Type VI: Devil's staircases and non-self-averaging  \label{sec:type6}}

\begin{figure}[b!]
\includegraphics[width=0.55\linewidth]{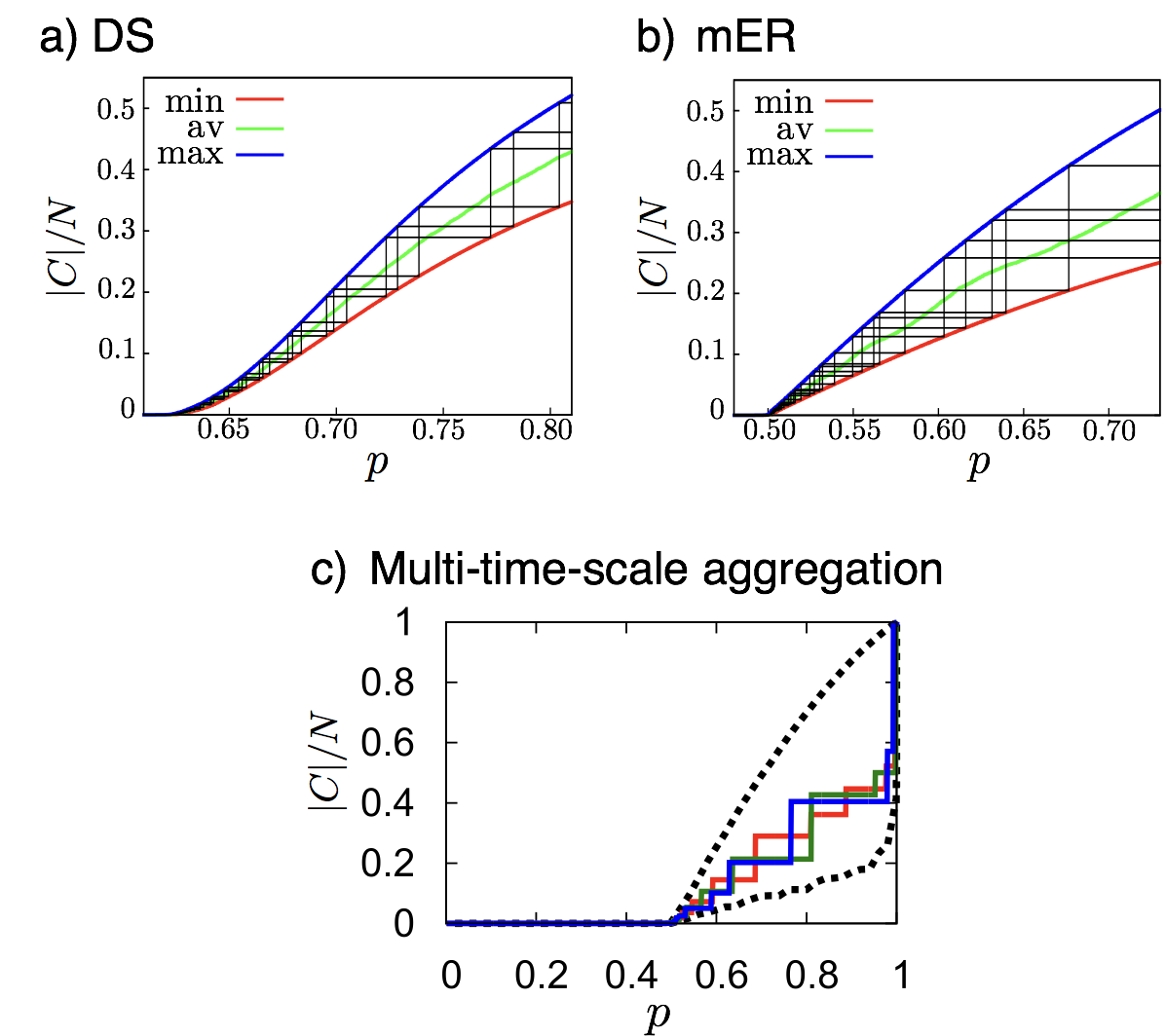}
\caption{\label{fig:C_cmp}
Stochastic staircases.
Models of EP show non-self-averaging behaviors~\cite{RDJN2015}. 
(a) Devil's staircase (DS) model~\cite{naglerPRX}. 
(b) Modified \ER\ model (mER)~\cite{riordanPRE2012}.
(a)-(b), Shown are realizations of staircases of genuinely discontinuous jumps with the relative size of the largest component $|C| /N$
 as a function of $p$ for several distinct realizations (black lines), together with ensemble average (green line), minimum (red line) and maximum (blue line). 
(c) Three realizations for a model of multi-time-scale aggregation~\cite{nagler2016a}, for $\alpha=1$, $\beta=0$,
 envelops displayed in dotted curves, for $N=10^7$. 
 In (a) and (b), realizations are deterministic in the supercritical regime,
 that means, a realization deterministically switches back and forth between the envelope curves.  Still, the models show large sample-to-sample fluctuations and the ``phase"  of a staircase is stochastic.
 In (c), realizations are stochastic in the supercritical regime.
}
\end{figure}

Perhaps the most unusual feature of the models leading to EP transitions is that their behavior can be non-convergent and non-self-averaging. Likewise, continuous percolation and discontinuous percolation can coexist.  These aspects were discussed briefly in Sec.~\ref{subsubsec:APvertex} with respect to the ``Devil's Staircase" (DS) model which is defined in Fig.~\ref{fig:DS}. Here we provide the details and broader context. 

Similar behaviors to the DS model are also found for the Nagler-Gutch (NG) model and the modified ER model (mER), 
as first reported on in~\cite{riordanPRE2012}.
The main mechanism of both the NG and mER models is that the largest component is prevented from growing as long as 
the two sampled largest components are not of {\em exactly} the same size.
Although this mechanism  is not based on suppressing the growth of large components,
it  leads to anomalous percolation features. The order parameter is blurred in the supercritical regime and does not converge
to a function of $p$ in the thermodynamic limit~\cite{riordanPRE2012}, see Fig.~\ref{fig:EP_classes}.
These models  
show tremendous
variation from one realization to another in the supercritical
regime~\cite{riordanPRE2012, RDJN2015}, see Fig.~\ref{fig:C_cmp}(a, b).

\begin{figure}[t]
\includegraphics[width=0.6\linewidth]{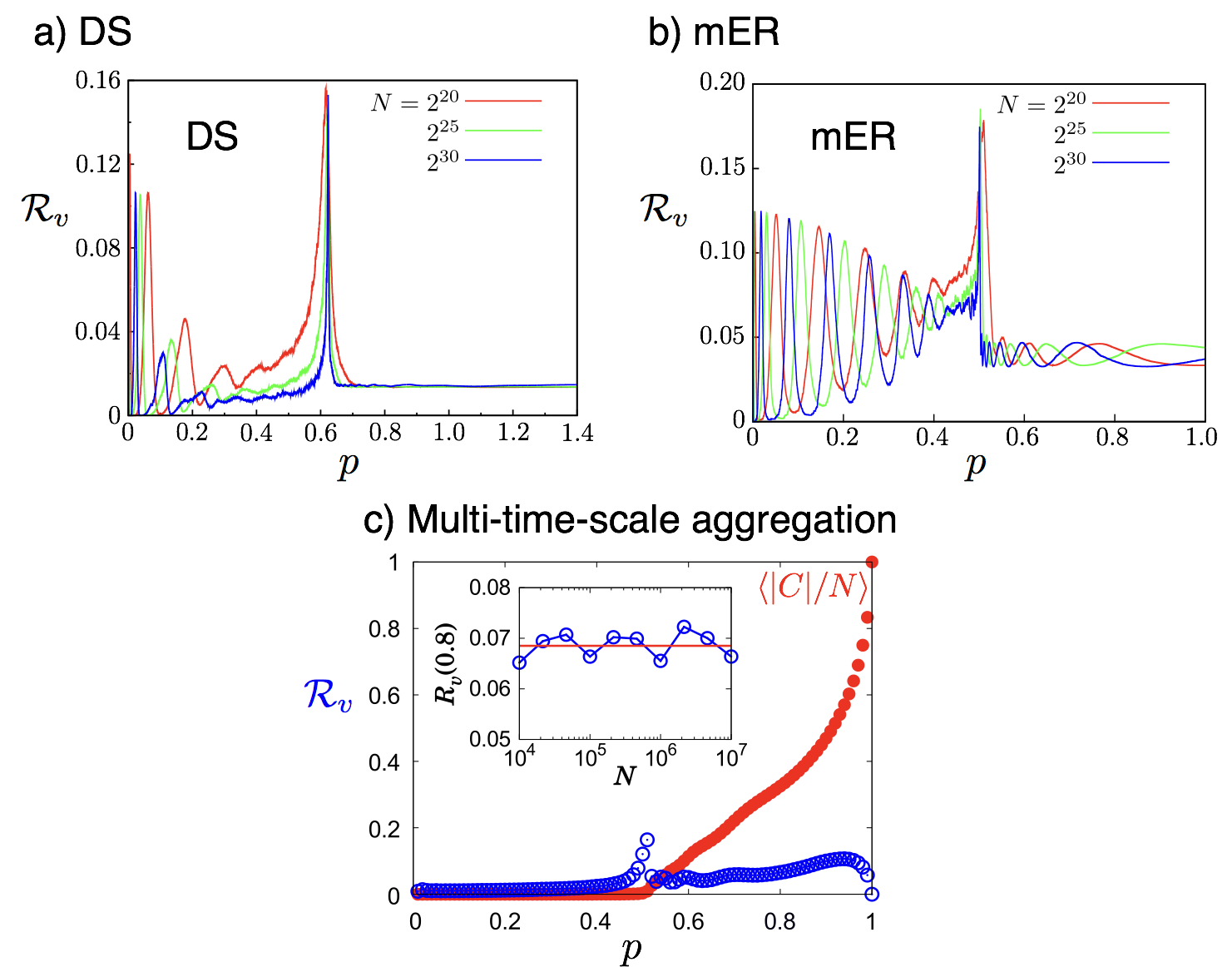}
\caption{\label{fig:Rvs}
Non-Self-averaging of EP models.
Relative variance of the size of the largest component $\Rv(|C|)$ as a function of $p$
show for (a)  the devil's staircase (DS) model, (b) the modified \ER\ (mER) model~\cite{riordanPRE2012}, 
and (c) the multi-time-scale aggregation model~\cite{nagler2016a}. 
Panels (a)-(b) show that DS and the mER display intricate patterns and oscillations~\cite{RDJN2015}.
(c) The ensemble average of the order parameter, $|C|(p)$($\bullet$, red) is shown together with the relative variance $\Rv$ ($\circ$, blue) 
as a function of time.
Inset shows testing for non-self-averaging, $\Rv$ vs $N$ at $p=0.8$.
$\Rv$  oscillates near $0.069$, suggesting that $\Rv$ does not shrink to zero in the thermodynamic limit.
All models are non-self-averaging in the supercritical region, $\Rv(|C|)>$0 for $p>p_c$.
Data are obtained for $N = 10^6$  averaged over $2 \times 10^4$  realizations. 
}
\end{figure}

The sample-to-sample fluctuations of the NG, mER, and DS models have one ``artifact" in common.
For single realizations of the NG, mER, and DS model, $\X$ necessarily
 jumps between the lower and upper bound envelopes~\cite{naglerPRX}, see Fig.~\ref{fig:C_cmp}.
 This rather particular behavior is a consequence of
  the strict impossibility for $\X$ to grow unless the second largest cluster has {\em exactly} the same size as $\X$,
  and is thus built-in to the models~\cite{riordanPRE2012, naglerPRX, schroder2013crackling}.

Note that the onset of the staircase, which is denoted as its ``phase", is determined by stochastic critical fluctuations at $p_c$. Yet, immediately after the largest component becomes macroscopic in size, the successive jumps in the order parameter are best described as a (deterministic) incomplete devil's staircase.
By contrast, the multi-time-scale aggregation models studied in Ref.~\cite{nagler2016a} can exhibit  
staircases that 
are {\em fully} stochastic. 
This means that not only the ensemble but also single realizations are stochastic (even for $N\rightarrow \infty$).

This anomalous behavior results from non-self-averaging, whose key property is quantified by the relative variance
of the order parameter over an ensemble of realizations,
\begin{equation} \label{Rv}
 \Rv(|C|)  = \frac{\langle |C|-\langle |C| \rangle \rangle^2}{\langle |C| \rangle^2} 
\end{equation}
where the brackets denote ensemble averaging.
A system is said to be self-averaging if $\Rv(|C|)\rightarrow 0$ for $N\rightarrow\infty$.
For non-self-averaging systems, however, the observable (here $|C|$) 
does not converge to a sharp value 
for large systems nor for large sample sizes.
Systems that lack self-averaging therefore lack the collapse of ensemble average, minimum, and maximum as well.
Non-self-averaging properties play an important role in the statistical physics of disordered systems, for instance in spin glasses~\cite{spinglasses}, neural networks, polymers, and population biology~\cite{derrida81,mezard87}.

For second-order phase transitions it is well
known that large fluctuations in $\Rv(|C|)$ are observed only
in the critical window and that they collapse to a singular
peak at $p_c$ in the thermodynamic limit.

The lack of self-averaging for the DS and the mER models 
is shown in Fig.~\ref{fig:Rvs}.
One observes elevated values of $\Rv(|C|)$  in the supercritical regime, which do {\em not} vanish with increasing $N$.

In the DS model, clusters that are most similar in size are preferentially merged in the process of percolation.
This 
leads to a non-finite number of discontinuous phase transitions immediately after the first (continuous) phase transition.
Single realizations for large systems are Devil's staircases in the supercritical regime~\cite{naglerNJP}. 
Even in the thermodynamic limit, fluctuations at the critical point determine the ``phase" of the staircase~\cite{schroder2013crackling}. 

This stochasticity implies power law distributed crackling noise, reminiscent of the Barkhausen effect of ferromagnetism.
The stochastic nature, and hence non-convergence, of the order parameter was investigated in Ref.~\cite{schroder2013crackling}
and is another example of non-self-averaging in percolation. The existence of several discrete ``mini-transitions" at points that randomly
fluctuate between realizations is not only true for the Barkhausen effect, but is also observed in other models of network evolution, such as the sudden emergence of ``cluster cores"~\cite{foster2010}, or in networks with core-periphery organization~\cite{colomer2014}.

\subsection{EP phenomena in real-world systems}\label{subsec:realworld}

Here we review studies that have explored the paradigm of EP in real-world systems
including growth of social networks, 
disordered materials, electric breakdown of substrates and emergence of molecular life. 
In many cases, the evolution of the cluster dynamics of the real system resembles those of the graph evolution models that delay formation of macroscopic components. 
Thus models leading to EP, and their underlying mechanisms, offer analogies and insights for real systems and allow for surprising
phenomena like modular structures
or non-self-averaging 
to arise from percolation processes. 
Likewise, understanding suppression mechanisms in real systems may provide future insights for the study of EP or how the paradigm might be applied to real-world systems.

\subsubsection{Non-Self-Averaging and the emergence of molecular life}

In 1975, Sherrington and Kirkpatrick~\cite{spinglasses} 
introduced a mean field model for a spin glass that exhibits unusual magnetic behaviors such as non-self-averaging.
Non-self-averaging has become an ubiquitous phenomenon in statistical physics of disordered systems 
most dominantly studied initially in spin glasses 
but soon the effect appeared also in
neural networks, polymers, and population biology~\cite{derrida81, mezard87, panchenko2012}.
As discussed in Sec.~\ref{sec:type6}, non-self-averaging  and
phase transitions are usually defined only in the thermodynamic limit~\cite{pastur1991}.
In~\cite{zimmer2018}, however, they argue that 
early molecular life is a matter of the sustained emergence of large but finite molecules and
its dynamics is governed by finite size effects from the viewpoint of statistical physics. 
In particular, they observe anomalous behaviors of the molecule concentrations, stochastic staircases and finite-size non-self-averaging.
This establishes a link between self-replication that underlies every species of living beings and an array of anomalous percolation features.

\subsubsection{Finite size of real systems}

In terms of the thermodynamic limit, the sizes of real-world networks, with typical ranges from millions to tens-of-billions of nodes, are quite small.
The rigorous proof  in~\cite{RWscience2011} showing the lack of a powder keg implies that in the limit $N\rightarrow\infty$ the scaling window $\Delta_{N}$ is linear in system size $N$. But numerical evidence on systems up to size $N\sim10^7$ indicates the window is sublinear, with $\Delta_{N} \sim N^{2/3}$ for the Product Rule~\cite{EPScience}. Thus, there must exist a crossover length, $N^* \gg 10^7$, where the system becomes large enough that actual realizations show convergence to the asymptotic limiting behavior. 

The dCDGM model introduced in~\cite{daCosta2010}, illustrated in Fig.~\ref{fig:dCDGMmodel}, enables a rough estimate of a crossover length as discussed in~\cite{RDJN2015}. 
The fractional size of the largest component of the dCDGM model obeys a scaling relation 
$|C|/N \approx (t - t_c)^ {\beta}$
 for $t$ just above $t_c$. 
For a discrete time process on a system of size $N$ the smallest increment of the control parameter possible, denoted $\Delta t$, is for one edge addition, with $\Delta t = 1/N$.  As the dCDGM model passes through the $t_c$, the next edge added causes growth
\be
|C|/N \approx (t - t_c)^ {\beta} = (1/N)^ {\beta}.   
\ee
We can specify a minimum jump in size of $|C|/N$ 
that would be observed for any finite realization of size $N$, which we denote with the parameter $0 < \Delta < 1$.  
A jump of size at least $\Delta$ is achieved for all systems of size 
\be
N \le \left(\frac{1}{\Delta}\right)^{1/\beta}.
\ee
 For the dCDGM model, $1/\beta = 1/0.0555 \approx 18$.  Setting $\Delta=0.1$ means that systems of as large as $N=10^{18}$ would show a discrete jump when passing through $t_c$ on the order of 10\% of the system size. Thus, for the dCDMG model a crossover length requires $N^* > 10^{18}$, far beyond the realm of most real world networks.

\subsubsection{Modular network structure in the real-world}
The microscopic evolution dynamics that suppress the growth of a giant component, as seen in process that lead to EP, gives rise to component structures that capture aspects of modules in a range of real-world systems. 

\begin{figure}[tb]
  \begin{center}
  {\small (a)\ \ \ }\fbox{\includegraphics[width=.35\textwidth]{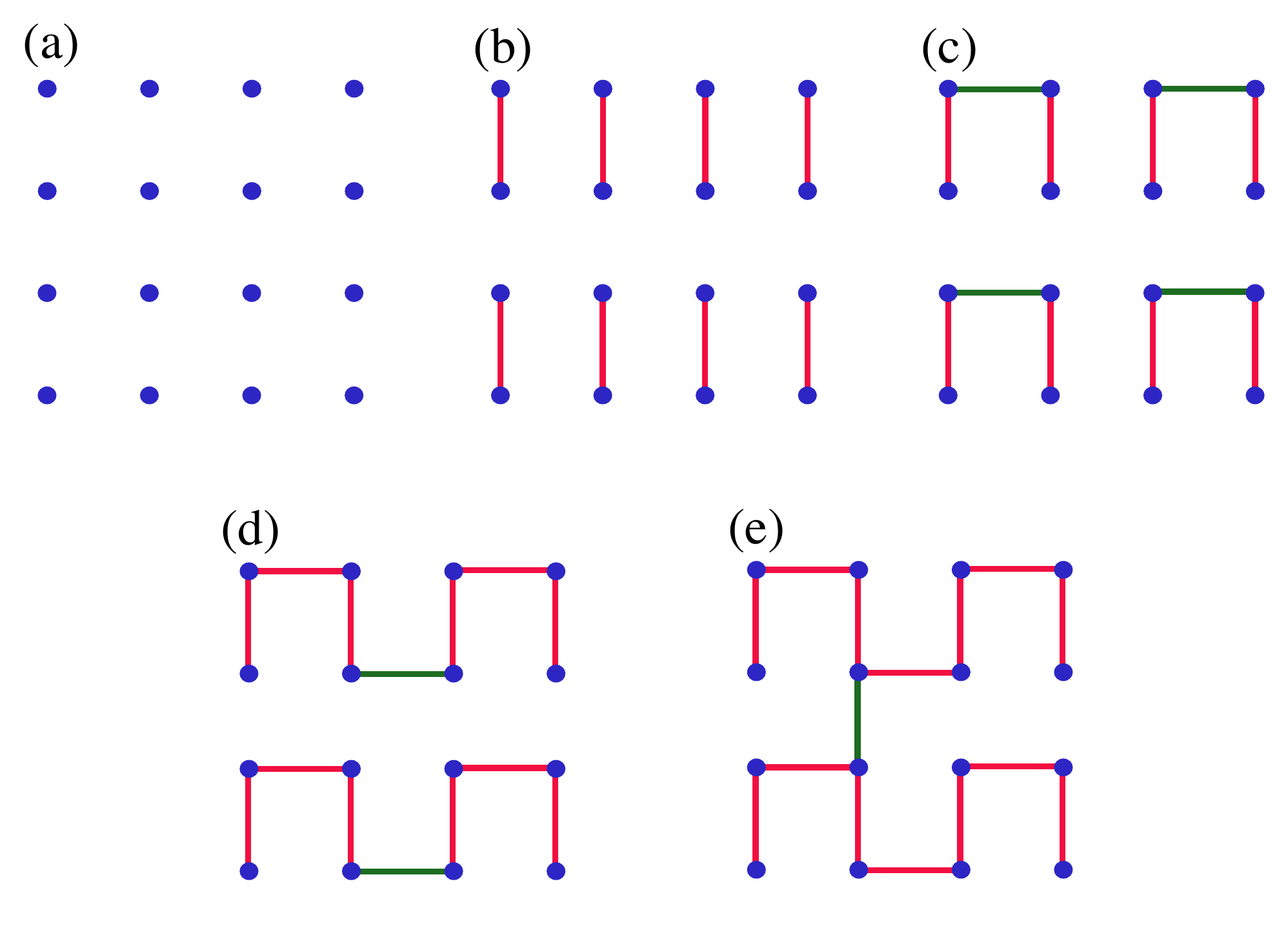}}\\[5mm]
    {\small (b)}\includegraphics[width=.42\textwidth]{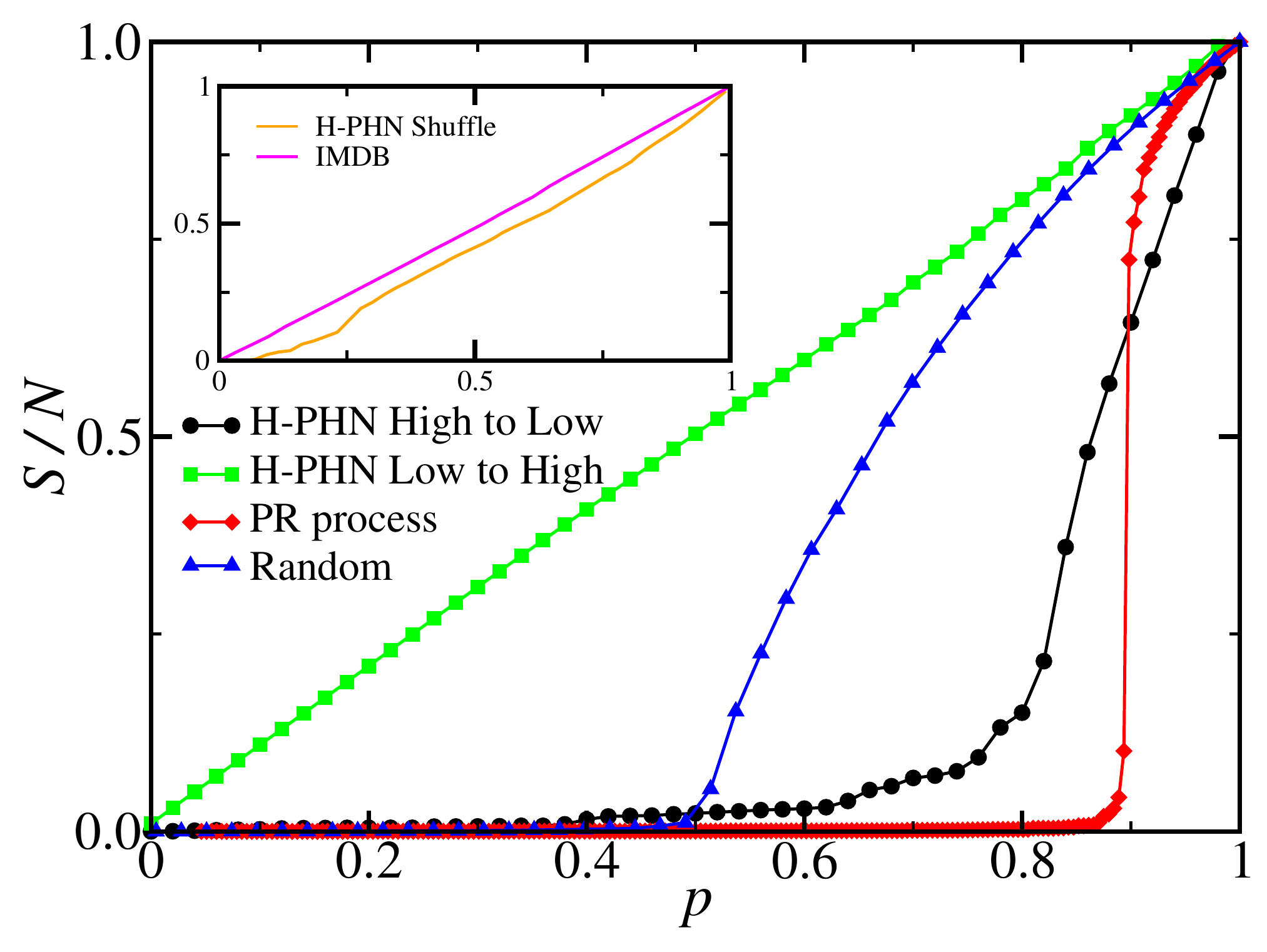}
  \end{center}
  \vspace{-0.2in}
  \caption{(a) As first discussed in~\cite{RozenEPJB2010} an $m$-edge rule AP with $m\rightarrow \infty$ is deterministic.  For a random graph evolution, this process proceeds in stages with creating clusters of only size $2^l$ until all such possible clusters are formed and the process moves onto forming clusters of size $2^{l+1}$.  
 (b) Using the Product Rule AP to favor linking similar proteins, as inspired by biological ideas about homology and evolution, leads to EP.   [Reprinted from~\cite{RozenEPJB2010}.]}
\label{fig:Rozen}
\end{figure}

\vspace{0.03in}
{\it Protein networks:} An early example connecting EP and real-world processes considers an evolutionary process on the Human Protein Homology Network~\cite{RozenEPJB2010}. Using data on human protein homology, Rozenfeld {\it et al.} initialize each  
protein as disconnected and then edges between the most similar (\ie, homologous) proteins are added sequentially. (This is equivalent to an AP with global edge choice, $m=N$.) This captures the general belief that proteins evolve via duplication-mutation events from ancestral proteins, and that more similar proteins organize into network modules~\cite{MediniPLOS2006}.  This process leads to the emergence of many large isolated components of tightly connected nodes, which eventually link together with the addition of just a few intercomponent edges, with global connectivity emerging in an explosive manner.  Details are shown in Fig.~\ref{fig:Rozen}.
As the authors of~\cite{RozenEPJB2010} remark, the emergent structure reflects the dense connectivity within a community and weak links between communities, suggested by Grannoveter for social systems, enabling phenomena such as the ``strength of weak ties"~\cite{granovetter1973}.

\vspace{0.03in}
{\it Social networks:} In Ref.~\cite{PanPRE2011} the authors show that the dynamic evolution of a network following a process leading to EP can reveal underlying community structure in social networks.  They analyze empirical data from two distinct real-world social networks: one is a mobile phone call network, the other is a co-authorship networks of scientists. They build up the aggregate interaction network in each case from the data and use it as the underlying substrate for a percolation process.  Initially, all the edges are considered ``unoccupied" and an AP is used to sequentially ``occupy" edges.  They show that, at $t_c$, the component structure reflects the underlying community structure of the network.

\vspace{0.05in}
{\it Global trade network:}
Schroder and coworkers investigate
 how large-scale connectivity emerges from decisions made by rational agents that individually minimize costs for satisfying their demand~\cite{nagler2018a}. 
 They establish an exact mapping of the resulting  nonlinear trading optimization problem to a local model that leads to EP via a large-cluster-suppression mechanism. Due to existing transportation costs, nodes satisfy their demand locally as long as it is feasible. Only at the percolation threshold are the products  purchased and shipped from the globally best producer.
They find that genuinely discontinuous percolation
and hystereses effects can result from decision making of rational agents in the global trade network.

\subsubsection{Disordered materials}

{\it Diffusion limited aggregation:} The suppression of a largest component captures the evolution of several physical systems. For example, consider the seminal model of diffusion-limited cluster aggregation~\cite{witten1981}.  Here clusters move via Brownian motion so that the velocity of a cluster is inversely proportional to the square root of its size and thus larger clusters move considerably more slowly.  In Ref.~\cite{ChoKhangPRE2011}, the authors show that diffusion-limited cluster aggregation can be mapped onto the framework of EP.  They consider clusters moving on an underlying two-dimensional lattice via Brownian motion, and whenever two clusters become nearest neighbors, they merge into one larger cluster. 
They show that Brownian motion suppresses of the mobility of the largest clusters, impeding their growth, and leading to the discontinuous emergence of a giant cluster as a function of
the number of aggregation events.
They also consider a generalized Brownian motion where the velocity is inversely proportional to the mass of the cluster to a 
power $\eta$ and map out the tricritical point separating discontinuous from continuous emergence as a function of $\eta$. 

\vspace{0.03in}
{\it Nanotubes:} Another area where the paradigm of EP is providing new modeling opportunities is phase transitions in nanotubes. The transition from insulator to conductor is often modeled by the emergence of percolating paths in bundles of nanotubes~\cite{explosiveNanotubesPRE2010}. But, in Ref.~\cite{explosiveNanotubesPRE2010}, the authors show that processes leading to EP offer more realistic models 
as experimental studies show that the sizes of the nanotube bundles are uniform.  Similar to the subcritical dynamics for EP (and unlike regular percolation), the growth of larger bundles is suppressed and the transition becomes extremely abrupt. The transition shows hysteresis, as is expected for first order transitions~\cite{explosiveNanotubesPRE2010}.  

\vspace{0.03in}
{\it Explosive electric breakdowns:} An exciting experimental realization of explosive percolation exhibiting macroscopic electric breakdowns 
is a system of highly conducting particles that are adsorbed and desorbed on an insulating substrate~\cite{HerrmannPRL2014}.
This theoretical study links abrupt macroscopic electric breakdowns to genuine discontinuous percolation transitions.

\vspace{0.03in}
{\it Crackling noise:}
Schr\"oder and coworkers generalized the 
devil's staircase model where 
the merging of components with substantially different sizes are
systematically suppressed  
and merging those whose size ratio is close to a fixed target ratio $f$ is enhanced~\cite{schroder2013crackling}.
For any given target ratio $f$ 
a series of multiple discontinuous jumps in the supercritical regime is observed. 
The sizes and locations of the jumps are randomly distributed, similar to crackling noise observed in materials, such as when a sheet of paper is crumpled. 
This framework links non-self-averaging explosive percolation with phenomena that exhibit crackling noise, power-law fluctuations, and 
Barkhausen noise in ferromagnets.

\vspace{0.03in}
{\it Multiple-time-scale aggregation:}
Anomalous supercritical behaviors are expected in percolation and cluster aggregation with
a separation of the reaction time scales, in particular due to mass segregation~\cite{nagler2016a}.
Examples include systems  where 
a force opposes diffusion in one spatial direction,
as for aggregation processes  affected by gravity, 
or in rotating (planetary ring) systems such as Saturn's rings coalescence dynamics~\cite{saturn2015}.

\subsubsection{Explosive Immunization}

The focus of this review thus far has been on delaying the onset of large-scale connectivity.  Yet an AP can also be used to efficiently destroy network connectivity once it exists.  In~\cite{clusella2016}, they introduce ``Explosive Immunization" for targeted destruction of a network on which is a disease is spreading.
They use an AP to identify ``superblockers", nodes who's removal is most efficient at destroying the network, and immunize these nodes to reduce the possibility of large epidemic outbreaks. Two different selection criteria are used depending upon whether the fraction of vaccinated nodes  
is sub- our supercritical,
and the number of candidate edges for the AP is $m=1000$.  The time complexity of the model is linear in the number of nodes $N$ (up to logarithmic corrections) allowing for efficient implementation.  Testing the method on many model networks and datasets of real-world networks suggests that explosive immunization is one of the 
most effective strategies that exist for curtailing the spread of large-scale infection, which may prove applicable in real-world scenarios.

\subsubsection{Optimal Dismantling } 
Dismantling of a network considers efficient node removal processes that fragmentize a given network as much as possible, subject to a fixed number of nodes or other cost functions~\cite{Braunstein2016}. In contrast, the problem of how to efficiently remove edges for the same objective is usually referred to as an interdiction problem (see, e.g., Ref.~\cite{Zenklusen2014} and references therein). Both dismantling and interdiction processes typically lead to discontinuous percolation transitions. Dismantling algorithms based on reverse EP~\cite{Zhao2018} belong to the most efficient methods studied~\cite{Marone2016, Braunstein2016, Zdeborova2016}.

\subsubsection{Information spreading} 
In~\cite{zhu2018} they use the paradigm of EP to model outbreaks of information sharing in social networks, arguing that abrupt and large-scale bursts of information sharing, especially about sensitive topics, can lead to social instability. This is particularly the case if the information being spread is false.  In~\cite{zhu2018}, they combine models of epidemic spreading on an $L \times L$ lattice with selection rules used to determine to which neighboring node a node spreads its information.  The hypothesis is that when the authenticity of the information cannot be verified, a node chooses to spread the information to the smallest group it has access to. Using the Sum Rule to implement this process in simulations, they show this approach increases the latency period when the information stays localized in small groups but at a critical stage in the evolution a global outbreak abruptly occurs, and the information is globally shared.

\subsection{Other interesting EP phenomena}

\subsubsection{Node arrival mitigates abruptness}

Two distinct studies now show that node arrival can mitigate the abruptness of an AP percolation transition, yet while still significantly delaying the onset of percolation~\cite{vikramPRE2013,growthEPL2013}. In the models of percolation discussed thus far, the number of nodes $N$ is a fixed quantity and the graph evolves via edge arrival.  In seminal work appearing in 2001, the impact of node arrival on the \ER~process was analyzed~\cite{callaway2001}.  Starting from a few seed nodes, a new node arrives at each discrete time step and, with probability $\delta \le 1$, an edge selected uniformly at random is added to the graph.  This leads to a infinite order percolation transition similar to the Berezinskii-Kosterlitz-Thouless phase transition in condensed matter physics~\cite{callaway2001}.  

The random graph model of~\cite{callaway2001} is extended in~\cite{vikramPRE2013}, so that when an edge is added to the graph (which happens with probability $\delta$),  
the ``Adjacent Edge" rule
~\cite{RDMMclustagg} is used for selecting between two candidate edges. 
Thus, an $m=2$-choice AP is used for edge selection, instead of choosing an edge uniformly at random. 
The cluster aggregation analysis in~\cite{vikramPRE2013} shows that this   
process  
considerably delays the onset of the percolation transition but retains the 
smooth, {\it infinite} order transition of the original~\cite{callaway2001} model. 

In~\cite{growthEPL2013} the authors study a similar variant of the growing random graph model of~\cite{callaway2001}. Here, 
when it is time to add an edge to the graph,  
rather than adding a randomly selected edge, instead the dCDGM rule~\cite{daCosta2010} for edge selection is used. Using the power-law behavior of the order parameter and analysis of the crossing of the fourth-order cumulant at the critical point, they similarly establish that the percolation transition is significantly delayed. However for this model, although the transition is smooth, it does not seem to be an {\it infinite} order transition.  So the Adjacent Edge rule of~\cite{vikramPRE2013} and the dCDGM model~\cite{growthEPL2013} both mitigate the abruptness of the transition, but lead to different universality classes.

\subsubsection{APs with directed edges}

Many important real-world networks have directed edges, where each edge points from one node to another.  For instance, consider a food web where a directed edge from node $a$ to node $b$ means that species $b$ consumes species $a$.  Likewise we see directed edges in gene regulation networks, scientific citation networks, the world-wide web, follower relations in on-line social networks, and many others. Yet, up until now we have only discussed networks with undirected edges.  

In~\cite{SquiresPRE2013} the authors explore a generalization of an AP to directed networks and study the scaling properties of the process. Many of the features of an AP on an undirected network are observed, but are less pronounced, so~\cite{SquiresPRE2013} refer to AP on directed networks as  ``weakly explosive". (Note in~\cite{Manna} the authors also introduce this term ``weakly explosive", but use it to refer to the impact of a single edge discussed in Sec.~\ref{subsubsec:APedge})  

Of course on a directed network defining a ``component" is more complicated as each node participates in an in-component (of paths to node $i$), an out-component (of paths to other nodes starting from node $i$), and potentially participate in a strongly connected component. At the percolation point for an \ER~random graph with directed edges a giant strongly connected component, giant in-component, and giant
out-component all emerge simultaneously. (See for instance~\cite{NSW2001}). 

In~\cite{SquiresPRE2013} the authors introduce an AP (referred to as a ``directed competition process") which minimizes the product of the in-component of the node located at the tail of the edge and the out-component of the node located at the head of the edge.  They analyze the critical exponents for this directed process when compared to the undirected process.  In particular they find the order parameter scaling exponent $\beta\approx 0.086$ for the undirected AP but for the directed analogue the giant out-component has $\beta \approx 0.34$ and the giant strongly connected component has $\beta \approx 1.2$.  

Directed networks and APs are also studied in~\cite{Waagen2017}. But, in addition, they consider that nodes have a rank ordering with edges preferentially directed from lower to higher-ranked individuals. This 
captures an essential feature of on-line social networks such as Twitter, where less popular individuals tend to follow more popular individuals (but not vice-versa).  They introduce a simple percolation process (with no edge choice) on an ordered, directed network. As edges are added between disconnected nodes, it is directed monotonically with respect to the rank ordering. Large-scale connectivity occurs at very high density compared to standard percolation processes, for both the strongly connected and the weakly connected component structure, with a dominant strongly connected component appearing in a seemingly discontinuous transition. 
Finally, by adding a competitive percolation rule with a small bias to link users of similar rank, they show this leads to formation of two distinct components, one of high-ranked users, and one of low-ranked users, with little flow between the two components.  Thus information flow is segregated into two-classes of users.

\subsubsection{Degree-based rules}\label{subsubsec:WaagenChoice}

As discussed in Sec.~\ref{subsubsec:APedge}, in the seminal work~\cite{RWscience2011} they show that for an AP on a random graph, if the number of random candidate edges, $m$, is allowed to increase in any way with system size $N$,  so that $m\rightarrow \infty$ as $N\rightarrow \infty$ then this is sufficient to allow for a discontinuous percolation transition.  In~\cite{Waagen2014} the authors show that given enough edge choice, simple local rules show EP.  More specifically, they define $m(N)$ to be function describing how $m$ increases with $N$,  and consider a purely local AP rule that uses only the degree of the nodes (rather than the sizes of components). They explicitly calculate the critical point, showing that $t_c = \left(1 - \Theta(\frac{1}{m})\right)N$ and that the critical widow is of size $O\left(\frac{N}{m(N)} \right)$. Thus the scaling window is sublinear if  $m$ depends on $N$, for instance $m=\log N$. 
Any finite realization of the process shows a stochastic jump size~\cite{Waagen2014} as discussed in more depth in Sec.~\ref{sec:type4},
exemplifying the Type IV transition in Fig.~\ref{fig:EP_classes}.
They also present complementary arguments to 
to those in~\cite{SpencWorm} that APs with bounded size rules lead to continuous percolation transitions. 

Degree based rules are of great interest because they require only local knowledge of the node degree, whereas cluster-based rules require knowledge of a more extensive property, namely the component sizes.  
Recent research has focused on amplifying the effects of a local rule with the ``Degree-base Product Rule" (DPR)~\cite{Trevelyan2018}.  Here $m$ candidate edges are chosen and the one that minimizes the product of the degrees of the two nodes that would be joined is selected.  Remarkably, for DPR increasing the number of choices from an initial $m=2$ does not increase the abruptness of the transition but does increasingly delay percolation. The locality of the information used constrains the evolution of the degree distribution rather than creating a powder keg of the component size distribution.  Similar to node arrival~\cite{vikramPRE2013,growthEPL2013}, this effectively delays the onset of percolation while mitigating the abruptness of the transition.  In the limit of global information, $m=N$, DPR reduces to a model with infinite choice exemplifying the Type III transitions shown in Fig.~\ref{fig:EP_classes}.

In~\cite{Hooyberghs2014} they also consider degree-based network growth, but without any competition. They consider starting from a fixed number of isolated nodes, with edges added using a degree-dependent linking probability, $p_k \propto k^{-\alpha}$, where $k$ is the degree of the node that would gain the edge. They can analyze the model using rate equations and show that for $\alpha > 0$ nodes with low degrees are connected preferentially asymptotically approaching an EP transition.  (For negative $\alpha$ incoming links are most likely attached to high-degree nodes.) 

\begin{figure}[tb]
  \begin{center}
  \includegraphics[width=.5\textwidth]{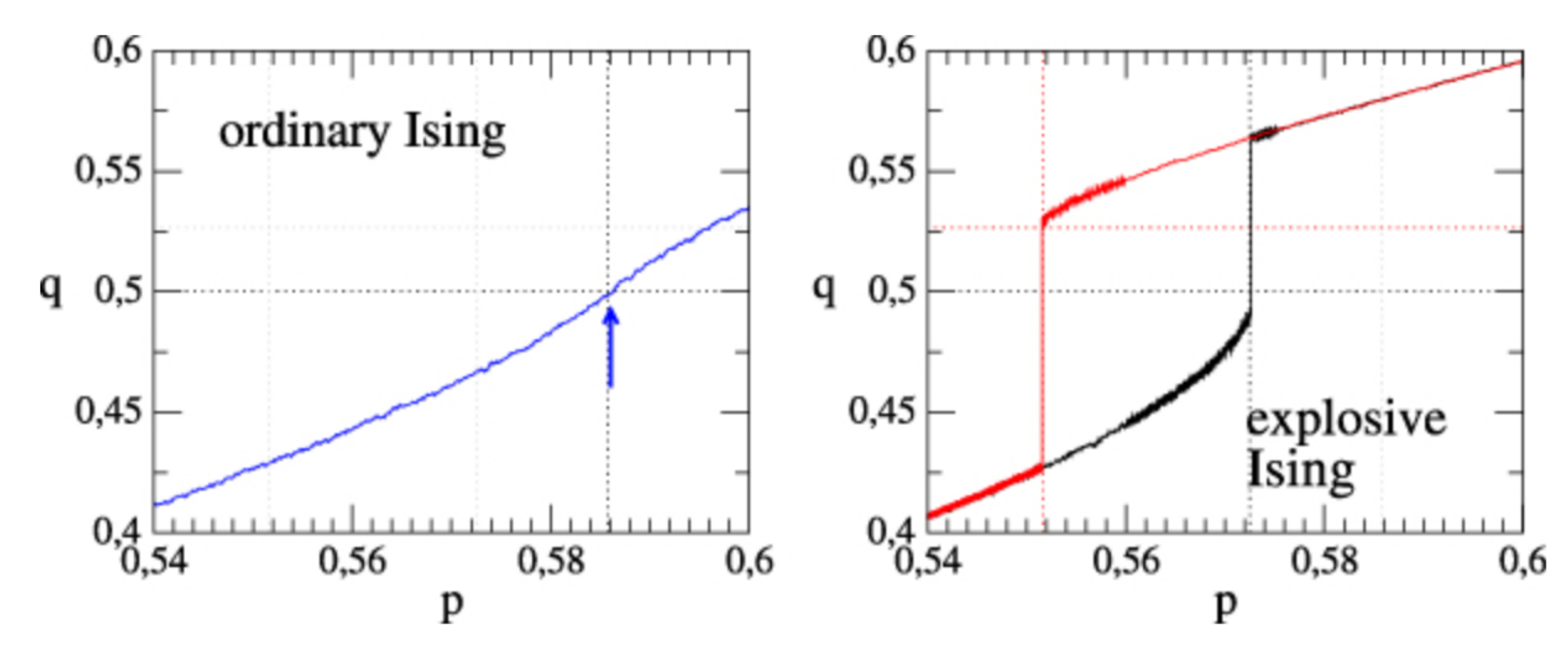}
  \end{center}
  \vspace{-0.2in}
  \caption{Hysteresis curve exhibited by adding an AP for cluster flipping in a kinetic Ising model. [Reprinted from ~\cite{angst2012explosiveIsing}.]}
\label{fig:Ising}
\end{figure}

\subsubsection{Explosive Ising from an AP}
In~\cite{angst2012explosiveIsing} the authors show that the concept of explosive cluster growth can also be applied successfully to kinetic spin models with cluster dynamics. They show that the use of an AP process for cluster dynamics turns the continuous Ising transition into a discontinuous transition.  Explicitly they consider the two-dimensional Ising model and a modification of Swendsen-Wang dynamics.  In the traditional Swendsen-Wang cluster dynamics, bonds are occupied between neighboring spins of the same orientation with probability $p$.  Here, rather than occupying bonds at random, they use the Product Rule to choose between two candidate bonds.  This leads to a discontinuous phase transition in magnetization that shows corresponding hysteresis behavior shown in Fig.~\ref{fig:Ising}.  Interestingly, they demonstrate that the modification to the dynamics is expected to break detailed balance, thus the stationary state of such an explosive Ising model will no longer be an equilibrium state.

\subsubsection{Discrete Scale Invariance} 

As discussed in Sec.~\ref{sec:type6}, the relative variance of EP transitions (Eq.~(\ref{Rv})) can show exotic behaviors.
 It is shown in~\cite{ChenPRL2014} 
 that peaks in the relative variance can announce the location of the critical point. 
 In particular, the generalized BFW process (see Sec.~\ref{subsubsec:BFW} and Eq.~(\ref{eqn:BFWalpha})) exhibits peaks in relative variance at well-defined values of link density $p_i$, where $i$ is an integer, which survive in the thermodynamic limit. 
 The peaks then become infinitely sharp and their 
 positions, $p_i$, follow a discrete scale invariance (DSI) pattern~\cite{Sornette1998}.
This pattern is described by a scale invariance that holds only for a discrete set of magnification factors.
As shown in~\cite{ChenPRL2014} the critical point can be predicted through $p_i\rightarrow p_c$ from the discrete scaling relation. 
The non-self-averaging behaviors displayed by some EP transitions can thus provide a predictive tool for determining the location of phase transitions in these models of graph evolution. 
How this may be extended to real-world systems is an open problem.

\subsection{General insights}\label{subsec:insights}
\subsubsection{Mechanisms and processes underlying EP}\label{subsubsec:mechanisms}

Many types of processes have now been identified that implement the delay in formation of macroscopic components necessary for EP transitions.  
A unifying feature of these processes is that they break the mechanism of ``multiplicative coalescence" (see Sec.~\ref{subsubsec:ClusterAgg}) that describes standard random graph percolation. 

We know from rigorous analysis that if the largest component is strictly prohibited from direct growth, meaning $P_{Gr} = 0$, this necessarily leads to growth by overtaking and the discontinuous emergence of a giant component. Furthermore the minimum discontinuous jump in relative size is $\Delta C_{\rm max} \ge 1/3$~\cite{naglerNP}.  The BFW process~\cite{BohFriezWormRSA2004} comes close to achieving this mechanism. It was found that the largest component can grow directly in BFW by merging with  isolated nodes, but any significant growth is from overtaking~\cite{ChenEPL2012}. It is also an important observation that $k$-vertex edge choice APs can achieve $P_{Gr} = 0$ but that $m$-edge choice APs cannot.  Models that achieve $P_{Gr} = 0$ show additional discrete jumps in $|C|/N$ in the supercritical regime after the first percolation transition~\cite{naglerPRX,schroder2013crackling,ChenPRE2013a,ChenPRE2013b}.

Suppressing the growth of the largest component is a key ingredient of many processes that lead to EP. The BFW model, similar to the model of~\cite{nagler2017}, achieves this by using a ``budget" for edge rejection. The Gaussian model of~\cite{Araujo}  suppresses direct growth in favor of creating components that are similar in size to the average component size.  More generally, as shown in~\cite{Moreira}, dynamical evolution processes the keep clusters similar in size as they grow is key for obtaining a discontinuous percolation transition. 
The Devil's Staircase rule~\cite{naglerPRX} is an example of such a process.

The creation of a ``powder keg" underlies many models that lead to EP~\cite{powderkeg09}.  Yet, finite size effects complicate its detection.  For $m$-edge APs the evolution of the cluster size distribution in the subcritical regime seems to indicate the build-up of a powder keg, yet it disappears as approaching the critical point as shown in Fig.~\ref{fig:dCDGMscaling}. The BFW process, in contrast, seems to create a true powder keg, as shown in Fig.~\ref{fig:BFW1}.

There is also an interplay between local and global information that underlies many processes that result in EP. 
In general we find, from the examples reviewed here, that the more non-local the dynamical evolution rule, the easier it is to achieve EP, in that simpler rules enable EP or more abrupt transitions result.

\subsubsection{The role of order parameter}

In statistical mechanics the choice of the order parameter is crucial and often non-trivial~\cite{stanley1971}.
In percolation, traditionally, the order parameter is the fractional size of the largest connected component of the system,
giving rise to a continuous phase transition from small scale components to a {\em unique} giant component. 

Many models of explosive percolation have adopted this choice of the order parameter, 
even if multiple macroscopic components emerge.
The impact that a particular choice of the order parameter may have has been an important and controversial issue.
Whenever possible, the control and order parameters should be chosen to be conjugate to each other,
which is a standard practice in (canonical) statistical mechanics and thermodynamics.
In percolation, however, a canonical approach is only possible for very specific models
such as Hamiltonian graph models~\cite{bizhani2012}.

If the order parameter is taken as the total relative size of all the macroscopic components, models of EP with an initial continuous transition but with multiple giant components, 
(such as the Type VI models discussed in Sec.~\ref{sec:type6}),
may be globally continuous. This perspective neglects mesoscopic structures,
yet shows that 
many of the observed discontinuities and non-self-averaging behaviors may be a consequence of the choice of the order parameter.
In Ref.~\cite{daCostaPRE2014} the authors argued that ``for continuous phase transitions the order parameter cannot be chosen in an arbitrary way, 
by demanding only that it is 0 in the normal phase and nonzero in the ordered phase. 
For these transitions the order parameter must satisfy several strict conditions that are well known in the theory of phase transitions". 
In order to satisfy a number of constraints such as the hyperscaling relation, however,
the order parameter  needs to be 
based on the model details (e.g., the choice for parameter $m$)~\cite{daCostaPRE2014},
even for similar models such as related $m$-edge processes.

Likewise, the choice of control parameter crucially determines which features are measured. 
Traditionally in percolation, the link density $p$ is taken as the control parameter. 
If time is used instead, for reversible percolation processes,  
the link density $p$ and time $t$ are equivalent. 
In particular, for 
the time-discrete case, 
a step forward means adding a link, whereas a step backwards means removing a link. As links are chosen entirely at random,
the process of standard percolation is reversible and can be equivalently described as a reversible kinetic approach.

In EP the matter is much more intriguing because the models (continuous or discontinuous) are usually irreversible.
The 
diffusion-limited cluster aggregation model of Ref.~\cite{ChoClusterAgg} (cf.\ Sec.~\ref{sec:cluster}) 
can exhibit a discontinuous phase transition (type III) when the order parameter is studied as a function of $p$
(the relative number of aggregation events) but not as a function of $t$, defined as the physical time of the Smoluchowski equation.
Thus, choosing, from the perspective of statistical mechanics, an atypical control parameter,
is sometimes delicate and can crucially alter the critical behavior and phase transition type.

Another example includes cascades in interdependent networks~\cite{bppsh10},
which {\em can} give rise to  discontinuous breakdowns.
In many models, failure of a single  node may cause many or even a macroscopic fraction of  all  nodes 
to fail as well.
In that case, a large number of links may be removed in a single ``time" step. If the largest connected component is studied as a function of time,
cascading failures may show discontinuous dismantling of the network. If in each time step only a single link can be removed (as in standard percolation),
cascades may show a continuous transition in the thermodynamic limit. In vivid debates it has been argued
whether the link removal process occurs instantaneously with node removal, or whether the link removal processes may belong to a substantially slower time scale than the node failure time scale. 
In addition, Son and coworkers demonstrated that the 
percolation is actually continuous for fully interdependent diluted lattices. Moreover, 
making networks interdependent can actually smoothen the transition~\cite{son2011}.

Many processes based on link culling assume instantaneous multi-link removal. Examples are numerous and
include the prototypical example of discontinuous $k$-core percolation~\cite{kcore83, kcore84, schwarz2006onset, cellai2011tricritical}.
Generally, changing the control parameter to the relative number of removed links, thereby excluding multiple-link culling in a single step,
can smoothen the transition in such a way that it becomes continuous in thermodynamic limit.

\subsubsection{Reversible versus irreversible transitions}

Traditionally, processes that exhibit a second order phase transition are reversible whereas first-order phase transitions
generically emerge from irreversible processes. When transitions are reversible, it makes no difference whether the control parameter is
tuned forwards or backwards, the order parameter shows the same continuous transition at the same critical point.
Standard percolation, and models in this universality class, are of that kind. 

We would like to emphasize that the occurrence of a continuous phase transition does not necessarily imply reversibility.
Examples include a great majority of EP models. 
Typically, for explosive percolation on random networks  
it is impossible to construct the  reversed process from the forward link-addition process, 
as the microscopic rule is based on cluster information which is lost during the process.
For the same reason the forward process for cluster aggregation is also known to be statistically irreversible.
In principle, it is possible to construct statistically reversible cluster aggregation models~\cite{kolb1986}, for example when 
linear molecules merge and break up independently of their size.
Likewise, percolation in one dimension is an example of a reversible process that shows a discontinuous transition of type III.

Discontinuous phase transitions are usually hysteretic and often show other interesting history-dependent phenomena.
Thus it is helpful to consider here an example for a reversible discontinuous percolation model in more detail.
Consider the process of successively merging the two smallest clusters in the system~\cite{RozenEPJB2010, naglerNP}
as shown in Fig.~\ref{fig:Rozen}(a). 
For the system size $N=2^n$, where $n$ is an integer,
the order parameter, the size of the largest component, is given by $C_1(T)=\frac{N}{N-T}$, where $T$ denotes the number of added links (or MC steps).
This global  rule
generates the following sequence of the largest cluster size as 
1, 
$\underbrace{2,\ldots, 2}_{N/2 \; \text{times}}$, 
$\underbrace{4,\ldots, 4}_{N/4 \; \text{times}}$, 
$\ldots$, with the last steps
$C_1(N-3)=N/2$, $C_1(N-2)=N/2$,
and finally $C_1=N$ at $T=N-1$.
Clearly, this {\em deterministic} process can be microscopically reversed:  
Successively split up the largest cluster in two equal parts, 
which leads to the reverse sequence: $N$, $N/2$, $N/2$,  $N/4$, $N/4$, $N/4$, $N/4$, $\ldots$, 
$\underbrace{2,\ldots, 2}_{N/2 \; \text{times}}$, 1.
In the thermodynamic limit, this process leads to a discontinuous jump from zero to one at $p_c=1$~\cite{naglerNP}.

Schr\"oder {\it et al.}\
 discuss a special form of random fragmentation~\cite{schroder2013crackling},
where, if possible, 
clusters are fragmented exactly into even parts
(if the cluster size is not a power of two, it is split up into clusters of size $(s+1)/2$ and $(s-1)/2$, respectively).
For large system size,  this process becomes the backward process of the Devil's staircase (DS) model
where only similarly sized components merge~\cite{naglerPRX}. 
Yet, as the order parameter does not 
converge to a function in the thermodynamic limit (Sec.~\ref{sec:type6}),
the DS model should not be considered reversible.

%
%
%
\section{Explosive Synchronization (ES)}\label{sec:ES}

Explosive Synchronization (ES) describes the abrupt onset of collective synchronized behavior in a network of oscillators which arises from the interplay of the network architecture and the distribution in the natural frequencies of the oscillators. The interplay conspires to impact the threshold for the onset of synchronization and has been shown to have dynamics with hysteresis.  This section introduces the standard process and variations of ES phenomenon in complex networks.  In contrast to EP, which focuses on network growth models from statistical physics, ES is grounded in systems of dynamical equations. 
Most of the analytic studies 
are based on the seminal 
and widely studied non-linear phase oscillator model, the Kuramoto model (KM)~\cite{Kuramoto75,Kuramoto84}.
This has allowed for the development of new tools and insights about the mechanisms underlying the ES process in many distinct cases.  We present the detailed formulations and synthesize the classes of frameworks that lead to ES, summarizing the classes in Fig.~\ref{fig:summaryES} in analogy to Fig.~\ref{fig:EP_classes} showing the classes of transitions for EP.

In addition to analytic formulations, experimental platforms~\cite{rossler,kvpb15} have now implemented ES in real-world systems, allowing the confirmation of several theoretical predictions. Furthermore, in light of the developments of ES, many abrupt transitions and hypersensitive responses to perturbations observed in neuroscience have been recently revisited from the perspective of ES~\cite{kmmvt16,kim17,wtdl17,pgnl16,lklkckmh18}, and the results suggest ES can be a useful paradigm for understanding neural phenomena from epilepsy to the conscious-unconscious transition when awaking from anesthesia.

Note that discontinuous synchronization transitions are known to show up also in mean field models but arising from mechanisms different to those discussed here. For instance, in the mean field Kuramoto model,~\cite{tlo97a,tlo97b} showed that adding inertial terms can lead to a discontinuous transition and
Pazo showed that a discontinuous transition emerges when the frequency distribution becomes sufficiently broad~\cite{p05}. However, explosive network transitions are rooted in mechanisms that involve both the local connectivity among nodes and the dynamical process. In particular, from the diverse mechanisms leading to ES that are summarized later in Fig.~\ref{fig:summaryES}, it is easy to note that most of the corresponding ES frameworks have no counterpart in mean-field models since their application demands the existence of a networked structure connecting dynamical units.

\subsection{Introduction to the Kuramoto model}
\label{sec:KM_original}

Before presenting the recent advances on ES phenomena starting in Sec.~\ref{sec:ES_networks}, let us briefly review here the basics of the Kuramoto model (KM) which is the dynamical model analyzed in most of the ES studies. The KM implemented on networks of coupled units is reviewed in Sec.~\ref{sec:KM_networks}.
Collective synchronization arises when a large number of interacting units oscillate in phase together uniformly regardless of the differences in their intrinsic dynamics. This phenomenon has attracted the attention of scientists 
for quite some time 
due to its beauty and its widespread presence in many diverse fields of science. This ubiquity ranges from networks of pacemaker cells in the heart and flashing fireflies to many examples in physics and engineering such as arrays of lasers and superconducting Josephson junctions~\cite{s00,sg04,abprs05}. The pioneering work by~\cite{winfree67} on collective synchronization spurred the field and called for mathematical approaches to tackle the problem.  One of these approaches considers that the system is made up of a huge population of coupled, nearly identical, interacting limit-cycle oscillators, where each oscillator exerts a phase dependent influence on the others and changes its rhythm according to a coupling function~\cite{s00,abprs05}.

\begin{figure*}
\begin{center}
\includegraphics[width=.85\textwidth,angle=0,clip=1]{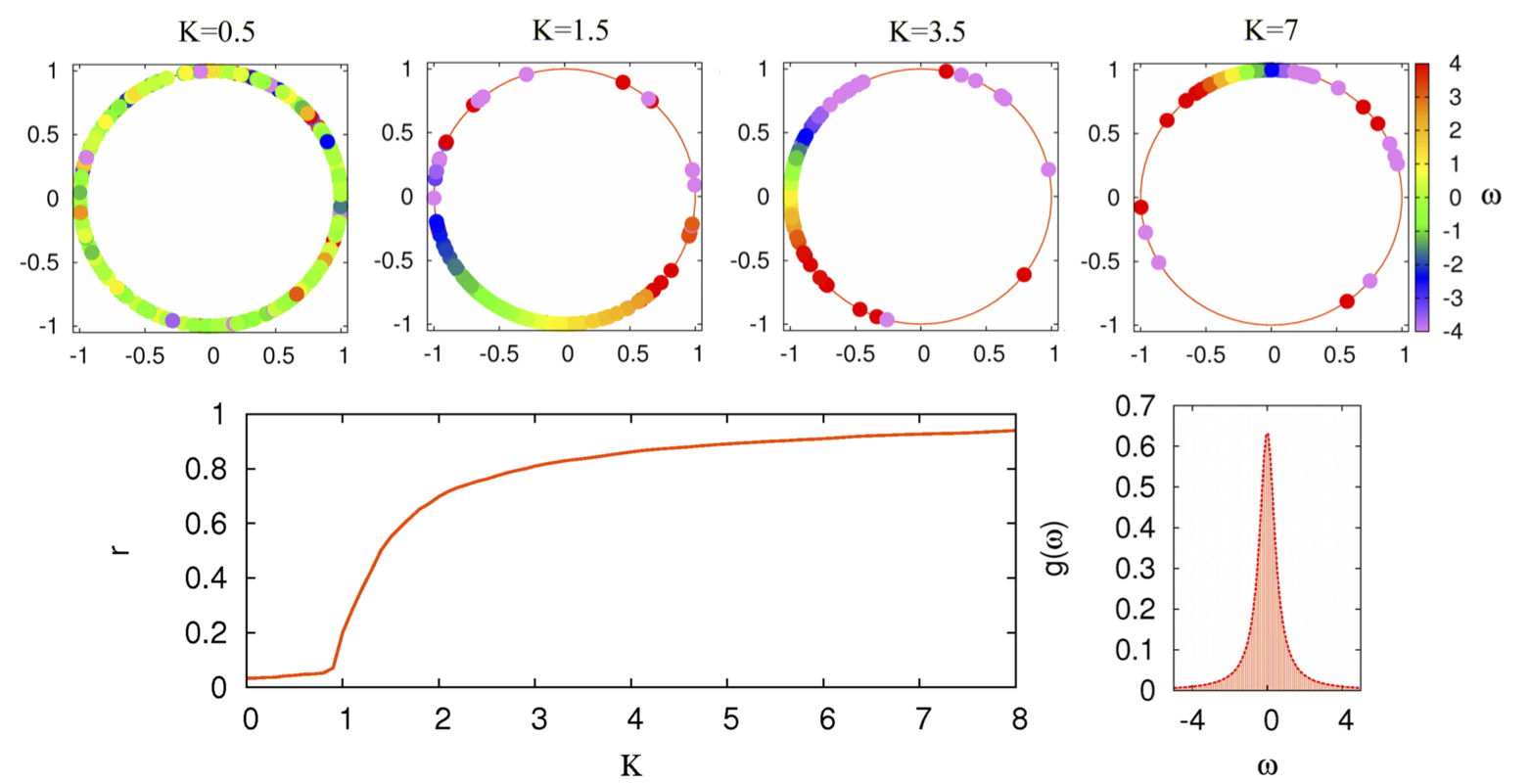}
\end{center}
\caption{
Synchronization in the classical Kuramoto model. Each panel on the top shows the collection of oscillators situated in the unit circle (when each oscillator $i$ is represented as $e^{{\mbox{i}}\theta_i} (t)$). The color of each oscillator represents its natural frequency. From left to right we observe how oscillators start to concentrate as the coupling $K$ increases. In the panels below we show the synchronization diagram, i.e., the Kuramoto order parameter $r$ as a function of $K$. It is clear that $K_c \simeq 1$, as obtained by using the Lorentzian distribution $g(\omega)$ shown in the right panel. [Reprinted from~\cite{hpgz14}].}
\label{fig:Hermoso2014}
\end{figure*}

Even if the previous simplifications seem to be very crude, the phenomenology of the synchronization problem can be obtained. 
When the natural (intrinsic) frequencies of the oscillators are too diverse compared to the strength of the coupling, they are unable to synchronize and the system behaves incoherently, showing a range of phases whose average is nearly zero. However, if the coupling is strong enough, the behavior of all the oscillators aligns and they beat in synchrony. The transition from incoherent to coherent behavior takes place
smoothly at a certain threshold where synchrony starts to gradually emerge.  At the transition point 
some set of elements lock their relative phase forming a cluster of synchronized nodes marking the \emph{onset of synchronization}.
Beyond this value, the population of oscillators is split into a partially synchronized population
made up of oscillators locked in phase and a group of other oscillators whose natural frequencies are too different 
to allow them to join the coherent cluster. Finally, after further increasing the coupling, more and more elements get entrained around the mean phase of the collective rhythm generated by the whole population and the system settles into the completely synchronized state, see Fig.~\ref{fig:Hermoso2014}.

Kuramoto~\cite{Kuramoto75,Kuramoto84} worked out a mathematically tractable model to describe this phenomenology, which we denote the Kuramoto Model (KM). He noticed that the most suitable case for analytical treatment should be the mean field approach, which corresponds to equally weighted, all-to-all coupling, taken to be purely sinusoidal for simplicity. In particular, Kuramoto proposed the governing equations for each of the $N$ oscillators $i$ in the system:
\be
\dot{\theta}_{i}=\omega_{i}+\frac{K}{N}\sum_{j=1}^{N}
\sin{(\theta_{j}-\theta_{i})}
\hspace{0.5cm} (i=1,...,N)\;,
\label{ekuramodel}
\ee
where the factor $1 / N$ is included to avoid divergences in the thermodynamic limit 
($N\rightarrow\infty$), $\omega_i$ stands for the natural frequency of
oscillator $i$, and $K$ is the coupling constant. The frequencies $\omega_i$ are distributed according to some function $g(\omega)$, that is usually assumed to be unimodal and symmetric about its mean frequency $\Omega$. The most commonly used frequency distributions are the
 Lorentzian (Cauchy) ($g(\omega)=\frac{1}{\pi}\frac{\gamma}{(\omega-\Omega)^2+\gamma^2}$) and Gaussian ($g(\omega)=\frac{1}{\sqrt{2\pi\sigma}}{\mbox{exp}}\left(-\frac{(\omega-\Omega)^2}{2\sigma^2}\right)$) distributions. Due to the rotational symmetry in the model, we can use a rotating frame and redefine $\omega_i\rightarrow\omega_i+\Omega$ for all $i$ and set $\Omega=0$, so that the $\omega_i$'s denote deviations from the mean frequency.

The whole population's collective dynamics is measured by the \emph{macroscopic} complex order parameter,
\be
r(t)\text{e}^{\text{i}\phi(t)}=\frac{1}{N}\sum_{j=1}^{N}\text{e}^{\text{i}\theta_{j}(t)}\;,
\label{ekuraorderparam}
\ee
where the modulus $0\le r(t) \le 1$ measures the phase coherence of
the population and $\phi(t)$ is the average phase. The values $r\simeq 1$ and $r\simeq 0$ (where $\simeq$ stands for fluctuations of size $O(N^{-1/2})$) describe the respective limits in which all oscillators are phase locked and in which they all move incoherently. Multiplying both parts of Eq.~(\ref{ekuramodel}) by $\text{e}^{\text{i}\theta_i}$ and substituting the imaginary part into Eq.~(\ref{ekuramodel}) yields
\be
\dot{\theta}_{i}=\omega_{i}+Kr\sin{(\phi-\theta_{i})}
\hspace{0.5cm} (i=1,...,N)\;.
\label{ekuramf}
\ee
Equation\ (\ref{ekuramf}) states that each oscillator interacts with all the others only through the mean field quantities $r$ and $\phi$. The second term 
provides a positive feedback loop to the system's collective rhythm. Namely, as $r$ increases the population becomes more coherent, so the coupling between the oscillators is further strengthened and more of them can in-turn be recruited to take part in the coherent set.
Moreover, Eq.~(\ref{ekuramf}) allows for the calculation of the critical coupling parameter $K_c$ and the characterization of the order parameter
$
\lim_{t\rightarrow\infty}r_t(K)=r(K).
$
Looking for steady-state solutions, one assumes that $r(t)$ is constant and $\phi(t)$ rotates uniformly at frequency $\Omega$. Without loss of generality we can set in Eq.~(\ref{ekuramf}) $\phi=0$ and go into the rotating frame~\cite{Kuramoto75,Kuramoto84,abprs05}. 
This reveals two different long-term behaviors when the coupling is larger than the critical value, $K_c$. On the one hand, when $|\omega_i|\leq Kr$, a group of oscillators are phase-locked at frequency $\Omega$ in the original frame according to the equation $\omega_{i}=Kr\sin{(\theta_{i})}$. On the other hand, the rest of oscillators for which $|\omega_i| > Kr$ is verified, are drifting around the circle, sometimes accelerating and sometimes rotating at lower frequencies. Demanding that some conditions hold for the stationary distribution of drifting oscillators with frequency $\omega_i$ and phases $\theta_i$~\cite{abprs05}, a self-consistent equation for $r$ can be derived as
\be
r=Kr\int_{\frac{-\pi}{2}}^{\frac{\pi}{2}}\cos^2\theta g(\omega)d\theta,
\nonumber
\ee
which admits a non-trivial solution corresponding to the existence of partially synchronized oscillators given by
\be
K_c=\frac{2}{\pi g(0)}.
\label{ekurakc}
\ee
Equation \ (\ref{ekurakc}) corresponds to the mean field approach for the critical coupling at the onset of synchronization. Moreover, near the onset, the order parameter, $r$, obeys the usual square-root scaling law for mean field models: $r\sim(K-K_c)^{\beta}$
with $\beta=1/2$. Numerical simulations of the model have proven that the mathematical approach developed by Kuramoto matches the numerics. However, in~\cite{cdj17} the authors prove that the coupling range $\delta K$ over which this scaling is valid shrinks like $\delta K \sim N^{-\alpha}$ with $\alpha \approx 1.5$ as $N \rightarrow \infty$. Away from this interval, the order parameter exhibits the scaling behavior $r-r_c\sim(K-K_c)^{2/3}$ predicted by~\cite{p05}. In this latter work, a first-order phase transition was revealed from imposing a uniform frequency distribution on the Kuramoto model. The transition is second-order for other distributions of frequencies that are not strictly uniform (Lorentzian Gaussian, parabolic, etc.).

The KM approach to synchronization was a great breakthrough for the understanding of the emergence of synchronization in large populations of oscillators. Even in the simplest case of a mean field interaction, there remain unsolved problems that have resisted any analytical attempt. This is the case, e.g., for finite populations of oscillators with respect to some questions regarding global stability results~\cite{Bonilla1992,abprs05}. In what follows, however, we focus on another aspect of the model's assumptions, namely that the connection topology of real systems~\cite{n03a,blmch06} usually do not show a homogeneous pattern of interconnections as required by the mean field approach, but instead posses complex patterns encoded as a network, see reviews~\cite{adkmz08,db14}. Note, the KM is still a subject of relevance in understanding biological systems, for example, very recently~\cite{hfb18} introduced a macroscopic reduction for networks of coupled oscillators motivated by a structure found in experimental measurements of circadian protein expression and in several mathematical models for coupled biological oscillators. They provide the first evaluation of the suitability of the Ott-Antonsen~\cite{oa08} reduction procedure for extracting macroscopic models of real biological networks.

\subsection{The Kuramoto model in networks}
\label{sec:KM_networks}

The KM was also one of the first dynamical processes to be analyzed on 
a complex network topology.
However, to do so, it was necessary first to reformulate Eq.~(\ref{ekuramodel}) to include the network connectivity between the oscillators:
\be
\dot{\theta}_i=\omega_i + \sum_{j}
\lambda_{ij}A_{ij}\sin(\theta_j-\theta_i) \hspace{0.5cm} (i=1,...,N)\;,
\label{ekurageneral}
\ee
where $\lambda_{ij}$ is the coupling strength between pairs of connected oscillators and $A_{ij}$ are the elements of the adjacency matrix encoding the connections between the nodes of the network.  
Note that the original KM is recovered by letting $A_{ij}=1, \forall i\neq
j$ (all-to-all) and $\lambda_{ij}=K/N, \forall i,j$. In what follows, we will also consider $\lambda_{ij}=\lambda$, so that the coupling strength between any pair of connected nodes is identical.

Studies of synchronization on complex network topologies, where each node is considered to be a Kuramoto oscillator, were first reported for Watts-Strogatz (WS) networks~\cite{w99,hck02} and Barab{\'a}si-Albert (BA) scale-free (SF) graphs~\cite{mp04,mvp04}, see~\cite{rpjk16} for a review. These works are mainly numerical explorations of the onset of synchronization, with the main goal being 
identifying the critical coupling above which sets of nodes beating coherently first appear.  In~\cite{hck02}, the authors considered oscillators with intrinsic frequencies distributed according to a Gaussian distribution with unit variance that are arranged in a WS network. 
They explored how the order parameter, Eq.~(\ref{ekuraorderparam}), changes upon addition of long-range links for varying rewiring probability, $p$. Moreover, they assumed a normalized coupling strength $\lambda_{ij}=\lambda/\langle  k \rangle$, where $\langle  k \rangle$   is the average degree of the graph. Numerical integration of the equations of motion (\ref{ekurageneral}) under variation of $p$ shows that collective synchronization emerges even for very small values of the rewiring probability.

The results confirm that networks obtained from a regular ring with just a small 
fraction of rewired links ($p\gtrsim 0$) can be synchronized with a finite $\lambda$. Moreover, in contrast with the arguments provided in~\cite{hck02}, we notice that their results had been obtained for a fixed average degree and thus the KM's critical coupling cannot be recovered by simply taking $p\rightarrow 1$ (which produces a random ER graph with a fixed minimum connectivity).  This limit is recovered by letting $\langle  k \rangle$   increase. 
More specifically, numerical simulations of the same model in~\cite{w99} showed that the Kuramoto limit is approached as the average connectivity grows.

The existence of a critical point for the KM on SF networks came as a surprise. This is one of the few cases in which a dynamical process shows a critical behavior when the substrate is described by a power-law connectivity distribution with an exponent $\gamma\leq 3$~\cite{n03a,blmch06,dgm07}.
In principle it could be a finite size effect, but results from numerical simulations show that this is not the case. To determine the exact value of $\lambda_c$, one can make use of standard finite-size scaling analysis. At least two complementary strategies have been reported. The first one allows bounding the critical point and is computationally more expensive. Consider a network of size $N$, for which
no synchronization is attained below $\lambda_c$, where $r(t)$ decays to a small residual value of size $O(1/\sqrt{N})$. Then, the critical point may be found by examining the $N$-dependence of $r(\lambda,N)$. In the sub-critical regime ($\lambda < \lambda_c$), the stationary value of $r$ falls off as $N^{-1/2}$, while for $\lambda >\lambda_c$, the order parameter reaches a stationary value as $N\rightarrow\infty$ (though still with $O(1/\sqrt{N})$ fluctuations). Therefore, plots of $r$ versus $N$ allow us to locate the critical point $\lambda_c$.

Before turning our attention to some theoretical attempts to tackle the onset of synchronization, it is worth briefly summarizing other numerical results that have explored how the critical coupling depends on other topological features of the underlying SF graph. Recent results have shed light on the influence of the topology of the local
interactions on the route to and the onset of synchronization. In particular, the authors in~\cite{mm05,mm07,gm07} explored the Kuramoto dynamics on networks in which the degree distribution is kept fixed, while the
clustering coefficient ($C$) and the average path length ($\ell$) of the graph change. The results suggest that the onset of
synchronization is mainly determined by $C$, namely, networks with a high clustering coefficient promote synchronization at lower values of the coupling strength. On the other hand, when the coupling is increased beyond
the critical point, the effect of $\ell$ dominates over $C$ and
the phase diagram is smoothed out (a sort of stretching), delaying the
appearance of the fully synchronized state as the average shortest path length increases.

In a series of papers~\cite{i04,i05,roh05a,roh05b,l05,roh06a}, the onset of synchronization in large networks of coupled oscillators has been analyzed from a theoretical point of view under certain (sometimes strong) assumptions. Despite these efforts no exact analytical results for the KM on general complex networks are available up to date. Moreover, the analytical approaches predict that for uncorrelated SF networks with an exponent $\gamma\leq 3$, the critical coupling vanishes as $N\rightarrow\infty$, in contrast to numerical simulations on BA model networks. It appears that the strong heterogeneity of real networks and the finite average connectivity strongly hampers analytical solutions of the model.

The analytical insights discussed so far can be formulated in terms of a mean field approximation for complex networks~\cite{i04,l05,roh05a,i05}. This approach (valid for large enough $\langle  k \rangle$) considers that every oscillator is influenced by the local field created in its neighborhood. To this aim, a local order parameter defined as:
\begin{equation}
r_i\text{e}^{\text{i}\phi_i}=\sum_{j=1}^{N}A_{ij}\la\text{e}^{\text{i}\theta_{j}}\ra_t\;,
\label{ekrlocal}
\end{equation}
is introduced,  where $\la\cdot\cdot\cdot\ra_t$ stands for a time average. Considering that the degree of a node $i$ is large enough, and that the network is close to the onset of synchronization, $\lambda\sim\lambda_c$, it is possible to write the local order parameter $r_i$ as:
\begin{eqnarray}
r_i&=&\sum_{j}A_{ij}\int_{-\lambda r_j}^{\lambda r_j}g(\omega)\sqrt{1-\left(\frac{\omega}{\lambda r_j}\right)^2}d\omega\nonumber\\
&=&\lambda\sum_{j}A_{ij}r_j\int_{-1}^{1}g(x\lambda r_j)\sqrt{1-x^2}dx\;.
\label{erestrepo6}
\end{eqnarray}

The definition of $r_i$ allows us to define a new global order parameter $r$ to measure the global coherence.
Moreover, assuming that  $r_i$ is proportional to the degree of the corresponding node $k_i$, i.e., $r_i\sim k_i$, one can write $r$ as:%
\be
r=\frac{r_i}{k_i}=\frac{1}{k_i}\left |\sum_{j=1}^{N}A_{ij}\langle  \text{e}^{\text{i}\theta_j}\ra_t\right |\;.
\label{elee1}
\ee
After summing over $i$ and substituting $r_i=rk_i$ in Eq.~(\ref{erestrepo6}) we obtain:
\be
\sum_{j}^{N}k_j=\lambda\sum_{j}^{N}k_j^2\int_{-1}^{1}g(x\lambda r k_j)\sqrt{1-x^2}dx\;.
\label{elee2}
\ee
The above relation, Eq.~(\ref{elee2}), was independently derived in~\cite{i04}, who first studied analytically the problem of synchronization in complex networks, though a different approach. Taking the continuum limit and considering the limit $r\rightarrow 0^{+}$ ($\lambda\rightarrow \lambda_c^{+}$),  Eq.~(\ref{elee2}) becomes
\bea
\int kP(k)dk&=&\lambda_c\int k^2P(k)dk\int_{-1}^{1}g(0)\sqrt{1-x^2}dx\nonumber\\
&=&\frac{\lambda_c g(0)\pi}{2}\int k^2P(k)dk\;,
\label{elee4}
\eea
which leads to the condition for the onset of synchronization: 
\be
\lambda_c=\frac{2}{\pi g(0)}\frac{\langle  k\ra}{\langle  k^2\ra}=K_c\frac{\langle  k\ra}{\langle  k^2\ra}\;.
\label{elee5}
\ee
The mean field result, Eq.~(\ref{elee5}), gives us a surprising result that the critical coupling $\lambda_c$ in complex networks is nothing else but the one corresponding to the all-to-all topology $K_c$ re-scaled by the ratio between the first two moments of the degree distribution, regardless of the many possible differences between the patterns of interconnections of the underlying network. Precisely, it states that the critical coupling strongly depends on the topology of the underlying graph.
In particular, $\lambda_c\rightarrow 0$ when the second moment of the distribution $\langle  k^2\rangle$   diverges, which is the case for SF networks with $\gamma\le 3$. Note, the contrast with the result for $\gamma >3$, when the coupling strength does not vanish in the thermodynamic limit. On the other hand, if the mean degree is kept fixed and the heterogeneity of the graph is increased by decreasing $\gamma$, the onset of synchronization occurs at smaller values of $\lambda_c$. Note that Eq.~(\ref{elee5}) is mathematically correct only for locally tree-like networks, however as pointed out in~\cite{i05}, the approximation works well even for clustered networks, depending on the probabilities of the different link forming triangles. This surprising effect was scrutinized later to unveil the unreasonable effectiveness of tree-based theory for networks with clustering~\cite{mhpmg11} where the authors proved that such a theory works well as long as the mean intervertex distance {\it l} is close to the expected value of  {\it l} in a random network with negligible clustering and the same degree-degree correlations. 

Interestingly enough, the dependence derived in Eq.~(\ref{elee5}) has the same functional form for the critical points of other dynamical processes such as percolation and epidemic spreading processes~\cite{n03a,blmch06, dgm07}. While this result is still under numerical scrutiny, it would imply that the critical properties of many dynamical processes on complex networks are essentially determined by the topology of the graph, no matter whether the dynamics is nonlinear or not. The corroboration of this last claim will be of extreme importance in physics, probably changing many preconceived ideas about the nature of dynamical phenomena.

\begin{figure}[t!]
\begin{center}
\includegraphics[width=.45\textwidth,angle=0,clip=1]{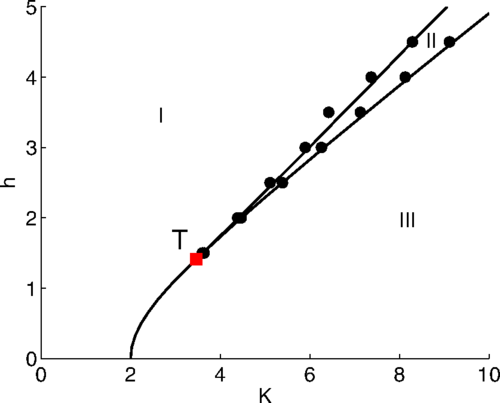}
\end{center}
\caption{
Phase diagram of the Kuramoto model on a complete graph in the presence of homogeneous random fields of magnitude h, as a function of the coupling K.  The solid line corresponds to the theoretical prediction and the points to MonteCarlo simulations. 
There are three regions: (I) asynchronous state, (II) region of hysteresis, and (III) partially synchronous state. 
The red symbol is a tricritical point below which there is a second-order phase transition, whereas above it there is a first-order phase transition 
[Reprinted from~\cite{llymg16}].}
\label{fig:Lopes2016}
\end{figure}

Within the mean field theory, it is also possible to obtain the behavior of the order parameter $r$ near the transition to synchronization. 
Another theoretical study in~\cite{olkk07} is worth mentioning here. They  have extended the mean field approach to the case in which the coupling is asymmetric and depends on the degree. In particular, they studied a system of oscillators arranged in a complex topology whose dynamics is given by
\be
\dot{\theta}_i=\omega_i+\frac{\lambda}{k_i^{1-\eta}} \sum_{j=1}^{N} A_{ij}\sin(\theta_j-\theta_i).
\ee
Here, $\eta=1$ corresponds to the symmetric, non-degree dependent, case. Extending the mean field formalism to the cases $\eta\neq1$, they investigated the nature of the synchronization transition as a function of the magnitude and sign of the parameter $\eta$. By exploring the whole parameter space $(\eta,\lambda)$, they found that for $\eta=0$ and SF networks with $2 <\gamma<3$, a finite critical coupling $\lambda_c$ is recovered in sharp contrast to the non-weighted coupling case for which we already know that $\lambda_c=0$. This result seems phenomenologically meaningful, since setting $\eta=0$ implies that the coupling in Eq.~(\ref{ekurageneral}) is $\lambda_{ij}=\lambda/k_i$, which, as discussed earlier~\cite{mzk05b}, may have the effect of partially destroying the heterogeneity inherent to the underlying graph by normalizing all the contributions $\sum_{j=1}^{N} A_{ij} \sin(\theta_j-\theta_i)$ by $k_i=\sum_{j=1}^{N} A_{ij}$.

In a more complex scenario~\cite{llymg16} the authors addressed the impact of random fields on the emergence of synchrony in the Kuramoto model on complete (fully-connected) graphs and uncorrelated random complex networks. Considering random fields with uniformly distributed directions and homogeneous and heterogeneous (Gaussian) field magnitude distribution, they  prove that the critical behavior of the order parameter presents a tricritical point above which a second-order phase transition gives place to a first-order phase transition when the network is either fully connected or scale-free with the degree exponent $\gamma > 5$. The tricritical point is at the intersection of three regions: (I) asynchronous state, (II) region of hysteresis, and (III) partially synchronous state. Below the tricritical point there is a second-order phase transition, whereas above it there is a first-order phase transition, see Fig.~\ref{fig:Lopes2016}.

Moreover, for scale-free networks with exponents $2 < \gamma <5$, the phase transition is of second-order at any field magnitude, except for degree distributions with $\gamma = 3$ when the transition is of infinite order at $\lambda_c = 0$ independent of the random fields.

\subsection{Explosive Synchronization in networks}
\label{sec:ES_networks}

We now introduce the basics of ES phenomena as a genuine effect of the interplay between the structure of the interaction networks and oscillators' dynamics. As already mentioned, discontinuous transitions in the usual (all-to-all) Kuramoto model were found in~\cite{p05} by using broad and homogeneous frequency distributions $g(\omega)$. However, the abrupt transitions observed in the synchronization diagram $r(\lambda)$ do not lead to any bi-stable region in which, for a range of coupling strengths, the synchronized and the incoherent states coexist. In addition, in degree heterogeneous networks the use of this kind of frequency distributions does not alter the nature of the transition. On the contrary, the use of homogeneous frequency distributions in scale-free networks~\cite{mp04,gma07b} corroborates through finite size scaling the second-order nature of the transition, with $\beta\sim 1/2$.  This finding was obtained together with the expectation that the onset of synchronization is enhanced as degree heterogeneity increases (controlled by the second moment of the degree distribution $\langle k^2\rangle$) as reported in Eq.~(\ref{elee5}).

For the great majority of abrupt ES phenomena there are no rigorous proofs for the occurrence of a genuine discontinuity. Yet, there is often compelling numerical evidence for it. In other words, the phase transition categorization of ES, in analogy to that sketched-out in Fig.~\ref{fig:EP_classes} for EP, is unknown. As a result, we will call explosive synchronization transitions which may be genuinely discontinuous by the term {\em abrupt}, to avoid unnecessary overstatements.

\subsection{The correlated degree-frequency (CDF) framework}

As introduced above, the ability of scale-free networks to synchronize relies on the role of hubs in centralizing the growth of a single synchronized connected component~\cite{gma07a}. Exploiting the idea that by delaying the onset of a collective behavior, such as synchronization, it is possible to change the nature of the corresponding transition, G\'omez-Garde\~nes {\it et al.}~\cite{ggam11}  proposed a modified Kuramoto model in which the frequency of each oscillator is correlated with its degree, $\omega_i=f(k_i)$. This degree-frequency correlation introduces an interplay between structural and dynamical attributes of the nodes. In particular, they chose to investigate the simplest linear correlation, $\omega_i=k_i$, thus 
ofsetting the topological ability of hubs to synchronization (provided by their large connectivity) by assigning them a very disparate natural frequency.  Under this framework the Kuramoto evolution equations become
\begin{equation}
\dot\theta_{i}(t)=k_i+\lambda\sum_{i=1}^{N}A_{ij}\sin(\theta_j(t)-\theta_i(t))\;.
\label{explosive_or}
\end{equation} 
In addition, the perfect correlation between degrees and natural frequencies automatically sets: $g(\omega)=P(k)$ and, consequently, $\langle k\rangle=\Omega$.

\begin{figure}[!t]
\begin{center}
\includegraphics[width=.5\textwidth,angle=-90,clip=1]{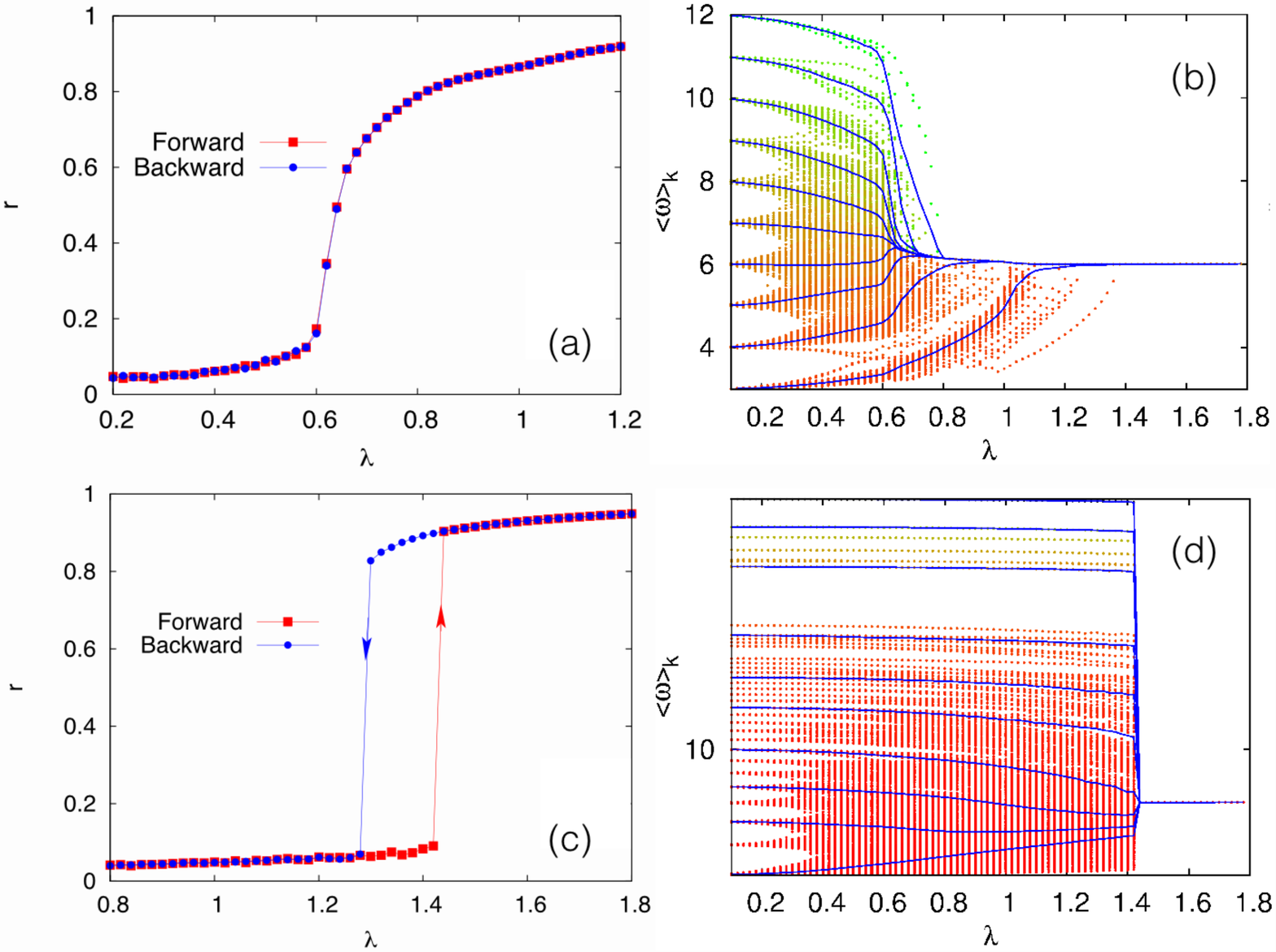}
\end{center}
\caption{Backward and forward synchronization curves $r(\lambda)$ for an ER (a) and SF (c) networks of $N=10^3$ and $\langle k\rangle=6$. Panels (b) and (d) show the evolution of the effective frequencies of nodes (dots) and the average effective frequency of nodes belonging to different degree classes (lines) as a function of $\lambda$ in the forward continuation branch. [Reprinted and combined from~\cite{ggam11}].}
\label{fig:ggam11}
\end{figure}

The numerical study of this model was done in homogeneous and scale-free networks by monitoring the synchronization transition $r(\lambda)$ though continuation, {\em i.e.} by increasing (forward) and decreasing (backward) the coupling $\lambda$, in steps of length $\delta\lambda\ll 1$. This way, the initial condition for each $\lambda$ value is exactly the final configuration, $\{\theta_{i}\}$, obtained for the previous value of the coupling strength. In Fig.~\ref{fig:ggam11}(a) the forward and backward branches of the synchronization diagram are shown for homogeneous ER networks. In this case the transition is smooth and both branches are equal thus indicating that the transition is, as usual, of second order. In addition, the authors computed the evolution of the effective frequency of each node defined as
\begin{equation}
\omega^{\mbox{\scriptsize eff}}_{i}=\frac{1}{T}\int_{t}^{t+T}\dot{\theta_{i}}(\tau)\;{\mbox
    d}{\tau}\;,
    \label{effective1}
\end{equation}
with $T\gg 1$. The evolution with $\lambda$ of these effective frequencies  is shown in Fig.~\ref{fig:ggam11}(b) for the ER network. From this panel we observe that some of the first oscillators to converge to 
the average frequency ($\Omega=\langle k\rangle=6$ in this network) are those have the largest degree. This observation is confirmed by computing (see solid lines in Fig.~\ref{fig:ggam11}(b)) the average effective frequency of nodes having the same degree (and hence the same natural frequency) as
\begin{equation}
\langle \omega\rangle_{k}=\frac{1}{N_{k}}\sum_{[i|k_{i}=k]}\omega^{\mbox{\scriptsize eff}}_{i}\;,
\label{effective2}
\end{equation}
where $N_{k}=NP(k)$ is the number of nodes with degree $k$ in the network. 


The above scenario dramatically changes when considering SF networks (see Figs.~\ref{fig:ggam11}(c) and~\ref{fig:ggam11}(d)). In this case the synchronization diagrams show an abrupt phase transitions and, moreover, the forward and backward branches do not coincide, pointing out hysteresis phenomenon. The abrupt change of $r(\lambda)$ becomes evident by looking at the evolution (by increasing $\lambda$) of the effective frequencies as defined in Eqs. (\ref{effective1}) and (\ref{effective2}). At variance with the ER network, the effective frequencies of all the nodes remain roughly locked to their natural ones ($\omega_i=k_i$) until the critical (forward) coupling, $\lambda\sim 1.42$, for which all the effective frequencies collapse into the same value $\Omega=\langle k\rangle=6$.

\subsubsection{Analytical characterization}

The numerical finding of an explosive  synchronization transition in scale-free networks motivated the theoretical analysis of the Kuramoto model with degree-frequency correlations as proposed in~\cite{ggam11}. In particular, in~\cite{pr12} the expression of the critical coupling, $\lambda_c$ was analyzed making use of the framework introduced in~\cite{i04} for the analysis of the critical properties of the Kuramoto model in complex networks. This way, by considering the density of nodes with degree $k$ and phase $\theta$ at time $t$, $\rho(k;\theta, t)$ and taking the continuum limit for the degree variable, Eq.~(\ref{explosive_or}) can be rewritten as
\begin{equation}
\frac{{\mbox d} \theta}{{\mbox d}t}=k + \lambda k \int dk'\int d\theta' P(k'|k)\rho(k;\theta',t) \sin(\theta - \theta')\;,
\label{eq:kuramoto_continuum}
\end{equation}
where $P(k'|k)$ is the probability that, given a node of degree $k$, it has a neighbor 
of degree $k'$. For uncorrelated networks this is simply: $ P(k'|k)=k'P(k')/\langle k\rangle$. After some derivations in the same line of~\cite{i04}, the Kuramoto order parameter, $r$, is found to satisfy the following equation 
\begin{equation}
\left\langle k\right\rangle r=\int_{\left\langle k\right\rangle /\left(1+\lambda r\right)}^{\left\langle k\right\rangle /\left(1-\lambda r\right)}kP(k)\sqrt{1-\frac{1}{\lambda^{2}r^{2}}\left(\frac{k-\left\langle k\right\rangle }{k}\right)^{2}}dk.
\label{eq:r_real_part}
\end{equation}
The non-trivial solution $r\neq 0$ close to the synchronization onset ($r\rightarrow 0^{+}$) yields the critical coupling
\begin{equation}
\lambda_c=\frac{2}{\pi \langle k\rangle P(\langle k\rangle)}\;.
\label{eq:estimationpr12}
\end{equation}
Note that this expression (for the threshold for the degree correlated framework) is different from the generic one for the synchronization threshold of the Kuramoto model in a network  
shown in Eq.~(\ref{elee5}). The validity of Eq.~\ref{eq:estimationpr12} 
was numerically tested for different synthetics networks. 

In a later work~\cite{ps15}, the analytical derivation of the critical coupling was generalized to the case in which degree-frequency correlation only holds for oscillators with $k\geq k^{*}$, while for nodes with $k< k^*$ the frequency  is randomly assigned following a distribution $g(\omega)$ as in the usual Kuramoto model. Following again the approach introduced in~\cite{i04} the expression for the critical coupling becomes
\begin{equation}
\lambda_c=\frac{2\langle k\rangle}{\pi[ \alpha g(\beta)+\beta^2 P(\beta)]}\;,
\label{eq:estimationps15}
\end{equation}
where 
\begin{eqnarray}
\alpha&=&\int_{k_{min}}^{k^*}k^2P(k)dk\;,\\
\beta&=&\frac{\int_{k^*}^{\infty}kP(k)dk}{\int_{k^*}^{\infty}P(k)dk}\;.
\end{eqnarray} 
It is easy to check that when $k^{*}=k_{min}$, so that the degree-frequency correlation holds for all of the oscillators, then $\alpha=0$ and $\beta=\langle k\rangle$, recovering the expression for the critical coupling shown in Eq.~(\ref{eq:estimationpr12}).

\begin{figure}[!t]
\begin{center}
\includegraphics[width=.45\textwidth,angle=0,clip=1]{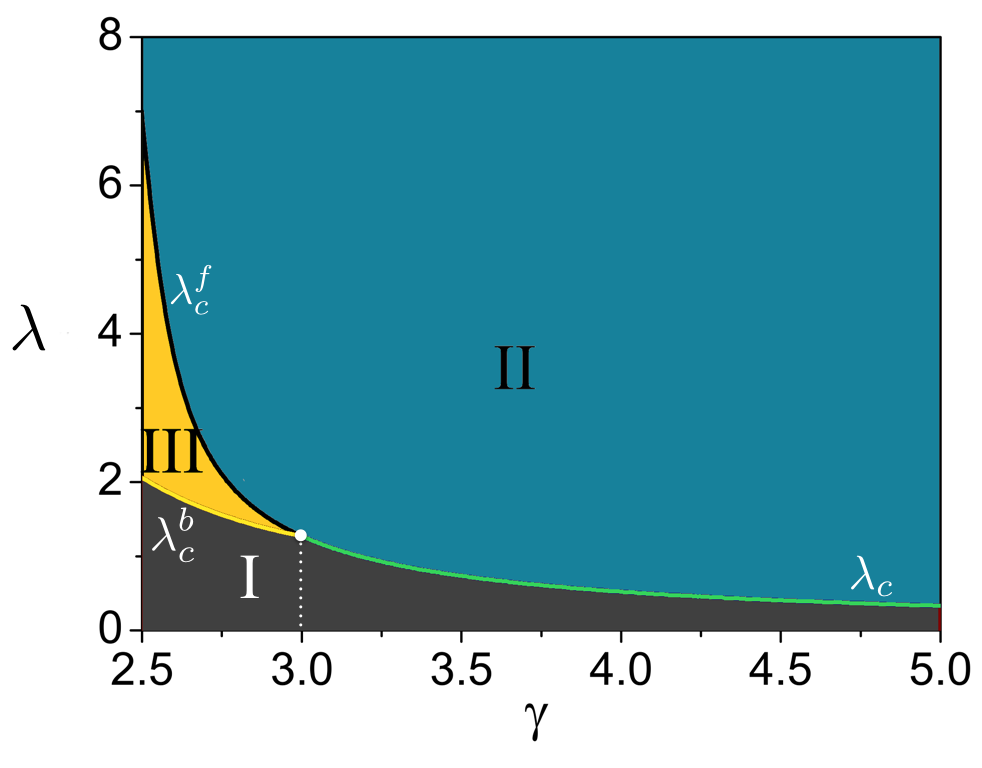}
\end{center}
\caption{Synchronization diagram in the $(\lambda,\gamma)$-plane. Phase I corresponds to the incoherent solution $r=0$ while phase II accounts for the synchronized regime. For $\gamma<3$ the synchronization transition in $\lambda$ becomes of first order and a new phase III appears. This phase indicates that the synchronized and incoherent solution coexist and thus corresponds to the region where hysteresis appears. Phase III is bounded by the lines corresponding to the critical couplings for the forward, $\lambda_c^f$, and backward, $\lambda_c^b$, continuations. [Reprinted and adapted from~\cite{cgdm13}].}
\label{fig:cgdm13}
\end{figure}

The dependence of the explosive transition on the degree heterogeneity of the underlying network was analyzed in detail in~\cite{cgdm13}. By using an annealed network approach, {\em i.e.} working with an effective adjacency matrix with elements $A_{ij}=k_{i}k_{j}/(2\langle k\rangle)$, the authors showed that the nature of the phase transition in scale-free networks depends on the exponent $\gamma$ of the degree distribution, $P(k)\sim k^{-\gamma}$. In particular, under the assumption of the annealed approach, scale-free networks with  $2<\gamma<3$ display a first order phase transition while for $\gamma>3$ the transition becomes 
second order. Interestingly, in the case $\gamma=3$, the CDF Kuramoto model undergoes an hybrid phase transition, {\em i.e.} the synchronization order parameter $r$ shows an abrupt change yet displays the critical properties typical of second order transitions. The authors show that for $\gamma=3$ the critical behavior close to the synchronization onset is: $r-r_c \propto (\lambda-\lambda_c)^{2/3}$.
Figure~\ref{fig:cgdm13} shows the phase diagram in the $(\lambda,\gamma)$-plane. This diagram shows, for $\gamma<3$, a phase (III)  where the fully synchronized state and the unsynchronized solution coexist and it is this which leads to the hysteresis phenomenon. Remarkably, the region associated with this phase increases as the exponent $\gamma$ of the degree distribution decreases, indicating that the hysteresis cycle becomes larger in this regime.

\subsubsection{The star network}
\label{sec:star}

A common way to explore collective phenomena found in scale-free networks is to analyze star-like networks. These networks capture in a minimal way the degree heterogeneity found in scale-free networks while allowing analytical calculations without strong assumptions. In the original article showing the existence of ES in the CDF Kuramoto model~\cite{ggam11} the authors  analyzed this toy graph composed of a central hub and $K$ peripheral nodes. Under the CDF scheme the frequency of the hub becomes $\omega_h=K$ whereas the leaves ({\it i.e.,} nodes of degree one) have $\omega=1$. This way, the Kuramoto evolution becomes
\begin{eqnarray}
\dot{\theta_h}&=&K+\lambda\sum_{i=1}^{K}\sin(\theta_{i}-\theta_{h}),
\label{eq:hub}
\\ \dot{\theta_i}&=&1+\lambda\sin(\theta_{h}-\theta_{i}),\;{\mbox{with}}\;i=1...K,
\label{eq:leaves}
\end{eqnarray}
and the average frequency is $\Omega=2K/(K+1)$. 

\begin{figure}[!t]
\begin{center}
\includegraphics[width=.45\textwidth,angle=0,clip=1]{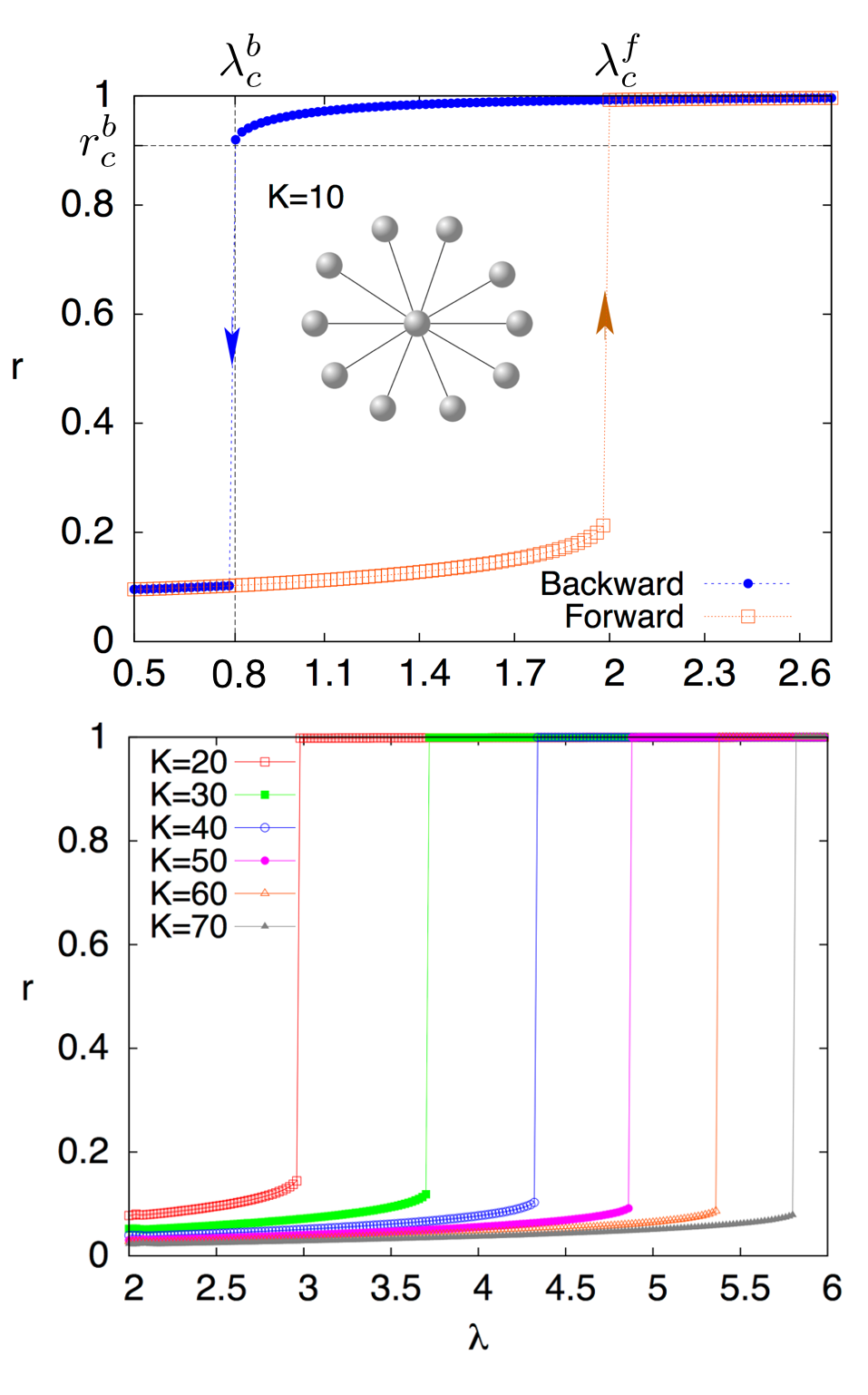}
\end{center}
\caption{(Top) Forward and backward synchronization diagrams for a star network with $K=10$ leaves. (Bottom) Forwad synchronization diagrams for star networks with different number of leaves. [Reprinted and adapted from~\cite{ggam11}].}
\label{fig:star}
\end{figure}

The numerical study of the star network is shown in Fig.~\ref{fig:star}. The top panel shows the backward and forward synchronization diagrams for a star network with $K=10$ leaves and illustrates the hysteresis cycle that is bounded by the two coupling values $\lambda_c^{b}$ and $\lambda_c^{f}$. In~\cite{ggam11} the authors derived the exact value of the critical coupling $\lambda_c^{b}$ by considering the backward branch and inspecting the necessary conditions for the existence of the locked solution as an attractor of the dynamics. To this aim, one can set the phase of the hub as the reference point 
by defining  $\phi_i=\theta_h-\theta_i$ so that the set of Eqs. (\ref{eq:hub}) and (\ref{eq:leaves}) transforms into
\begin{equation}
\dot{\phi_i}=(K-1)-\lambda\sum_{j=1}^{K}(1+\delta_{ij})\sin(\phi_{j}),\;{\mbox{with}}\;i=1...K.
\label{startransform}
\end{equation}
In the locked regime $\dot\phi_{i}=0$ and $\phi_i=\phi$ $\forall i$ so that the condition for the synchronized solution to exist is
\begin{equation}
\sin(\phi)=\frac{K-1}{\lambda(K+1)}\;.
\end{equation}
The minimum $\lambda$ value that satisfies the former equation is
\begin{equation}
\lambda_c^{b}=\frac{K-1}{K+1}\;.
\label{backward}
\end{equation}
Remarkably, the value $\lambda_c^{b}\rightarrow 1$ as $K$ increases, thus being almost independent of the number of leaves surrounding the hub. This result was also obtained in~\cite{zpslk14} by analyzing the local attractiveness of the synchronized solution, allowing them to derive as well the value of the synchronization parameter $r$ at $\lambda_c^{b}$, $r_c^{b}$ (see Fig.~\ref{fig:star}), {\em i.e.} when the locking manifold disappears,
\begin{equation}
r_c^{b}=\frac{\sqrt{K^2+1}}{K+1}\;.
\label{rub}
\end{equation}
An estimation of $r_c^{b}$ was also derived in~\cite{ggam11} yielding a slightly different result: $r_c^{b}=K/(K+1)$, which coincides with Eq.~(\ref{rub}) when $K\gg 1$.

The calculation of the upper bound, $\lambda_c^{f}$, of the hysteresis cycle is more complicated and was tackled in~\cite{zpslk14}. Making use of the theory developed in~\cite{pebtj13}, the authors estimate the value $\lambda_c^{f}$ by identifying the value of $\lambda$ for which the basin of attraction of the locked manifold passes from being local (and thus coexisting with the unlocked solution) to global. Their derivation yield a value of: $\lambda_c^{f}\simeq0.6989(K-1)/\sqrt{K})$. This value was later calculated in~\cite{xgshz15} yielding
\begin{equation}
\lambda_c^{f}=\frac{(K-1)}{\sqrt{2K+1}}\;,
\label{forward}
\end{equation}
that for $K\gg 1$ coincides with the one derived in~\cite{zpslk14}. These two results point out a scaling of the forward critical coupling as $\lambda_c^{f}\propto \sqrt{K}$, pointing out that the length of the hysteresis cycle increases with $K$, since $\lambda_c^{b}$ remains bounded ($\lambda_c^{b}\rightarrow 1$). This effect is numerically illustrated in the bottom panel of Fig.~\ref{fig:star} where the forward continuation branch $r(\lambda)$ is shown for several star graphs with different values of $K$.

Another interesting approach to the analysis of the star network was presented in~\cite{vzp15}. In particular,  Vlasov {\em et al.} noticed that the Kuramoto equations for the star, transformed as in Eq.~(\ref{startransform}), allow a direct application of the approach introduced in~\cite{watanabes93,watanabes94}. This way, the set of $K$ coupled equations can be studied by analyzing a single evolution equation for a complex number $z(t)$ whose modulus $|z|$ is the order parameter $r$
\begin{equation}
\dot{z}={\mbox i}\left[-(K-1)-K\lambda{\mbox{Im}}(z)\right]z+\lambda\frac{1-z^2}{2}\;.
\label{watanabe}
\end{equation}
By looking for the steady state solutions of Eq.~(\ref{watanabe}) the authors obtain three such steady state solutions, namely
\begin{eqnarray}
z^{b}&=&{\mbox{exp}}\left\{{\mbox i}\;\arcsin\left(-\frac{(K-1)}{(K+1)\lambda}\right)\right\}\;{\mbox{if}}\;\lambda>\lambda^{b}_c\;,
\nonumber
\\
z^{f}&=&-{\mbox i}\frac{(K-1)-\sqrt{(K-1)^2-(2K+1)\lambda^2}}{(2K+1)\lambda}\;{\mbox{if}}\;\lambda<\lambda^{f}_c\;,
\nonumber
\\
z^{u}&=&-{\mbox i}\frac{(K-1)+\sqrt{(K-1)^2-(2K+1)\lambda^2}}{(2K+1)\lambda}\;{\mbox{if}}\;\lambda\in(\lambda^{b}_c,\lambda^{f}_c)\;.
\nonumber
\end{eqnarray}
Notice that the moduli of these three solutions, $|z^{b}|$, $|z^{f}|$ and $|z^{u}|$ correspond to the three possible values for the order parameter $r$. After performing the stability analysis of the three solutions the authors show that only $z^{b}$ and $z^{f}$ are stable, so that $|z^{b}|$ corresponds to the backward (upper) synchronization branch and  $|z^{f}|$ to the forward (lower) one. Moreover, the existence bounds of these  solutions ($\lambda^{b}_c$ and $\lambda^{f}$) are exactly the expressions in Eq.~(\ref{forward}) and Eq.~(\ref{backward}) respectively, thus recovering the limits of the hysteresis cycle previously reported. 

\subsubsection{Effect of structural properties}
 
 Up to now we have reported the results of the CDF Kuramoto model in ER and SF networks, {\em i.e.} focusing on the effect of degree heterogeneity on the appearance of ES. However, real networks display features like degree-degree correlations  between adjacent nodes and modular architectures that may have strong influence in the dynamical processes at work. For this reason, additional research has addressed the role that these two structural properties have on the onset of ES.
 
{\em Assortativity.-} The role of degree assortativity has been studied in several works~\cite{lwxz13,lzxzs13,slnvawb15}. In particular, in~\cite{slnvawb15}, the authors made an exhaustive study by constructing scale-free networks with tunable degree-degree correlations. The degree-degree correlations were characterized using a Pearson correlation coefficient between the degrees of adjacent nodes as introduced in~\cite{n03b} 
\begin{equation}
\alpha=\frac{L^{-1}\sum_{i>j}A_{ij}k_{i}k_{j} -\left[  L^{-1}\sum_{i>j}A_{ij}(k_{i}+k_{j})\right]^2}
{\frac{1}{2}L^{-1}\sum_{i>j}A_{ij}(k_{i}^2+k_{j}^2) -\left[ \frac{1}{2} L^{-1}\sum_{i>j}A_{ij}(k_{i}+k_{j})\right]^2}.
\end{equation}
The networks were produced by a targeted randomization of graphs generated both from preferential attachment and configurational models. The randomization procedure rewires links of the original networks in order to achieve the desired assortativity value $\alpha$ and, importantly, preserves the degree distribution of the original graphs. This way, the authors generate ensembles of networks each one characterized by two parameters: $\gamma$ (the exponent of the power-law degree distribution) and the assortativity $\alpha$.

\begin{figure}[!t]
\begin{center}
\includegraphics[width=.6\textwidth,angle=0,clip=1]{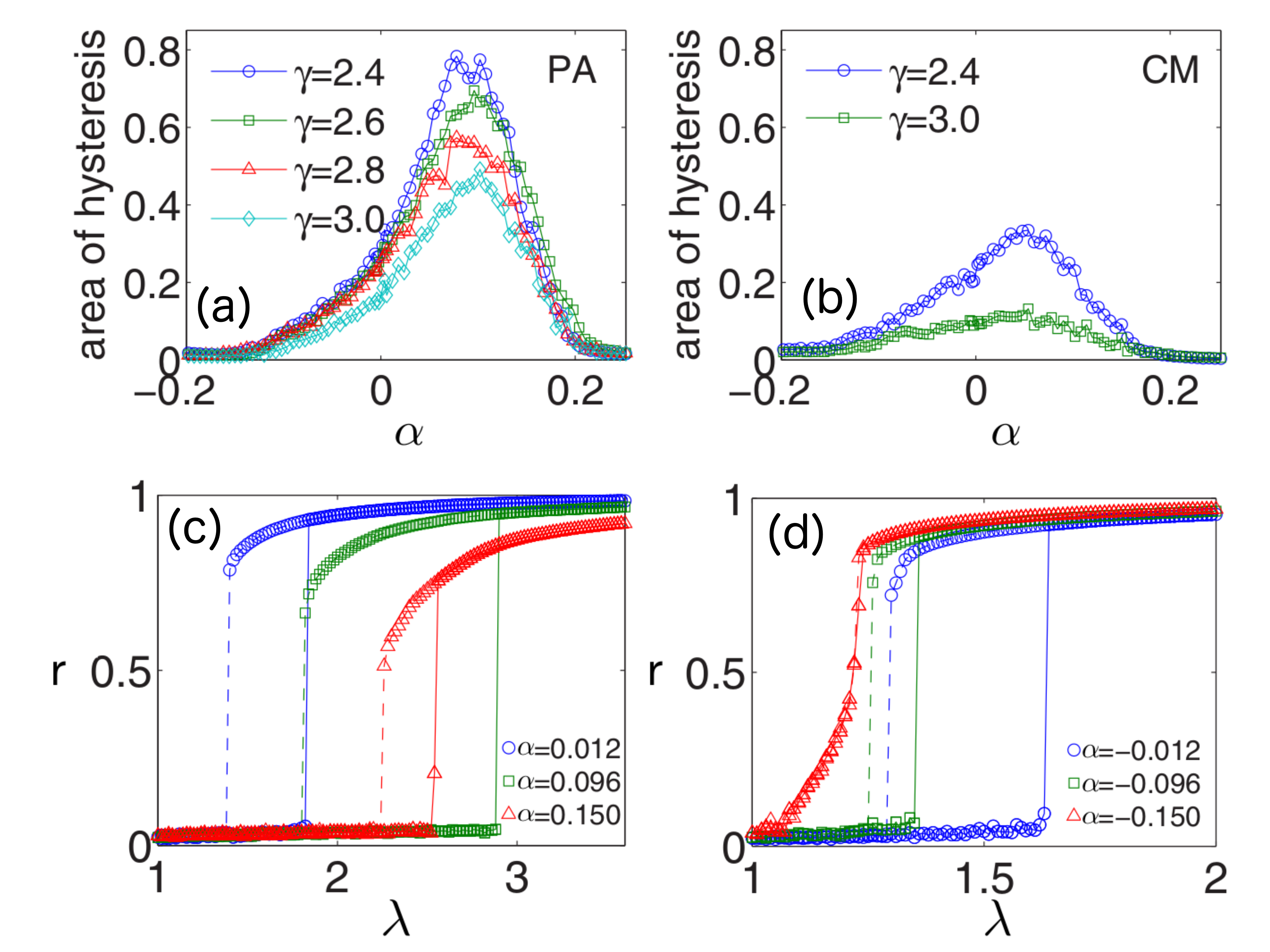}
\end{center}
\caption{(Top) Area of the hysteresis cycle as function of the assortativity $\alpha$ for different scale-free networks constructed from the preferential attachment model (a) and the configurational one (b). (Bottom)  Forward and backward synchronization diagrams $r(\lambda)$ for scale-free networks with $\gamma=2.4$ and positive (c) and negative (d) assortativity. [Reprinted and adapted from~\cite{slnvawb15}].}
\label{fig:correlations}
\end{figure}

The numerical simulation of the CDF Kuramoto model unfolding on these networks produces different synchronization diagrams $r(\lambda)$ depending on the values $\gamma$ and $\alpha$. In Fig.~\ref{fig:correlations} (a)-(b) the total area of the hysteresis cycle is reported as a function of $\alpha$ for different scale-free networks. The most important effect is that the assortative networks ($\alpha>0$) display, for moderate values of $\alpha$, much larger hysteresis cycles that their dissasortative counterparts. Interestingly, the increase of degree heterogeneity also magnifies the irreversible nature of ES. Apart from the effect on the hysteresis cycle, Fig.~\ref{fig:correlations} (c)-(d) reveal that the critical couplings,  $\lambda_c^{f}$ and $\lambda_c^{b}$, are also affected by the presence of degree correlations.
\medskip

{\em Modular networks.-} Reported in~\cite{ljmsz13} is the systematic study of ES transition in networks with communities. In this work, the authors study a model network composed of several densely connected modules in which the mixing between modules (the fraction of links connecting nodes belonging to different communities) can be tuned continuously. Their numerical simulations point out that when the modularity of the network is large (so that the mixing between modules is small) the onset of synchronization is 
enhanced or, equivalently, the value of $\lambda_c^{f}$ decreases. Interestingly, to explore in more detail the effects of mixing between different modules, the authors of~\cite{ljmsz13} 
study a system composed of two star graphs in which the two hubs share some fraction of the leaves surrounding them. The authors observe a double sharp synchronization transition whose position and synchronization gaps can be tuned by  controlling the mixing ({\it i.e.,} the fraction of common leaves) between the two stars.

\subsubsection{More general CDF scenarios}

\begin{figure}[!t]
\begin{center}
\includegraphics[width=.6\textwidth,angle=0,clip=1]{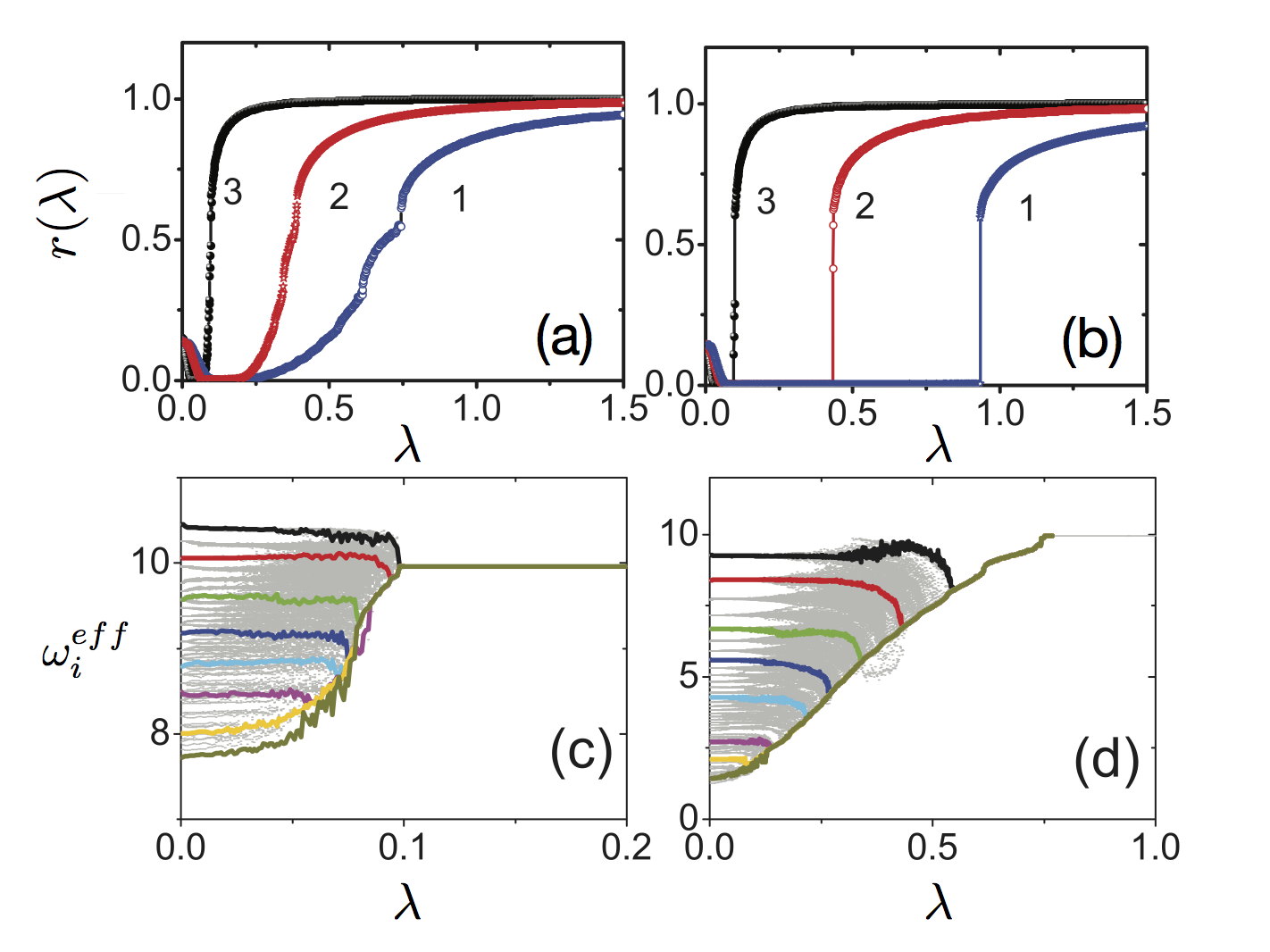}
\end{center}
\caption{(Top) Synchronization  diagrams $r(\lambda)$: (a) for $\beta=-0.8$ (1), -0.4 (2) and -0.1 (3), and (b) for $\beta=0.8$ (1), 0.4 (2), 0.1 (3). (Bottom) Evolution with $\lambda$ of the effective frequencies of nodes for $\beta=-0.1$ (c) and $\beta=-0.8$ (d). [Reprinted and adapted from~\cite{sstr13}].}
\label{fig:liu}
\end{figure}

The investigations reported on above have assumed a linear correlation between degree and frequency, $\omega_i=k_i$. However, different correlation functions $\omega_i=f(k_i)$ can be used. In~\cite{lwxz13} the authors explore a function of the form: $\omega_i=\alpha k_{i}^{\beta}$, so that correlations different from linear can be analyzed. The authors focused on the exponents in the range $\beta\in[-1,1]$, thus covering the case of negative correlations between frequencies and degrees. At first sight, the results show that for $\beta<0$ the abrupt transition is suppressed, becoming
second order, and the onset of synchronization is enhanced 
(see Fig.~\ref{fig:liu}(a)-(b)). Interestingly, the microscopic view of the synchronization transition for $\beta<0$ unveils a very different scenario than that of the the typical second-order transitions found in scale-free networks. Instead of finding a sequential entrainment of oscillators into the synchronized set moving  
at the average frequency (similar to that shown in Fig.~\ref{fig:ggam11}(b)), the authors observe for $\beta<0$ what they term an {\em hierarchical synchronization transition}.

This hierarchical synchronization transition is illustrated  in Fig.~\ref{fig:liu}(c)-(d) for $\beta=-0.1$ and $-0.8$ by monitoring the evolution with the coupling $\lambda$ of the effective frequencies of nodes, Eq.~(\ref{effective1}), and the average effective frequencies of nodes of the same degree, Eq.~(\ref{effective2}). These panels show that the oscillators do not evolve monotonously from their natural frequency to the average frequency $\Omega$. On the contrary, the synchronization cluster is governed by those nodes with the largest degrees (and lower natural frequencies). Thus it is  
the nodes of lower degree that, as $\lambda$ increases, adapt their effective frequencies to join this cluster. As $\lambda$ increases, the average frequency of this synchronized cluster increases, finally reaching the frequency $\Omega$.

A further generalization of the CDF function is analyzed in~\cite{sstr13}. In this case the authors extend the frequency-degree correlation proposed in~\cite{ggam11} to $\omega_i=\pm \alpha k_i^{\beta}$, {\em i.e.} they consider the modulus of the natural frequency of a node $i$ as in~\cite{lwxz13}, but the sign of this frequency is randomly assigned. This way, the probability of finding a node of degree $k$ and frequency $\omega$ becomes
\begin{equation}
P(k,\omega)=P(k)\left[\frac{1}{2}\delta(\omega-\alpha k^{\beta})+\frac{1}{2}\delta(\omega+\alpha k^{\beta})\right]\;.
\end{equation}
The purpose of this CDF scheme is to have a symmetric frequency distribution, $g(\omega)=g(-\omega)$, as in the original Kuramoto model, which is a very convenient situation for most theoretical studies.

Under the correlation scheme proposed in~\cite{sstr13} no explosive transition to synchronization is observed. However, the authors report a two-stage synchronization transition. First, after crossing a first critical coupling, $\lambda_1$, a regime of partial synchronization, containing two fully-synhcronized clusters, appears. This is explained as a consequence of the bi-modal frequency distribution $g(\omega)$ obtained from the uni-modal degree distribution. This way, in this regime (termed {\em Standing Wave}) the two clusters pace at opposite frequencies, {\em e.g.} for $\beta=1$: $\Omega^+\simeq \alpha\langle k\rangle$ and $\Omega^-\simeq -\alpha\langle k\rangle$. When the coupling increases further, for $\lambda>\lambda_2$, the whole network synchronizes. Interestingly, for $\beta=1$ the synchronized regime that appears for $\lambda>\lambda_2=2\alpha$ shows that all the oscillators have become phase-locked.

\subsubsection{Noisy degree-frequency correlations}

The former results are obtained considering a perfect match between degrees and natural frequencies. However, in most real systems this is not the situation and it is interesting to study the CDF model when $\omega_i=k_i+\xi_i$, where $\xi$ is a random small perturbation, $\xi\in(-\epsilon,\epsilon)$. Several works have addressed this issue~\cite{zpslk14,skardala14}. 

\begin{figure}[!t]
\begin{center}
\includegraphics[width=.45\textwidth,angle=0,clip=1]{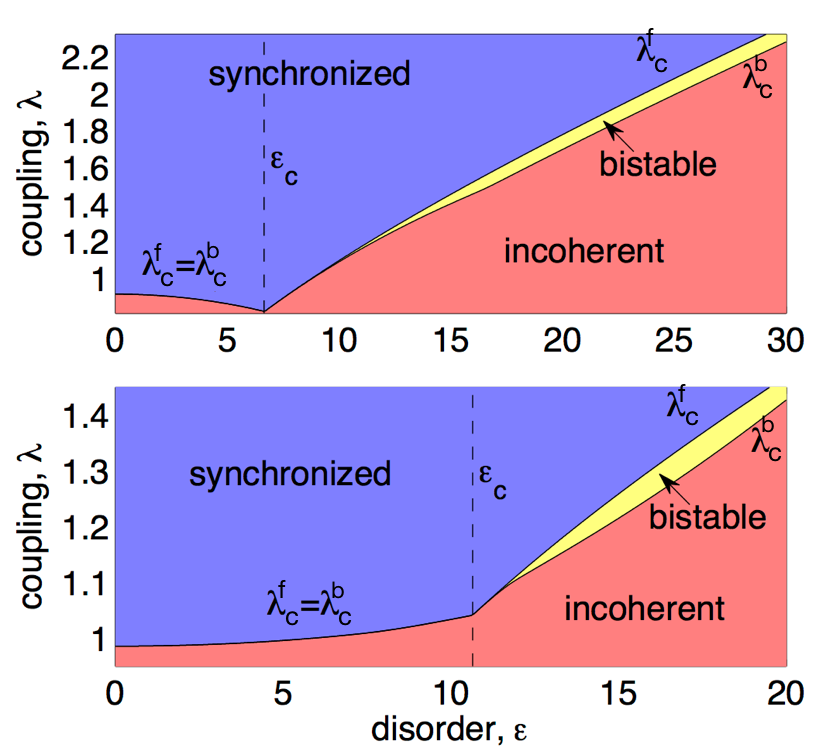}
\end{center}
\caption{Synchronization diagram as a function of $\epsilon$ and $\lambda$ for SF networks with $\gamma=3.5$ (top) and SE networks with $\beta=0.7$ and $\mu=0.5$ (bottom). The lines separate the incoherent and synchronized regions. The area between $\lambda_c^{f}$ and $\lambda_c^{b}$ corresponds to the bistable region. [Reprinted and adapted from~\cite{skardala14}].}
\label{fig:skardal}
\end{figure}

The most remarkable result in this direction was the stimulation of explosive transitions in networks that, due to their degree homogeneity, do not show this phenomenon in the CDF model. This result, presented in~\cite{skardala14}, was shown for one real network, the {\em C. elegans} neural network~\cite{achacoso1991ay}, and two synthetic networks: SF graphs with $\gamma>3$, and networks with a degree distribution given by a stretched exponential (SE), $P(k)=k^{\beta-1}{\mbox{exp}}[-(k/\mu)^{\beta}]$. These graphs show a mild heterogeneity and, as it is proven for SF networks with $\gamma>3$, one should not expect ES to happen. In Fig.~\ref{fig:skardal} the phase diagram for SF ($\gamma=3.5$) and SE networks is shown. In both cases, when the amount of noise in the degree-frequency matching is larger than some critical value $\epsilon_c$ ES appears.

\subsubsection{CDF with time delay and frustration}

Another avenue for testing the robustness of the CDF framework is to consider the effects of time-delay in the coupling between neighbors. Time-delay has been already considered in the original~\cite{yeung99} and networked~\cite{epba11} Kuramoto model. Here the question to address is to what extent ES can persist under the presence of time delay. 

In~\cite{peronr12} the authors analyze the following equations
\begin{equation}
\dot\theta_{i}(t)=k_i+\lambda\sum_{i=1}^{N}A_{ij}\sin(\theta_j(t-\tau)-\theta_i(t))\;,
\label{explosive_TD}
\end{equation} 
that for $\tau=0$ recover the original CDF Kuramoto model, Eq.~(\ref{explosive_or}). Interestingly, the authors observed a non-trivial dependence of the synchronization transition with the time delay. In particular, they implemented Eq.~(\ref{explosive_TD}) in a SF network for $\tau=0$, 0.1, 0.3, 0.5, 1.0, 1.5 and 2.0 and discover that only the cases $\tau=0$ and 1.0 lead to an explosive transition. In addition, while the rest of the $\tau$ values show a smooth transition the observed scaling of the order parameter $r$ is rather different. 

To gain insight about this anomalous behavior the authors analyze the star-graph. By performing a similar analysis as that in~\cite{ggam11} for the existence of phase-locked solutions, the authors find that the expression of the critical coupling $\lambda_c^{b}$ needed for the existence of the synchronized state
\begin{equation}
\lambda_c^{b}=|\omega-\Omega(\tau)|
\label{lambda_tau}
\end{equation}
where $\omega$ is the frequency of leaf-nodes and $\Omega(\tau)$ the average synchronization frequency. For $\tau=0$ this expression recovers that in
Eq.~(\ref{backward}). However, when $\tau>0$ the value depends strongly on $\Omega(\tau)$. By using $\omega=1$ for the leaves and $\omega_h=K$ for the hub, as in~\cite{ggam11}, the authors derive a self-consistent equation for $\Omega$
\begin{equation}
\Omega=\frac{2K}{K+1}\left[1-\lambda\sin(\Omega \tau)\right]\;.
\label{Omega_tau}
\end{equation}
From this equation it is clear that the non-monotonous behavior of $\Omega(\tau)$ (and thus the non-trivial behavior of $\lambda^{b}_c(\tau)$) is due to the periodicity of the sine function.

Related to these results, the Kuramoto model with frustration 
(referred to as the Sakaguchi-Kuramoto (SK) model~\cite{sakaguchi}), has been analyzed with the CDF framework~\cite{hgszx16,csgxz17,kkhp17}. The frustration term modifies Eq.~(\ref{explosive_or}) to become
\begin{equation}
\dot\theta_{i}(t)=k_i+\lambda\sum_{i=1}^{N}A_{ij}\sin(\theta_j(t)-\theta_i(t)-\alpha)\;,
\label{explosive_F}
\end{equation} 
where $\alpha$ is the frustration parameter. It is important to recall that the SK model was previously found to capture the dynamics of weakly coupled Josephson junctions arrays~\cite{wcs96,wcs98}.

Under synchronization conditions, it is easy to show that Eq.~(\ref{explosive_F}) is equivalent to Eq.~(\ref{explosive_TD}): when the oscillators pace at a common synchronization frequency $\Omega$ we can write $\theta_{j}(t-\tau)=\theta_j(t)-\Omega\tau$ and therefore, by comparing Eqs. (\ref{explosive_TD}) and (\ref{explosive_F}), we obtain $\alpha=\Omega\tau$.  Equation (\ref{explosive_F}) has been analyzed in the star graph in~\cite{hgszx16,csgxz17}.  In particular, by inspecting the synchronization conditions, in~\cite{hgszx16} the authors obtain expressions for both $\lambda_c^{b}$ and $\Omega(\alpha)$ similar to those in Eqs.~(\ref{lambda_tau})-(\ref{Omega_tau}) by changing $\Omega\tau$ by $\alpha$. The study of the SK model in complex networks was tackled in~\cite{kkhp17} finding that, in SF networks, the increase of the frustration $\alpha$ 
enhances the onset of synchronization at the expense of transforming the ES transition into the usual second order one.

\subsection{The correlated frequency-coupling (CFC) framework}

The ES transition obtained through the CDF framework motivated the question of whether other mechanisms correlating dynamical and structural aspects of oscillators can lead to same phenomenon. In this regard,~\cite{zhkl13} propose an alternative approach by correlating the natural frequency of an oscillator with the strength of the coupling with its neighbors. This framework, termed here as correlated frequency-coupling (CFC), transforms the Kuramoto model into
\begin{equation}
\dot{\theta_i}=\omega_i+\frac{\lambda|\omega_i|}{k_i}\sum_{j=1}^{N}A_{ij}\sin(\theta_{j}-\theta_{i})\;.
\label{CFCKuramoto}
\end{equation}
This framework allows us to get rid off the constraint that $g(\omega)=P(k)$ (as required in the CDF) so that one can tackle the study of different frequency distributions $g(\omega)$ using the same underlying network. Why should the CFC framework transform the usual second-order synchronization transition 
into an explosive one? If we focus on unimodal frequency distributions with mean $\Omega=0$, the effective coupling $\lambda|\omega_i|/k_i$ for those nodes close to the mean frequency $\Omega$ is very low, compared to that of the nodes which have  
$\omega_i$ that lies on the tails of the distribution $g(\omega)$. Therefore, the CFC mechanism punishes those oscillators that, in the usual Kuramoto model, are the first to get synchronized and, on the other hand, favors those that become phase-locked the last in the traditional setup. In a nutshell, the CFC homogenizes the ability of nodes to become synchronized, thus favoring the sudden emergence of synchronized components.

\subsubsection{Independence of network properties} 
\label{sec:CFCtheory}

\begin{figure}[!t]
\begin{center}
\includegraphics[width=.55\textwidth,angle=0,clip=1]{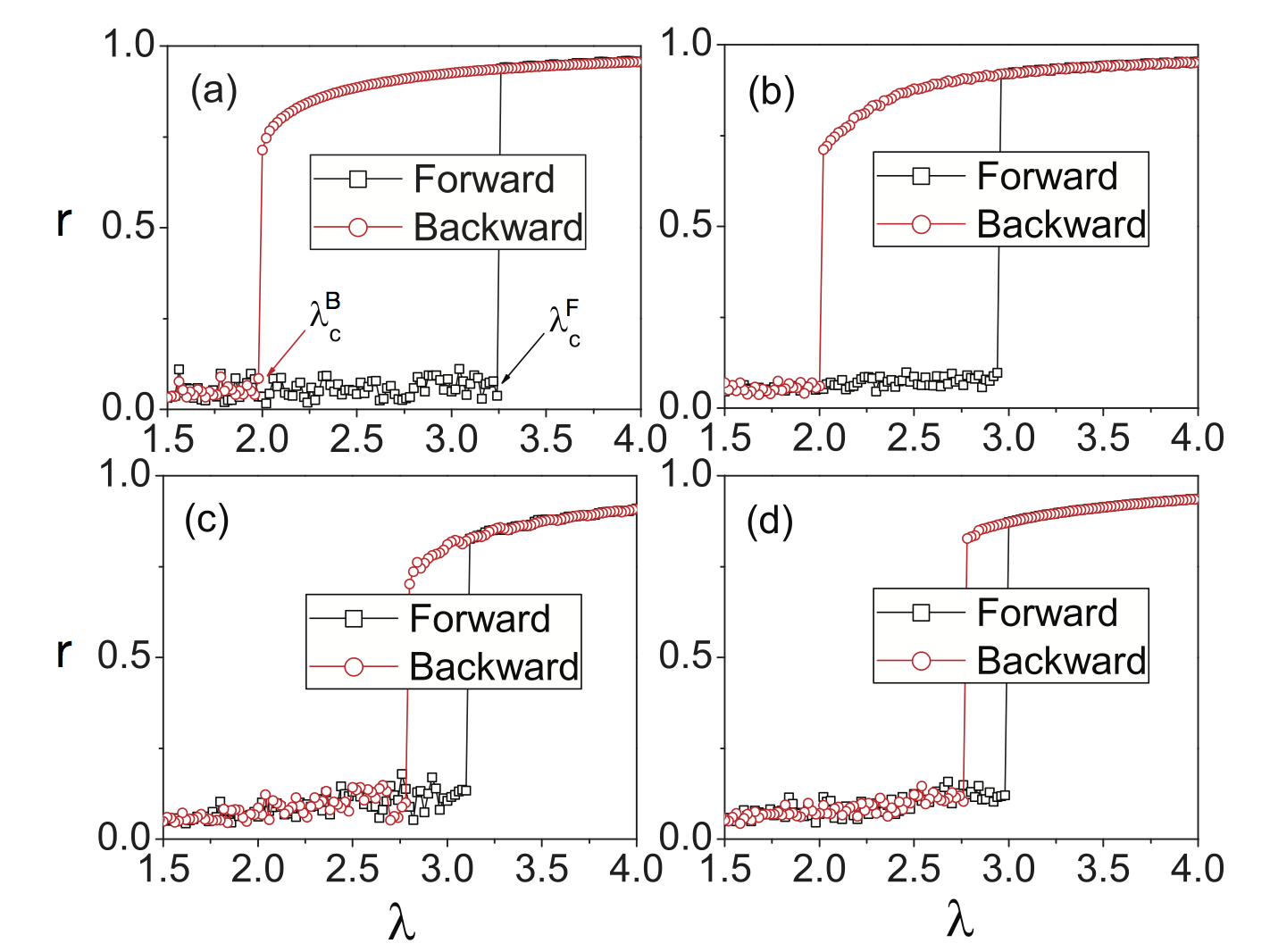}
\end{center}
\caption{Forward and backward synchronization  diagrams $r(\lambda)$: (a) all-to-all network with Lorentzian ($\gamma=0.5$) $g(\omega)$, (b) all-to-all network with Gaussian ($\sigma^2=1$) $g(\omega)$, (c) ER network with Lorentzian ($\gamma=0.5$)  $g(\omega)$ and (d) SF network with Lorentzian ($\gamma=0.5$)  $g(\omega)$. All the networks have $\langle k\rangle=6$. [Reprinted from~\cite{zhkl13}].}
\label{fig:CFC}
\end{figure}

The most remarkable result obtained under this framework~\cite{zhkl13} is that, for a variety of symmetric and unimodal $g(\omega)$ (centered in $\Omega=0$), Eq.~(\ref{CFCKuramoto}) displays ES transitions independent of
the structural patterns of the underlying network, even for the all-to-all coupling benchmark. In Fig.~\ref{fig:CFC} the forward and backward synchronization diagrams $r(\lambda)$ corresponding to three different network topologies are shown, namely: fully connected network (a)-(b), ER (c) and SF (d) graphs. For the fully connected case, two distributions $g(\omega)$ are considered, Lorentzian and Gaussian, while for the ER and SF graphs the plots refer to Lorentzian distributions. These four synchronization diagrams show the explosive and irreversible nature of the synchronization transition, thus confirming the homogenizing effects of the CFC mechanism.

\begin{figure}[t!]
\begin{center}
\includegraphics[width=.5\textwidth,angle=0,clip=1]{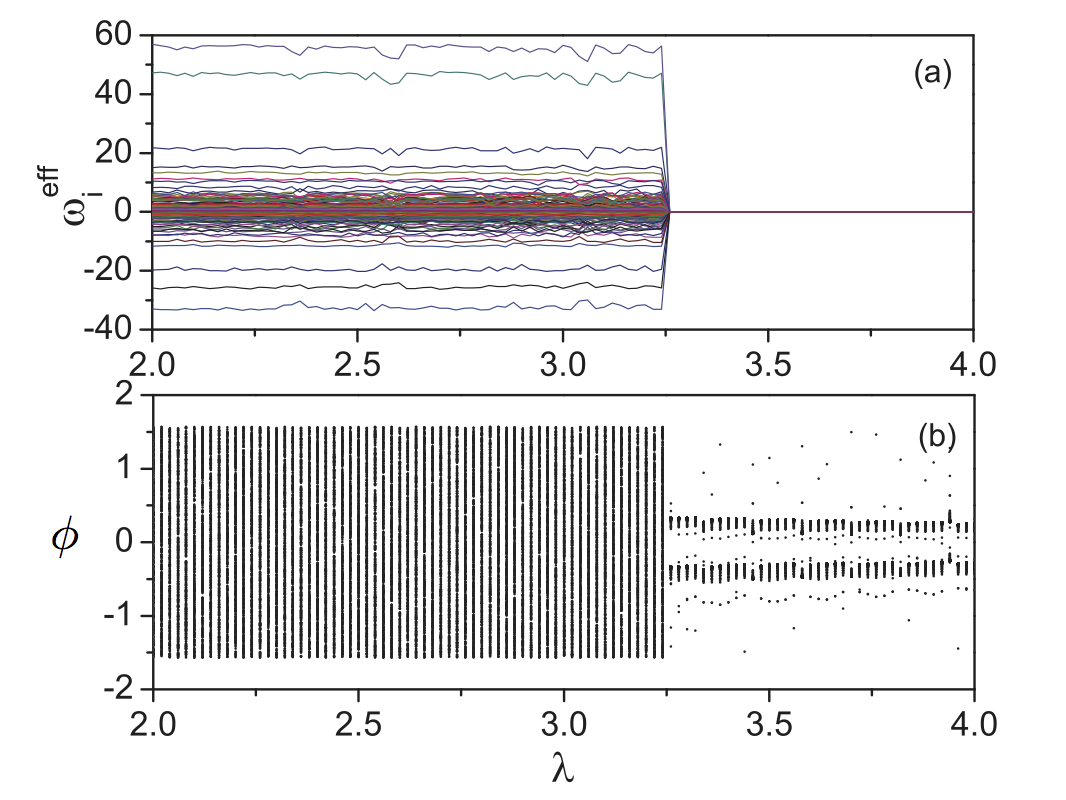}
\end{center}
\caption{Microscopic evolution of the (forward) synchronization transition for an all-to-all network with Lorentzian distribution $g(\omega)$ ($\gamma=0.5$). (a) Evolution with $\lambda$ of the effective frequencies of the nodes [see Eq.~(\ref{effective1})]. (b) Evolution of the values $\phi(t)$ with $\lambda$ revealing the emergence of two clusters of phase-locked oscillators just after frequency entrainment in panel (a).  [Reprinted and adapted from~\cite{zhkl13}].}
\label{fig:CFC2}
\end{figure}

To shed some theoretical light on the transition the authors in~\cite{zhkl13} propose to follow the approach introduced in~\cite{i04}. In a similar fashion as Eq.~(\ref{eq:kuramoto_continuum}) in the CDF framework, once starts by casting the set of evolution equations for the nodes, Eq.~(\ref{CFCKuramoto}) as
\begin{equation}
\frac{{\mbox d} \theta}{{\mbox d}t}=\omega + \lambda |\omega|  \int dk'\int d\theta' \frac{k'P(k')}{\langle k\rangle}\rho(k';\theta',t) \sin(\theta - \theta')\;,
\label{eq:CFC_continuum}
\end{equation}
where, as in Eq.~(\ref{eq:kuramoto_continuum}), $\rho(k;\theta, t)$ is the density of nodes with degree $k$ and phase $\theta$ at time $t$.  
Here we have neglected degree-degree correlations and the time fluctuations due to finite size effects. By writing the Kuramoto order parameter in terms of $\rho(k;\theta, t)$ Eq.~(\ref{eq:CFC_continuum}) can be rewritten as
\begin{equation}
\dot{\theta}=\omega+\lambda|\omega|r\sin(\Phi(t)-\theta(t))\;.
\label{effectiveCFC}
\end{equation}
Note that this equation is completely independent of the degree distribution $P(k)$ of the underlying network. This generality comes from the normalization by $k_i$ of the coupling between units  [see Eq.~(\ref{CFCKuramoto})] as discussed previously in Sec.~\ref{sec:KM_networks}. At this point it is convenient to work in a reference  frame rotating with the average frequency: $\phi(t)=\theta(t)-\Phi(t)=\theta(t)-\Phi_{0}-\Omega t$, so that two phase-locked, $\dot{\phi}=0$, solutions are possible
\begin{eqnarray}
\phi_1^{*}&=&\arcsin(\frac{1}{\lambda r})\;\;{\mbox{if}}\; \omega>0\;,
\label{sol1CFC}\\
\phi_2^{*}&=&\arcsin(-\frac{1}{\lambda r})\;\;{\mbox{if}}\; \omega<0\;.
\label{sol2CFC}
\end{eqnarray}
These two solutions reveal a novel feature of the CFC framework: the synchronized regime is described as two coexisting clusters of fully phase-locked units. From the form of the two solutions it becomes clear that these two clusters progressively approach each other as $\lambda$ increases. This microscopic picture was confirmed numerically in~\cite{zhkl13} and is shown in Fig.~\ref{fig:CFC2}. 

From the set of solutions in Eqs.~(\ref{sol1CFC})-(\ref{sol2CFC}) it is possible to evaluate the critical properties of the synchronization transition by expressing the Kuramoto order parameter $r$ in the synchronized regime as the sum of the contributions of the two sets of phase-locked oscillators
\begin{equation}
r=\frac{1}{2}\left(\cos\phi_1^{*}+\cos\phi_2^{*}\right)=\sqrt{\frac{\lambda+\sqrt{\lambda^2-4}}{2\lambda}}\;.
\end{equation}
This expression reveals that the phase-locked regime exists when $\lambda>2$, so that the limit of the backward (upper) branch of the synchronization diagram is $\lambda_c^{b}=2$. Additionally, the value of $r$ at this point, $r(\lambda_c^{b})=\sqrt{0.5}$, indicates the explosive nature of the synchronization transition. This theoretical prediction is, in principle, valid for any network topology. However, for complex networks with small $\langle k\rangle$ the assumptions made to derive Eq.~(\ref{effectiveCFC}) are too strong and, as observed in Fig.~\ref{fig:CFC}(c)-(d), fail to predict the correct value for $\lambda_c^{b}$.
   
The analysis of the CFC framework was tackled in several subsequent studies~\cite{zzbl14,hbhzlg14,zcbhlg15,blzg17,xjxlz17,loftird18}, see Sec.~\ref{sec:EPES} for the analysis of the results in~\cite{zzbl14}. In~\cite{hbhzlg14} the authors provide with the exact solution of the CFC model in the fully connected graph. Following a similar analysis as in~\cite{zhkl13} the authors reproduce the backward transition finding the two phase-locked solutions $\phi_1^{*}$ and $\phi_2^{*}$ as in Eqs.~(\ref{sol1CFC})-(\ref{sol2CFC}) and the critical coupling $\lambda_c^{b}=2$. In addition, they are able to solve the forward transition, unveiling how the incoherent state $r\simeq 0$ turns unstable at the value $\lambda_c^{f}$. In particular, for even frequency distributions $g(\omega)$, the expression to calculate $\lambda_c^{f}$ is the following transcendental equation
\begin{equation}
1=\frac{\lambda}{2}\int_{-\infty}^{\infty}\frac{z|\omega|}{z^2+\omega^2}g(\omega){\mbox d}\omega\;,
\label{trascendental}
\end{equation} 
with $\lambda_c^{f}$ being the value of $\lambda$ that satisfies Eq.~(\ref{trascendental}) when the real part of the complex number $z$, ${\mbox{Re}}(z)$, changes from negative to positive, {\em i.e.} when the incoherent state  loses its stability. Importantly, in contrast 
with $\lambda_c^{f}$,  the value of the critical coupling for the forward branch depends on the frequency distribution $g(\omega)$ at work. 

\subsubsection{The Sakaguchi-Kuramoto model under the CFC framework} 

{\color{black} The SK model was also tackled under the CFC framework in~\cite{xjxlz17}. In this work the effects of the frustration $\alpha$ over the synchronization transition is shown to strongly depend on the nature of $g(\omega)$. Remarkably, for both triangular and uniform distributions of $g(\omega)$,
the increase of frustration implies the inhibition of ES, pointing out the same result observed for the CDF framework in~\cite{kkhp17}. However, in the CFC scheme with frustration, the use of a Lorentzian frequency distribution delays considerably the onset of ES and, more importantly, increases the bistability region.}

Related to the SK model, the Kuramoto model in the CFC framework with time delays has been recently tackled in~\cite{wu2018}. At variance, with the CDF approach, in this case ES is not altered by the addition of time delays (even in the case when they are distributed across the units). Moreover the authors in~\cite{wu2018} show analytically and numerically that time delay can shift the critical coupling $\lambda_c{f}$, thus enlarging the bistability region.

\subsubsection{Loss of rotational invariance} 

The presence of the absolute value, $|\omega_i|$ in the effective coupling between oscillators in Eq.~(\ref{CFCKuramoto}) has very important consequences for the onset of ES. In the seminal article~\cite{zhkl13} the authors noted that replacing $|\omega_i|$ by $\omega_i$ in  Eq.~(\ref{CFCKuramoto}) may lead to second order transitions in some situations, thus losing the generality of the results described above. In particular, the transition in Fig.~\ref{fig:CFC}(b) changes to second order when the aforementioned replacement in the effective coupling is done. As a conclusion, the presence of the absolute value $|\omega_i|$ is key for the CFC framework.

As already mentioned in Sec.~\ref{sec:KM_original}, one of the key properties of the usual Kuramoto model, the invariance the equations under rotational transformations, $\theta^{'}_i=\theta_i-\omega t$ and $\omega_{i}^{'}=\omega_i-\omega$, is lost due to the inclusion of the absolute value $|\omega_i|$ in the effective coupling. Due to this, one should expect non-trivial effects when changing the center of the  (even and unimodal) frequency distribution $g(\omega)$ from $\Omega=0$ to $\Omega\neq 0$. This was the object of study in~\cite{zcbhlg15}. The main finding of the authors is that, under the CFC scheme, explosive transitions may turn second order when the center of $g(\omega)$ is translated to non-zero values of $\Omega$. 

Interestingly, although the transition becomes
second-order when $\Omega$ is large enough, the authors find that the microscopic synchronization patterns are very different from those expected in the traditional Kuramoto model. For small $\Omega$ (when the transition is still explosive) the authors find the two phase-locking clusters reported in~\cite{zhkl13}, Eqs.~(\ref{sol1CFC})-(\ref{sol2CFC}). However, for the second order transition they find (for the forward and backward continuation) two radically different synchronization patterns. Namely, for a small (but nonzero) degree of synchronization those oscillators with small natural frequencies become synchronized forming a cluster of phase-locked oscillators, ({\em i.e.}, a picture similar to that of the original Kuramoto model). Surprisingly, as $\lambda$ and $r$ increase, two phase-locking clusters show up involving those nodes with large natural frequencies, a pattern similar to that found for the explosive transition. The existence of these two regimes can be derived by writing Eq.~(\ref{effectiveCFC}) in a new reference frame rotating with $\Omega$, so that $\theta^{'}=\theta-\Omega t$ and $\omega^{'}=\omega-\Omega$. This way one obtains
\begin{equation}
\dot{\theta'}=\omega'+\lambda r |\omega^{'}+\Omega|\sin(\Phi-\theta^{'})\;,
\label{CFCKuramoto2}
\end{equation}
 implying that the phase-locked solutions must satisfy
\begin{equation}
\left|\frac{\omega'}{\omega'+\Omega}\right|\leq \lambda r\;.
\label{conditionCFC}
\end{equation}
Note that the function in the l.h.s has the following properties: (i) it diverges at $\omega'=-\Omega$, it has a single minimum at $\omega'=0$ and it tends asymptotically to $1$ when $\omega'\rightarrow \pm\infty$. The form of this function
implies that: {\em (i)} for small degree of synchronization $\lambda r<1$ only oscillators with $\omega'\simeq 0$ ($\omega\simeq \Omega$) will satisfy the condition (\ref{conditionCFC}), while {\em (ii)} when $\lambda r>1$ the condition is satisfied by those oscillators with positive and negatives frequencies far away 
from $\omega'=-\Omega$. The first condition captures the existence of a single synchronized cluster, while the second one shows the conditions for the emergence of two different clustered of phase locked units.

\subsubsection{Including frequency mismatch} 

An extension of the CFC method was introduced in~\cite{lsanob13} by weighting the coupling between adjacent oscillators  
by the absolute value of their frequency mismatch. In particular, the authors considered the following equation
\begin{equation}
\dot{\theta_i}=\omega_i+\lambda\sum_{j=1}^{N}A_{ij}|\omega_i-\omega_j|^{\alpha}\sin(\theta_{j}-\theta_{i})\;.
\label{CFCKuramoto_missmatch}
\end{equation}
Under this framework, and for $\alpha>0$, those connected units with close frequencies, that are prone to synchronize in the traditional Kuramoto model ($\alpha=0$ within this framework), become weakly coupled. In contrast, the coupling between neighbors with very different frequencies is strengthened 
by the new term. 

The authors studied the emergence of explosive phenomena for different values of $\alpha$ both for homogeneous and heterogeneous networks. For homogeneous, ER and random regular graphs, their results point out that ES naturally shows up for large enough values of $\alpha$ and for a variety of frequency distributions $g(\omega)$. However, by deforming ER into heterogeneous SF networks following the method introduced in~\cite{gm06}, they observed that the explosive transition is lost progressively as the SF limit is approached and heterogeneity is increased. 
           
\subsubsection{Effective Synchronization centrality}

To shed more light on how the CFC framework induces ES, in~\cite{nvlasb15} a method for analyzing how the modification of  the coupling between units affects the onset of synchronization clusters is introduced. This method relies on the computation of the eigenvector centrality of each node by considering the modified adjacency matrix corresponding to the effective coupling at work. For instance, in the case of Eq.~(\ref{CFCKuramoto_missmatch}) this effective matrix is $\tilde{A}_{ij}=A_{ij}|\omega_i-\omega_j|^{\alpha}$,  while for Eq.~(\ref{CFCKuramoto}) it reads $\tilde{A}_{ij}=A_{ij}|\omega_i|/k_i$. Once the effective matrix ${\bf{\tilde{A}}}$ is constructed, a subsequent transformation is performed to derive a new matrix as
\begin{equation}
C_{ij}=\tilde{A}_{ij}\left(1-\frac{\Delta|\omega_{i}-\omega_j|}{\Delta\omega_{max}}\right)\;, 
\end{equation}
with $\Delta\omega_{max}$ being the maximum value of frequency mismatch between connected nodes in the system. The components of the eigenvector of ${\bf C}$ corresponding to the largest eigenvalue, $\vec{v}_{max}$, reveal those nodes that act as synchronization seeds, {\em i.e.} those forming the first microscopic synchronization clusters. Thus, the ranking obtained from $\vec{v}_{max}$ of matrix ${\bf C}$ can be seen as a measure of synchronization centrality for nodes.

With this method the authors compare the centrality of nodes in the CFC framework with that of the network with the original adjaceny matrix, {\em i.e.} computing $\vec{v}_{max}$ from a matrix ${\bf C}$ derived from ${\bf A}$. The main finding is that, for those networks in which the CFC scheme leads to an ES transition, the CFC acts by homogenizing the synchronization centrality across nodes. This way, the CFC framework effectively suppress those {\em loci} originally provided by the network for the creation of synchronization seeds.

\subsection{The frequency gap conditioned (FGC) framework}
\label{subsec:FGC}

The rationale behind the last approach  to the CFC framework, Eq.~(\ref{CFCKuramoto_missmatch}), is to implement an effective decoupling of those connected nodes with similar natural frequencies. This idea is at the core of the next method to be discussed for obtaining ES transitions in networks: the frequency gap conditioned (FGC) approach. However, instead of acting through the evolution equations, as in the cases of the CDF and CFC frameworks, in 
\cite{lnsabzpb13} the authors propose to act directly on the design of the network. Namely, given a network topology encoded in an adjacency matrix ${\bf A}$ and a given frequency distribution $g(\omega)$, the aim is to assign natural frequencies to each node in such a way that there is a prescribed mismatch between them given by
\begin{equation}
|\omega_i-\omega_j|>A_{ij}\gamma\;.
\label{freqgap}
\end{equation}
Note that we have set the term $A_{ij}$ so that the mismatch only applies to neighboring nodes. 

Obviously, satisfying condition (\ref{freqgap}) for any given network topology and frequency distribution $g(\omega)$ is hard to achieve and one would face the problem of frustration. To avoid this problem, the authors propose a method to construct networks satisfying Eq.~(\ref{freqgap}). The procedure is as follows: {\em (i)} Start from an isolated set of $N$ nodes, {\em (ii)} assign randomly the frequencies of each node following the prescribed distribution $g(\omega)$, {\em (iii)} choose at random one of the $N(N-1)/2$ pairs of nodes and create a link provided they fulfill condition (\ref{freqgap}), and {\em (iv)} repeat step  {\em (iii)} until a number of $L$ links have been created. The networks constructed with this  
procedure display ER-like interaction backbones with $\langle k\rangle=2L/N$.

\begin{figure*}[t!]
\begin{center}
\includegraphics[width=.9\textwidth,angle=0,clip=1]{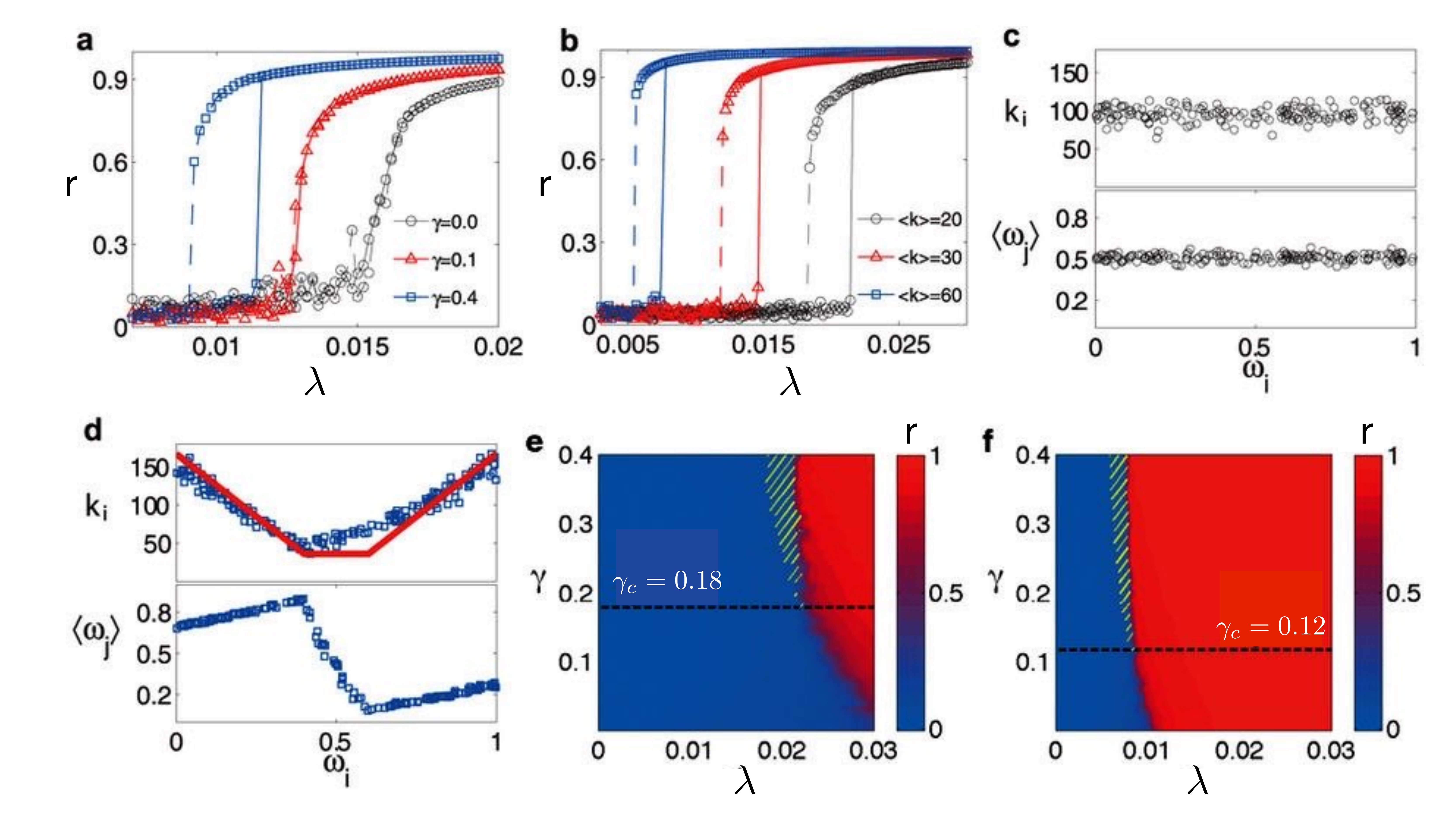}
\end{center}
\caption{ (a) Synchronization diagrams of FCG networks constructed with different frequency mismatch $\gamma$ and $\langle k\rangle=40$. (b) Synchronization diagrams for FCG networks with $\gamma=0.4$ and different average degree $\langle k\rangle$. Panels (c) and (d) report the correlation between natural frequencies of the nodes and their degrees (top) and the average natural frequency of their neighbors (bottom). Panels (e) and (f) show the diagrams $r(\lambda,\gamma)$. The stripped region corresponds to the metastable phase where hysteresis is found  [Reprinted and adapted from~\cite{lnsabzpb13}].}
\label{fig:FGC}
\end{figure*}

The authors analyze in detail the role of the frequency mismatch $\gamma$ for a wide variety of frequency distributions $g(\omega)$. In particular, Fig.~\ref{fig:FGC} shows the results corresponding to an homogeneous distribution in the interval $\omega\in[0,1]$ for several values of $\gamma$ (panel a) and $\langle k\rangle$ (panel b). From these plots the authors observe that when the frequency mismatch exceeds some critical value, $\gamma>\gamma_c$, ES is observed. The value of $\gamma_c$ is reported together with the hysteresis region in panels (e) and (f) for networks with $\langle k\rangle=20$ and $60$ respectively. 

Interestingly, when $\gamma>\gamma_c$ the resulting networks automatically show degree-frequency correlations, similar to
those imposed {\em ad hoc} in the CFD scheme. In the top panel of Fig.~\ref{fig:FGC}(d) the V-shaped curve reveals these frequency-degree correlations that are necessary to cause the explosive transition and that for $\gamma=0$ (see top panel of Fig.~\ref{fig:FGC}(c)) do not show up. In~\cite{zts13} the authors explore a very similar FGC framework, making an exhaustive study of how these V-shaped frequency-degree correlations appear exactly at $\gamma_c$ and, as a byproduct, they 
are accompanied by the onset of negative degree-degree correlations in the interaction network.  

\subsection{The adaptive framework}
\label{ssec:adaptive}
Up to now, the different frameworks (CDF, CFC and FGC) favoring the onset of ES have focused on coupling the (static) natural frequencies of the oscillators with the structural attributes of the interaction network, either at the level of nodes (CDF) or links (CFC and FGC). A different avenue was proposed in~\cite{zbgl15} by modifying the Kuramoto model to incorporate a new dynamical variable modulating the coupling of a unit $i$ with its neighbors. In particular, the  proposed model reads
\begin{equation}
\dot{\theta}_i=\omega_i+\lambda\alpha_i(t)\sum_{j=1}^{N}A_{ij}\sin(\theta_j-\theta_i)\;,
\label{eq:adaptive}
\end{equation}
where 
$\alpha_i(t)$ can take two different values. Namely, a fraction $f$ of randomly chosen nodes have
\begin{equation}
\alpha_i(t)=r_i(t)=\left|\frac{1}{k_i}\sum_{j=1}^{N}A_{ij}{\mbox{e}}^{{\mbox{i}}\theta_j(t)}\right|\;,
\end{equation}
{\em i.e.} the coupling is weighted by the instantaneous local field $r_i(t)$, defined as in Eq.~(\ref{ekrlocal}) and normalized by the degree $k_i$ so that $r_i(t)\in[0,1]$. The remaining $(1-f)N$ units have $\alpha_i=1$ as in the original Kuramoto model. Note that for $f=0$ the usual Kuramoto model is recovered.

The formulation of this framework seemingly contradicts the rationale behind the former schemes leading to ES scenarios. Namely, the former three frameworks established different suppressive rules for the formation of small clusters of phase-locked units that serve as the microscopic seeds leading to a large synchronized component. On the contrary, Eq.~(\ref{eq:adaptive}) apparently dictates that those $fN$ oscillators with $\alpha_i(t)=r_i(t)$ tend to reinforce their connections when the degree of local synchrony is large, thus boosting the creation of the aforementioned microscopic synchronization clusters. However, this reinforcing effect has as its counterpart the null coupling, regardless the value of $\lambda$, when a node is surrounded by dynamically incoherent units. Therefore, the reinforcement of synchronized neighborhoods is compensated by their isolation from the rest of the network.

\begin{figure}[!t]
\begin{center}
\includegraphics[width=.5\textwidth,angle=0,clip=1]{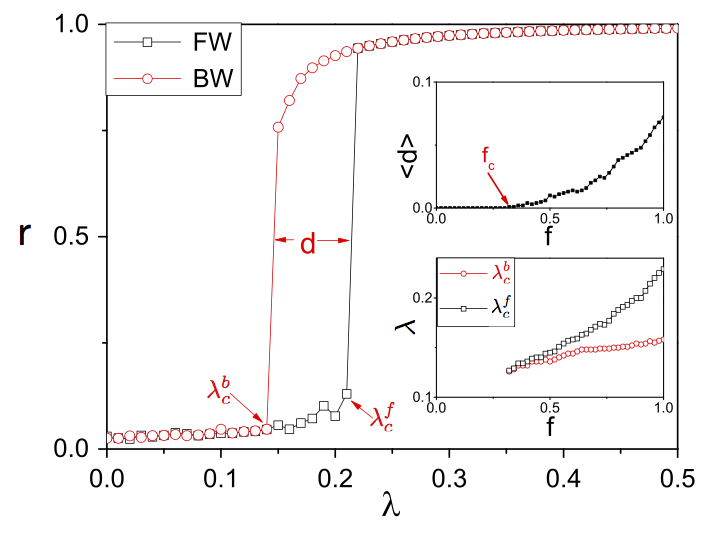}
\end{center}
\caption{Forward and backward synchronization diagram for an ER network of $N=1000$ nodes for $f=1$ and an homogeneous frequency distribution $g(\omega)$ in the interval $[-1,1]$. The insets show the dependence of the hysteresis cycle (top) and the critical couplings $\lambda_c^{f}$ and $\lambda_c^{b}$ (bottom) with $f$.  [Reprinted and adapted from~\cite{zbgl15}].}
\label{fig:Adaptive1}
\end{figure}

The numerical findings in~\cite{zbgl15} confirm the above reasoning. In Fig.~\ref{fig:Adaptive1} the forward and backward synchronization diagrams are reported for an ER network when $f=1$. Interestingly, the insets reveal the emergence of ES as a function of $f$ by reporting the average length of the hysteresis cycle $\langle d\rangle$ and the evolution of the two critical couplings $\lambda_c^{f}$ and $\lambda_c^{b}$ as $f$ varies from $0$ to $1$. From these plots it is clear that there exists a critical value $f_c$ so that for $f>f_c$ the second order transition
becomes instead abrupt.  

The theoretical explanation for the ES transition in the particular case $f=1$ was presented in~\cite{zbgl15}. However, in~\cite{dmkzhb16} the authors develop a theoretical treatment for the full adaptive model. To this aim, the authors write the effective equations for the phases of adaptive oscillators ($\alpha_i(t)=r_i(t)$), termed as class I, and of Kuramoto ones ($\alpha_i=1$), termed as class II
\begin{eqnarray}
\dot{\theta}^{I}_i&=&\omega_i+k_i\lambda r^2_i\sin(\phi_i-\theta_i)\;,
\label{dazinger1}\\
\dot{\theta}^{II}_i&=&\omega_i+k_i\lambda r_i\sin(\phi_i-\theta_i)\;,
\label{dazinger2}
\end{eqnarray}
where $\phi_i$ is the average phase of the neighbors of node $i$. The authors then make a mean-field consideration and approximate the local degree of synchronization to the global one, $r_i\simeq r$, and the average phase of the neighbors of $i$ to the global average phase, $\phi_i\simeq \Phi$. Note that this approximation is valid when dealing with homogeneous networks having relatively large $\langle k\rangle$. Under these assumptions and changing  to a reference frame rotating with the average frequency $\Omega$ (that can be placed at $\Omega=0$), $\psi_i=\theta-\Phi$, the phase-locked solutions of Eqs. (\ref{dazinger1})-(\ref{dazinger2}) satisfy
\begin{eqnarray}
\psi^{I,*}_i&=&\arcsin\left[\frac{\omega_i}{\lambda r^2 k_i}\right] \;,
\\
\psi^{II,*}_i&=&\arcsin\left[\frac{\omega_i}{\lambda r k_i}\right] \;.
\end{eqnarray}

Now, considering that in the new rotating frame $r=\sum_{i}\cos\psi_i$ and neglecting the contribution of the drifting oscillators (those not satisfying the former equations)  to $r$, it is possible to write the following self-consistent equation for the global degree of synchronization that, in the continuum limit, becomes
\begin{eqnarray}
r&=&\frac{f}{\langle k\rangle}\int_{-\lambda r^2 k}^{ \lambda r^2 k}g(\omega){\mbox d}\omega\int_{0}^{\infty} P(k)\sqrt{1-\left(\frac{\omega}{\lambda r^2k}\right)^2} {\mbox d}k\nonumber \\
&+&\frac{1-f}{\langle k\rangle}\int_{-\lambda r k}^{ \lambda r k}g(\omega){\mbox d}\omega\int_{0}^{\infty} P(k)\sqrt{1-\left(\frac{\omega}{\lambda r k}\right)^2} {\mbox d}k\;.
\label{theodazin}
\end{eqnarray}
The two integrals must be solved by specifying the frequency and degree distributions at work $g(\omega)$ and $P(k)$. 

To recover the example of Fig.~\ref{fig:Adaptive1} shown in~\cite{zbgl15} the authors of~\cite{dmkzhb16} solved the associated integrals numerically finding a surprising result: for any nonzero value of $f$  synchronization becomes explosive. This theoretical result contrasts with the results shown in the insets of Fig.~\ref{fig:Adaptive1}. However, the authors note that for small values of $f$ the explosive regime is very small so that, for finite networks is practically undetectable. 

The adaptive framework shows its efficiency in causing ES transitions in homogenous (ER) networks. However, its performance in SF networks is still under study.  In fact, the architecture of SF networks poses the following question: given the topology constraints imposed by degree heterogeneity and the associated presence of hubs,  
what is the possibility of achieving the dynamical isolation of small sets of synchronized clusters?

{\color{black} The adaptive framework has been recently reformulated by considering link adaptivity. In this case the connection of the adaptive scheme with the suppresive rule of the static framework is clear. In~\cite{aalbnlb18} the authors implement an anti-Hebbian rule that enhances the links connecting dynamical units poorly synchronized and weakens the links between coherent oscillators. This framework implements a suppressive rule aimed at delaying the synchronization onset that, as a byproduct, leads to an ES scenario. In addition, the dynamical evolution of link weights naturally evolves towards a CDF architecture, thus explaining the observation of ES transitions.}

\subsection{Other paradigmatic dynamical models}

Most of the studies on ES have been restricted to the Kuramoto model. However, some studies have extended the analysis of ES and the different frameworks addressed above to other well-known dynamical models. Here we report the most relevant ones.

\subsubsection{Second-order Kuramoto model}

The study of the usual Kuramoto model with inertia has been addressed in several studies~\cite{tlo97a,tlo97b,acebrons98,onbt14}.  Its implementation in complex networks is also of interest due to its important application to the emergence of synchronous states in power grids~\cite{rstw12}. The usual form of this model reads as follows
\begin{equation}
\frac{{\mbox{d}}^2\theta_i}{{\mbox{d}}t^2}=-\alpha\frac{{\mbox{d}}\theta_i}{{\mbox{d}}t}+\omega_i+\lambda\sum_{j=1}^{N}A_{ij}\sin(\theta_{j}-\theta_i)\;.
\label{KM_inertia}
\end{equation}
As briefly mentioned in Sec.~\ref{intro_transitions} the synchronization transition of the Kuramoto model with inertia is one of the few examples of first-order transitions obtained without the addition of terms coupling structural and dynamical features. Therefore, the relevant question to address is how the different frameworks leading to explosive phenomena in the networked Kuramoto model affect the second-order transition.

This question was tackled in the context of the CDF framework in~\cite{jpmrk13,jprk14,pjrk15,ctwzl18} by defining the natural frequency in Eq.~(\ref{KM_inertia}) as $\omega_i=D(k_i-\langle k\rangle)$. The first numerical observations pointed out that, at variance with the CDF Kuramoto model, ES is changed  by a cascade of sudden synchronizations of different group of nodes. The authors termed this  kind of transition as {\em cluster synchronization} and the main novelty is that the sets of nodes synchronizing together are those having the same number of neighbors, {\em i.e.}, the degree classes. 
This cluster synchronization is revealed by the sudden changes in $r$ shown in the forward branch of the synchronization diagram.

The cluster synchronization phenomenon is analytically explained in~\cite{jprk14}. As for the CDF Kuramoto model, see Eq.~(\ref{eq:kuramoto_continuum}), the authors consider the continuum limit approach by~\cite{i04}. Using the density of oscillators with degree $k$ that have phase $\theta$ at time $t$, $\rho(\theta,t;k)$, and assuming degree uncorrelated networks, Eq.~(\ref{KM_inertia}) can be expressed as
\begin{equation}
\ddot{\theta}=-\alpha\dot{\theta}+D(k-\langle k\rangle)+\lambda k r \sin(\phi-\theta)\;,
\label{KM_inertia_eff1}
\end{equation}
where $r$ is the Kuramoto order parameter and, as usual, $\phi$ is the average phase of the ensemble of oscillators.

\begin{figure}[!t]
\begin{center}
\includegraphics[width=.45\textwidth,angle=0,clip=1]{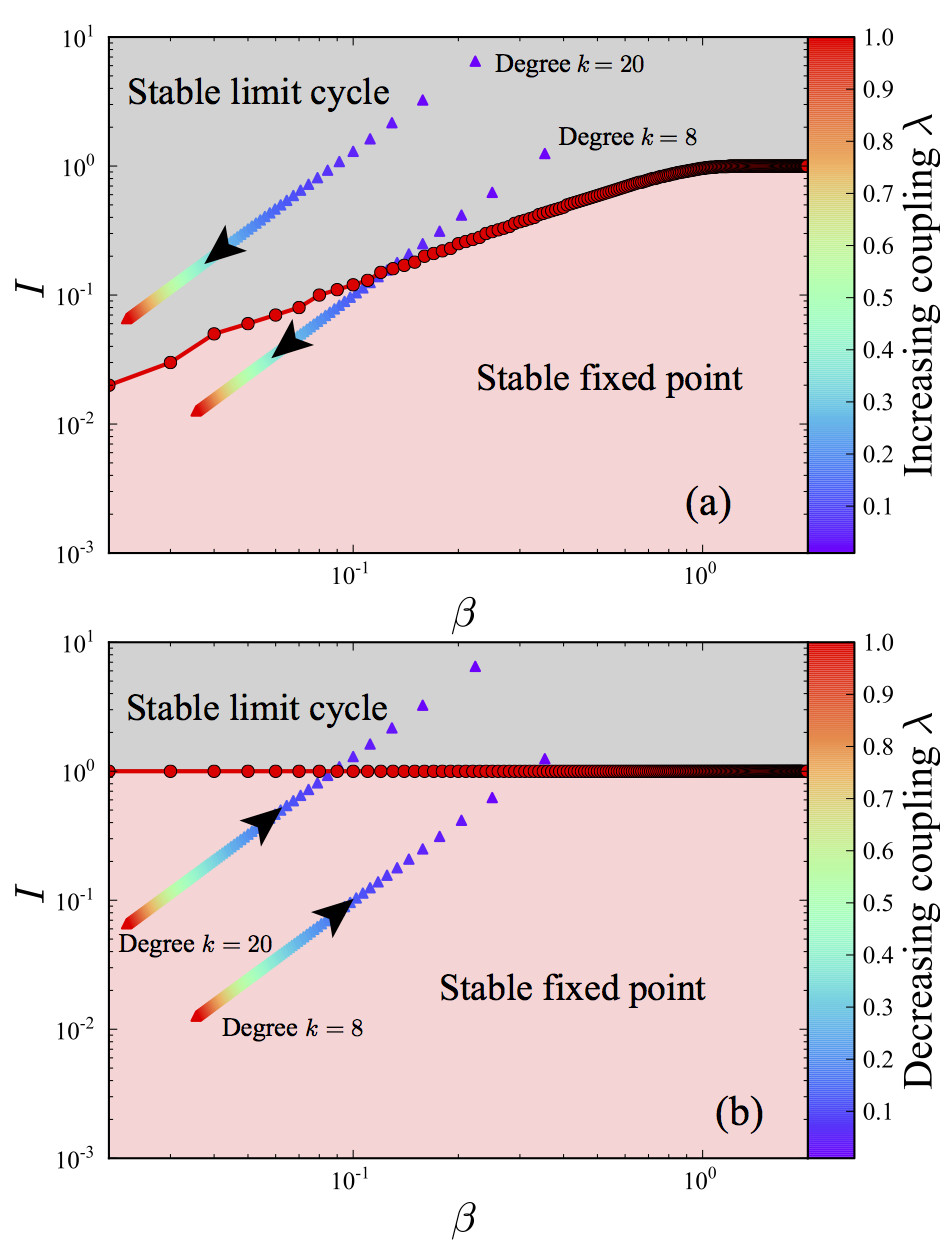}
\end{center}
\caption{Phase diagram for the driven pendulum in the $(\beta,I)$-plane for (a) increasing and (b) decreasing  $\lambda$. The two phase diagrams separates the stable limit cycle solution (incoherent state) and the stable fixed point (synchronized regime).  Both panels show the trajectories in the $(\beta,I)$-plane of oscillators with degree $k=8$ and $k=20$ as $\lambda$ increases (a) and decreases (b). The network is SF of $N=3000$ with $\gamma=3$ and $\langle k\rangle=10$. [Reprinted and adapted from~\cite{jprk14}].}
\label{fig:second2}
\end{figure}

After  changing to a rotating frame that oscillates with the same pace as 
$\phi$, with $\psi=\theta-\phi$, the non-dimensional version of Eq.~(\ref{KM_inertia_eff1}), reads
\begin{equation}
\frac{{\mbox{d}}^2\psi}{{\mbox{d}}\tau^2}+\beta\frac{{\mbox{d}}\psi}{{\mbox{d}}\tau}+\sin\psi=I
\label{KM_inertia_eff2}
\end{equation}
where $\tau=\sqrt{\lambda k r} t$, $\beta=\alpha/\sqrt{\lambda kr}$, $I=D[k-\langle k\rangle-C(\lambda r)](\lambda k r)$ and $C(\lambda r)=(\ddot\phi+\alpha\dot\phi)/D$. This equation is easily recognized as that of a damped pendulum driven by a constant torque $I$. This system is known to display hysteresis when, for instance, at a fixed damping value $\beta$, one increases and then decreases the value of $I$~\cite{strogatzbook}. In particular, the transition from the stable fixed point (here the synchronized state) to the limit cycle (here the incoherent state) and vice-versa do not  happen at the same value of $I$. 

Following the analogy with the driven pendulum, it is possible to understand the roots behind the abrupt synchronization transition of those group of nodes of the same degree, {\em i.e.}, the cluster synchronization transition. Recall that each degree class $k$ obeys a different Eq.~(\ref{KM_inertia_eff2}), since $\tau=\tau(k)$, $\beta=\beta(k)$ and $I=I(k)$. To illustrate the different behavior of each degree class, Fig.~\ref{fig:second2} shows the phase diagram of the pendulum in the $(\beta,I)$-plane for (a) increasing (forward) and (b) decreasing (backward) $\lambda$ (note that bi-stability appears in the region where the limit cycle occurs for increasing $\lambda$ while the fixed point is the local attractor when decreasing $\lambda$). The trajectories corresponding to two degree classes ($k$=8 and $k=20$) are superposed. From Fig.~\ref{fig:second2}(a) is clear that those nodes with degree $k=8$ reach the synchronized state for a smaller value of $\lambda$ in the forward diagram, while Fig.~\ref{fig:second2}(b) shows that nodes with $k=20$ leave the synchronized state before than those of $k=8$.

{\color{black} The scenario of cluster synchronization, however, is not robust under external perturbations. Interestingly, Cao {\em et al.}~\cite{ctwzl18} have recently shown that when the second-order Kuramoto model with CDF is subjected to stochastic perturbations cluster synchronization may be suppressed. Importantly, in these cases, cluster synchronization is  replaced by the usual ES of the first-order Kuramoto model.}

\subsubsection{Oscillators with amplitude dynamics}

We now move to the more realistic case of coupled units with both amplitude and phase as dynamical variables, $u_{i}(t)=\rho_{i}(t){\mbox{e}}^{{\mbox i}\theta_{i}(t)}$. A paradigmatic model for studying the connection between synchronization of purely phase oscillators with that of phase and amplitude ones is the Stuart-Landau (SL) model~\cite{ak02},
\begin{equation}
\dot{u}_i=\left(\alpha-{\mbox{i}}\omega_i-|u_i|^2\right)u_i+\lambda\sum_{j=1}^{N}A_{ij}(u_j-u_i)\;,
\label{SLmodel}
\end{equation}
where $\omega_i$ represents the natural frequency of the $i$-th oscillator. Interestingly, when parameter $\alpha$ is large (compared to the coupling strength $\lambda$) the amplitudes of the oscillators are steady $\dot{\rho_i}=0$ and Eq.~(\ref{SLmodel}) becomes that of the Kuramoto model, Eq.~(\ref{ekurageneral}). Thus, the SL model allows a direct connection with Kuramoto dynamics~\cite{ggf16} by tuning parameter $\alpha$. In this class of models two synchronization order parameters are at work, namely,
\begin{eqnarray}
r^{\theta}&=&\left|\frac{1}{N}\sum_{i=1}^N{\mbox{e}}^{{\mbox{i}}\theta_i}\right|\;,
\\
r^{u}&=&\left|\frac{1}{N}\sum_{i=1}^Nu_i\right|\;.
\end{eqnarray}
The first, $r^{\theta}$, measures the phase coherence while the second, $r^{u}$, measures the degree of global synchronization including the coherence of amplitudes.

The application of the CDF framework in Eq.~(\ref{SLmodel}) was studied in~\cite{bfsbnfk12,gcffgf13} for star graphs and complex networks. However, the match $\omega_i=k_i$ in Eq.~(\ref{SLmodel}) does not produce ES in any of the topologies studied. On the other hand, the authors observed that the CDF scheme produces a novel phenomenon called {\em remote synchronization} which refers to phase coherence between units that are neither directly connected nor connected by a finite path of units sharing phase coherence. 

\begin{figure}[!t]
\begin{center}
\includegraphics[width=.75\textwidth,angle=0,clip=1]{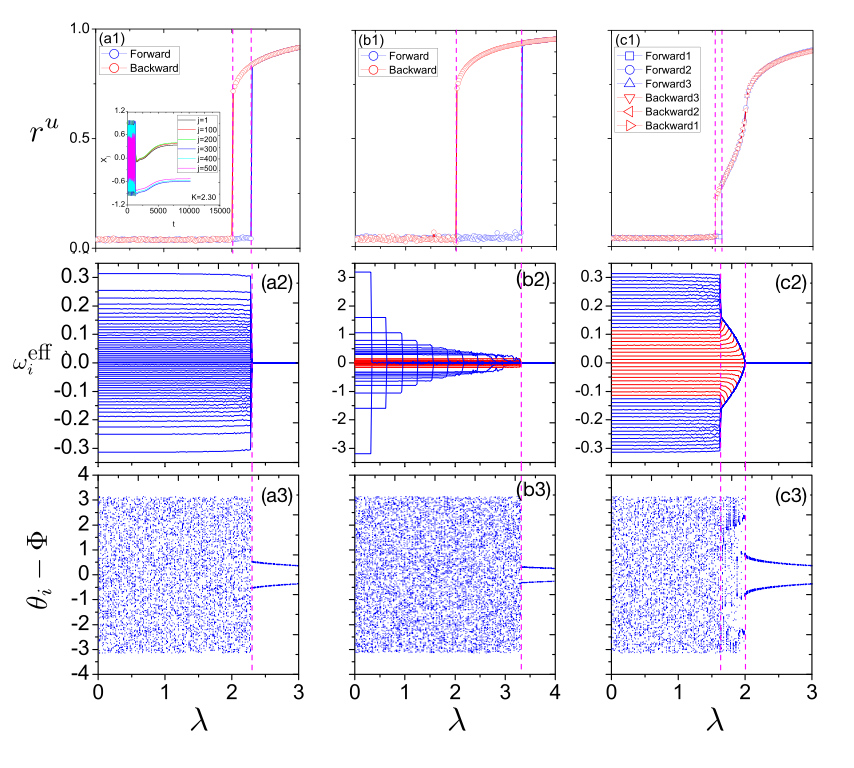}
\end{center}
\caption{Each column corresponds to a different $g(\omega)$: (a) triangular, (b) Lorentzian and (c) homogeneous. The first row reports the forward and backward global synchronization diagrams $r^{u}(\lambda)$. The second row shows the evolution with increasing $\lambda$  of the effective frequencies, Eq.~(\ref{effective1}). Finally, in the bottom row the evolution of the differences between the individual phases $\theta_i$ and the average one $\Phi$ is reported for increasing $\lambda$. [Reprinted and adapted from~\cite{bhzzzg14}].}
\label{fig:SL}
\end{figure}

At variance with CDF, the implementation of the CFC framework in the SL model gives rise to  explosive transitions. {\color{black} However, in this case the abrupt transition refers to the sudden death of oscillations (OD).} This result was presented in~\cite{bhzzzg14} where the authors implemented the frequency-coupling correlation in Eq.~(\ref{SLmodel}) by changing  $\lambda\rightarrow\lambda|\omega_i|/N$. The authors focused on fully connected graphs and analyzed different frequency distributions $g(\omega)$: triangular, Lorentzian and homogeneous. In the three cases the numerical simulations showed explosive transitions with bi-stability regions for both $r^{\theta}$ and $r^{u}$. However, the microscopic analysis of these explosive transitions yield some differences in terms of how the effective frequencies of oscillators approach to the mean $\Omega=0$. On the other hand, the common feature of these transitions is that, in the synchronized regime, the oscillators self-organize into two clusters of the same amplitude but with two different phases. In Fig.~\ref{fig:SL} the three explosive transitions are reported via the diagrams $r^{u}(\lambda)$, the microscopic evolution with increasing $\lambda$ of the effective frequencies $\omega^{\mbox{\scriptsize eff}}_{i}$, Eq.~(\ref{effective1}), and the difference between the phases of the oscillators and the average phase $\Phi$.

\begin{figure*}[!t]
\begin{center}
\includegraphics[width=.95\textwidth,angle=0,clip=1]{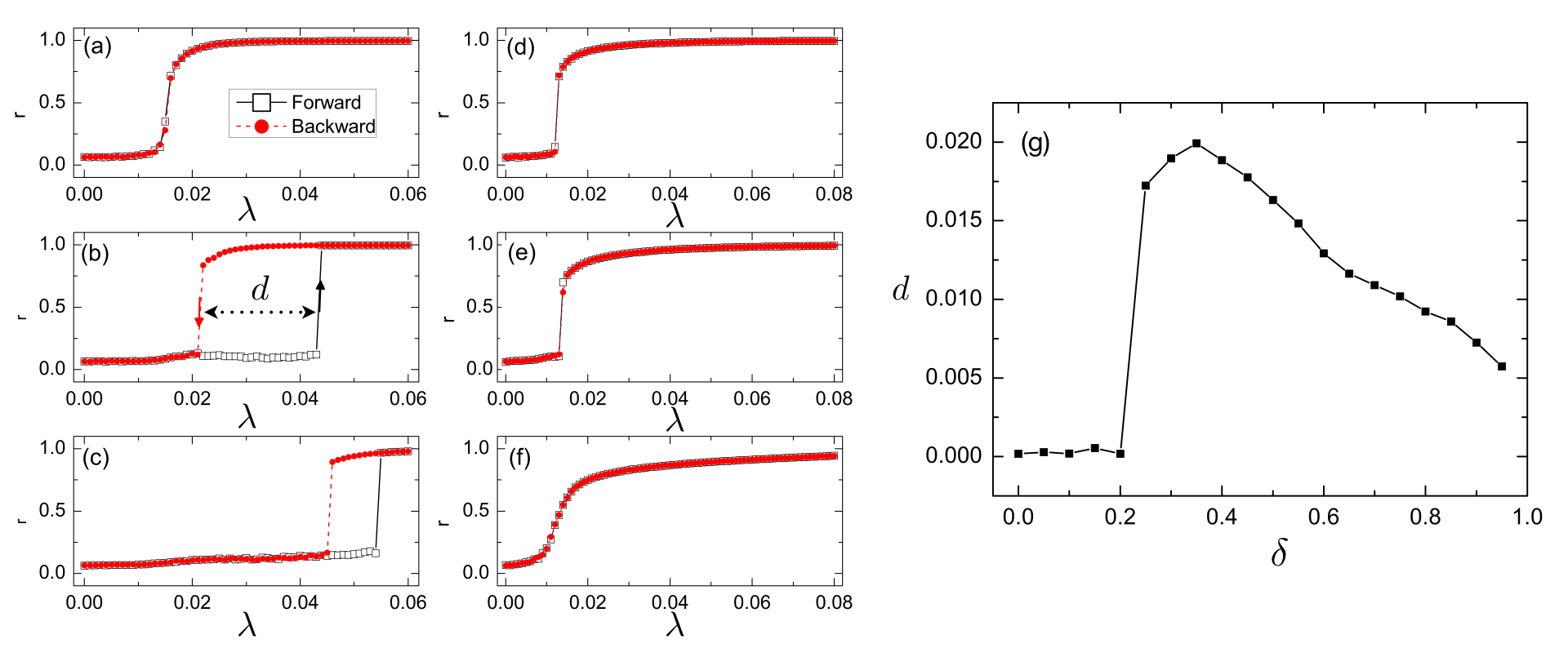}
\end{center}
\caption{(a)-(c) Forward and backward synchronization diagrams for a SF network ($\gamma=3$, $N=200$, $\langle k\rangle=6$) of coupled FN units. (d)-(f) Forward and backward synchronization diagrams for an ER network ($N=200$, $\langle k\rangle=6$). The values of $\delta$ are $\delta=0.1$ [(a) and (d)], $\delta=0.3$ [(b)-(e)] and $\delta=0.9$ [(c)-(f)]. Panel (g) shows the length of the hysteresis cycle, $d$, as a function $\delta$. [Reprinted and adapted from~\cite{chhsh13}].}
\label{fig:FN}
\end{figure*}

{\color{black} Apart from the SL, other paradigmatic models of coupled oscillators with phase and amplitude dynamics have been analyzed. In~\cite{ccwh15}, the Complex Ginzburg-Landau~\cite{ak02} is analyzed in a population of homogeneous ($\omega_i=\Omega$) oscillators interacting through a SF network. The authors show that, despite of the dynamical homogeneity of this system, explosive phase coherence shows up in a wide region of parameter space. In~\cite{Verma17} and~\cite{zsyx18} explosive transitions to amplitude death (AD) are found for coupled Van der Pol (VP) oscillators. In particular, in~\cite{Verma17} a set of mean-field coupled VP oscillators is analyzed while~\cite{zsyx18} tackles the networked version. Interestingly, in the latter work it is shown that the backward transition point is independent of the underlying network topology while the average degree has a strong impact on the nature of the explosive (AD or OD) transition found.}

\subsubsection{Neural models}

Following with equations for diffusively coupled oscillators with cubic nonlinearities, we now report the results in~\cite{chhsh13} for the FitzHugh-Nagumo (FN) model~\cite{Murraybook}. In this work the authors consider the system analyzed in~\cite{pikovskyk97} of coupled FN oscillators,
\begin{eqnarray}
\epsilon \dot{x}_i&=&x_i-x_i^3-y_i+\lambda\sum_{j=1}^{N}A_{ij}(x_j-x_i)\;,\label{FN1}\\
\dot{y}_i&=&x_i+\alpha_i+\xi_i(t)\;.\label{FN2}
\end{eqnarray}
In these equations $\epsilon=0.01$ so that, as usual in relaxation oscillators, we have a  fast ($x$) and a slow ($y$) variable. Finally, $\xi_i(t)$ is the term of Gaussian noise of unit $i$ satisfying $\langle \xi_i(t)\rangle=0$ and $\langle \xi_i(t)\xi_j(t')\rangle=2D\delta_{ij}\delta(t'-t)$. 

The relevant parameter governing the behavior of each unit is $\alpha_i$. In particular, for $|\alpha_i|>1$ the unit becomes excitable, while $|\alpha_i|<1$ leads to an oscillatory behavior. Since the frequency of unit $i$ is a decreasing function of $\alpha_i$ the authors in~\cite{chhsh13} applied the CDF framework by defining
\begin{equation}
\alpha_i=0.99-\delta\frac{k_i-k_{{\mbox{min}}}}{k_{{\mbox{max}}}-k_{{\mbox{min}}}}\;,
\label{corrFN}
\end{equation}
where the parameter $\delta$ controls the slope of the dependence of $\alpha_i$ with the degree of the node. Given that $\delta>0$, from Eq.~(\ref{corrFN}), is clear that the oscillation frequency of a node $i$ increases with its degree.

By solving numerically Eqs.~(\ref{FN1})-(\ref{FN2}) the authors study the synchronization diagrams for ER and SF networks for different values of $\delta$. These simulations, see Fig.~\ref{fig:FN}, reveal that only SF networks show ES transitions. However, a careful inspection of the role played by $\delta$ (see Fig.~\ref{fig:FN}(g)) highlights that there is critical value $\delta_c\simeq 0.2$ for the onset ES. At variance with most of the models reviewed, for $\delta\gtrsim \delta_c$ the length of the bistable region $d$ suddenly takes large values. Moreover, as $\delta$ increases the length of the bi-stable region decreases. 

\subsubsection{Rossler chaotic oscillators}

\begin{figure}[!t]
\begin{center}
\includegraphics[width=.5\textwidth,angle=-90,clip=1]{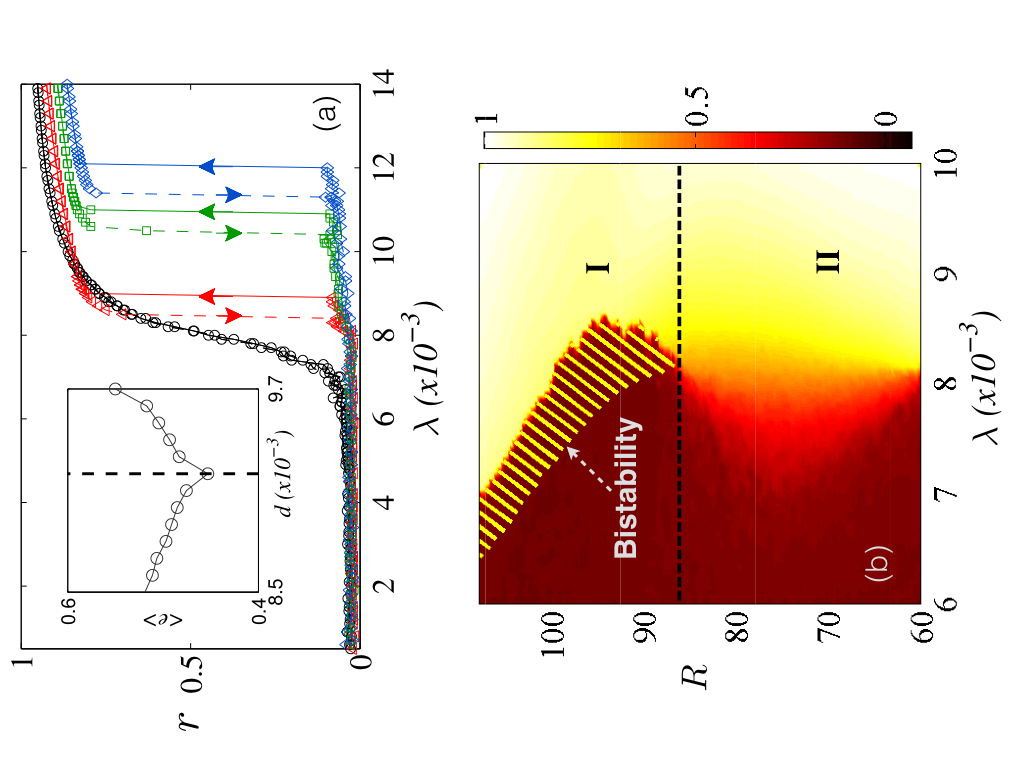}
\end{center}
\caption{(a) Synchronization diagram $r(\lambda)$ for different SF networks with $\gamma=2.2$ (red triangles), $\gamma=2.5$ (green squares), $\gamma=3.0$ (blue diamonds and black circles). In all the cases $N=1,000$ and $\langle k\rangle = 6$. The parameters are $\delta=6$ and $R=70$ (black circles) and $\delta=10$ and $R=100$ (other cases). The inset plot reports the average synchronization error $\langle e \rangle$ vs.~$\lambda$ close to the first-order transition for the network with $\gamma=3.0$, $R=100$ and
$\Delta\alpha=8.0$. (b) Average synchronization degree $r$ in the parameter space $\lambda-R$ for the SF network with $\gamma=3$ and $\delta=6$. The dashed line separates the region (I) where a second-order transition occurs and that for which the transition is explosive (II). This region contains area where the system is bistable. [Reprinted and adapted from~\cite{rossler}].}
\label{fig:Rossler1}
\end{figure}

To complete the discussion, 
we move to the domain of chaotic oscillators. In~\cite{rossler} the authors analyze a system of diffusively coupled Rossler-like units. In particular, the authors analyze the following Rossler system introduced in~\cite{Pisarchik06,Pisarchik09},
\begin{equation}
\dot{x}_i = -\alpha_i \left[ \Gamma \left( x_i-\lambda\sum_{j=1}^N A_{ij}(x_j-x_i) \right) + \beta y_i + \epsilon z_i \right],
\label{Rosslerx}
\end{equation} 
\begin{eqnarray}
\dot{y}_i &=& -\alpha_i(- x_i + \nu y_i) \;, 
\label{Rosslery}\\
\dot{z}_i &=& -\alpha_i(-g(x_i)+z_i) \;.
\label{Rosslerz}
\end{eqnarray}
As usual in Rossler units, nonlinearity enters in the $z$ variable of each unit and,  for this particular system, it is specified by a piecewise function defined as
\begin{equation}
g(x_i)=\left\lbrace \begin{array}{cc}
0 & \mbox{if $x_i\leq 3$} \\
\mu(x_i-3) & \mbox{if $x_i > 3$}
\end{array} \right. \ .
\end{equation}
In~\cite{rossler} some of the parameters are fixed ($\Gamma=0.05$, $\beta =0.5$, $\epsilon=1$ and $\mu = 15$) while $\nu = 0.02-\frac{10}{R}$, where $R$ is a tunable quantity that regulates the dynamical state of the system. In particular, for the values specified above, the system operates in the chaotic region when $R\in[55, 110]$~\cite{Pisarchik06}. The second tunable parameter is the coupling between units $\lambda$ that, in this case, is applied through the $x$ variable.

Since the natural oscillation frequency of a node $i$ depends linearly on the parameter $\alpha_i$, the authors apply the CDF framework by  imposing a positive correlation between $\alpha_i$ and the degree $k_i$ of each unit according to
\begin{equation}
\alpha_i=\alpha \left(1+\delta \frac{k_i-1}{N}\right)\,,
\label{corre}
\end{equation}
where $\alpha=10^4$ and, as in Eq.~(\ref{corrFN}), $\delta$ controls the slope of the function $\alpha_i(k_i)$. Note that, at variance with Eq.~(\ref{corrFN}) here $\alpha_i$ is not bounded and, therefore, for heterogeneous degree distributions the CDF framework implemented yields a very broad frequency spectrum.

The synchronization diagrams obtained from the numerical solution of Eqs.~(\ref{Rosslerx})-(\ref{Rosslerz}) are reported in Fig.~\ref{fig:Rossler1}(a) for three SF networks with different exponents $\gamma$. The results for the SF network with $\gamma=3$ point out that, depending on the dynamical region (controlled by $R$), this network shows explosive or continuous transitions. In Fig.~\ref{fig:Rossler1}(b) the synchronization diagram $r(\lambda,R)$ for this network confirms the former result. For $R\gtrsim 88$ (region II) ES of chaotic states of chaos appears, together with a region of bistability.

\subsection{Experiments and real-world applications}

The variety of mechanisms leading to ES allows for them to be implemented in controlled experimental setups and observed in real-world systems. Here we review the experimental observations of ES reported in the literature. 

\subsubsection{Implementations in the lab} 

\begin{figure}[!t]
\begin{center}
\includegraphics[width=.55\textwidth,angle=-90,clip=1]{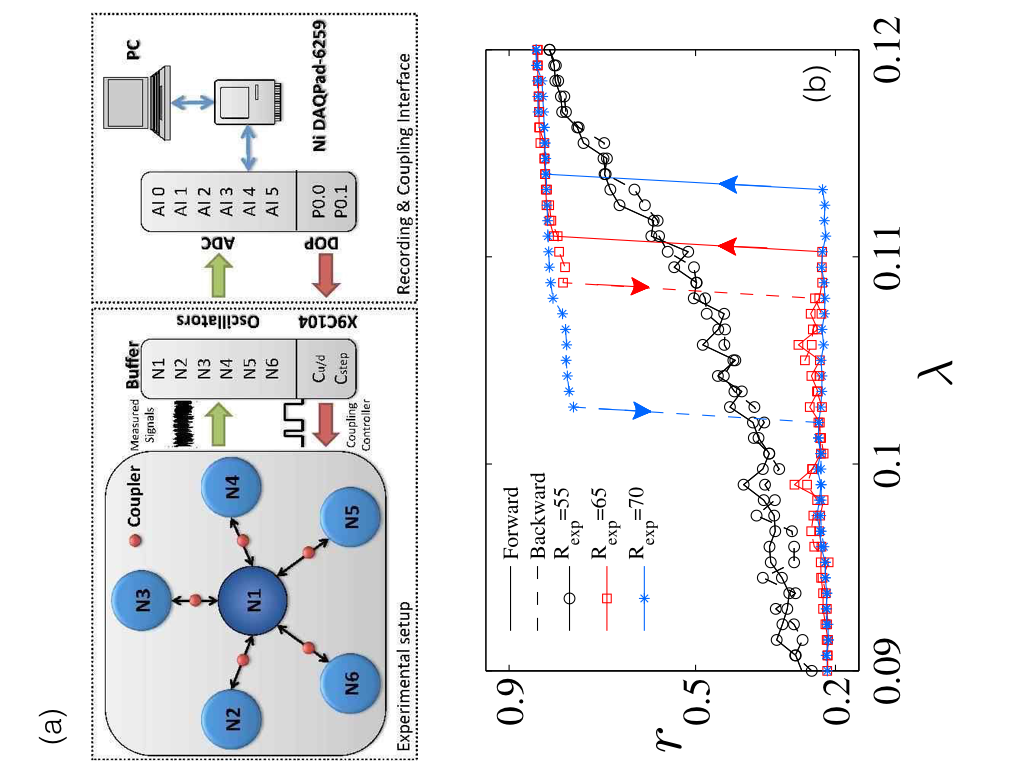}
\end{center}
\caption{(a) Schematic illustration of the experimental setup. The $6$ R\"ossler circuits are connected in a star configuration and the bidirectional coupling between the central node and each of the $5$ leaves is tuned by means of $5$ digital potentiometers. The outputs of the Rossler circuits are sent to a buffer and then used to control the coupling term of each node. The experiment is controlled by a PC with a the software Labview. (b) Forward and backward synchronization diamgram, $r(\lambda)$  for the star configuration shown in (a) for $3$ values of the internal resistance determining the chaotic state are $R_{\mbox{\scriptsize exp}}=55$ (black circles, second-order phase transition), $R_{\mbox{\scriptsize exp}}=65$ (red squares, first order phase transition with narrow hysteresis) and $R_{\mbox{\scriptsize exp}}=70$ (blue stars, first order phase transition with wide hysteresis). [Reprinted and adapted from~\cite{rossler}].}
\label{fig:Rossler2}
\end{figure}

The numerical observation of ES in the Rossler system allowed the authors of~\cite{rossler} to validate experimentally the existence of this phenomenon in a real setup subjected to stochastic fluctuations and parameter mismatches. In particular,  they implemented the electronic network shown in Fig.~\ref{fig:Rossler2}.a consisting of $N=6$ piecewise R\"ossler circuits (labeled as $N1$, $N2$, \dots, $N6$) having the same internal $R_{\mbox{\scriptsize exp}}$, so to ensure that they operate in an almost identical dynamical (chaotic) regime. These circuits were constructed as in introduced in~\cite{Pisarchik06,Pisarchik09} so that they obey the evolution described in Eqs.~(\ref{Rosslerx})-(\ref{Rosslerz}).

Due to the impossibility of constructing large-scale SF networks as those used in numerical simulations, the circuits were arranged in a star-like configuration, as this topology has been previously shown (Sec.~\ref{sec:star}) to display ES in the CFD Kuramoto model.
To implement the frequency mismatch between the central and peripheral nodes, the authors made a fine tuning of the values of the capacitors so that the central circuit ($N1$) oscillates with a mean frequency of $3,333$~Hz, and those circuits acting as leaves  ($N2$, \dots, $N6$) have frequencies in the range of $2,240$~$\pm$~$200$~Hz. Obviously, the frequencies of the circuits are slightly different  due to experimental variability.

The experiment was performed by coupling simultaneously the $6$ chaotic oscillators by means of $5$ digital potentiometers (red nodes) to ensure that the coupling strength $\lambda$ can be tuned simultaneously for all the pairs of connected nodes. This way, for each value of $\lambda$, the output signal from each circuit is stored and the time evolution of the phase of each circuit is reconstructed. These phases are then analyzed to compute the global order parameter $r$ of the system. 

In Fig.~\ref{fig:Rossler2}(b)  the experimental forward and backward synchronization diagrams, $r(\lambda)$, are reported for $3$ values of the parameter $R_{\mbox{\scriptsize exp}}$. In close analogy with the results for SF networks, Fig.~\ref{fig:Rossler1}(b) shows that the nature of the synchronization transition crucially depends on the value of $R_{\mbox{\scriptsize exp}}$. For  $R_{\mbox{\scriptsize exp}}=55$ (just in the limit  of the chaotic regime) the transition to synchrony is smooth while the diagram for $R_{\mbox{\scriptsize exp}}=65$ and $70$  shows a sharp increase of the synchronization parameter $r$ together with the hysteresis cycle. This experimental observation was the first one to put the theoretical predictions about ES into practice. 

Another interesting experimental set up has been proposed in~\cite{kvpb15} where the authors analyze a star network of mercury beating-heart (MBH) oscillators. A MBH is an electrochemical redox reaction between mercury, iron and chromium discovered by~\cite{l73}. The reaction causes a blob of mercury in water to oscillate. The authors implement the CDF framework in this system and show that, as the coupling strength is increased (forward direction) and decreased (backward direction), the forward and backward curves of the synchronization order parameter do not overlap, thus indicating the existence of a hysteresis (bistability) region, where explosive synchronization emerges. Moreover, it is given evidence that the dynamics of the oscillators can be conveniently and permanently switched from one phase to the other superimposing an instantaneous external signal, if the system operates in the bistable region.

\subsubsection{Spontaneous synchronization in real systems}

The first reported study on the phenomenon of spontaneous synchronization in networked real systems, was presented in~\cite{mman13} for the case of the electric power-grid. The authors show that the leading factor for stability is the relation between the specifics of the dynamical units and the network structure. To this aim they represent each dynamical unit using the swing equation~\cite{anderson2008power}, and couple them in a topological network 
in accordance with real configurations of the power-grid around the world. The object of analysis, however, is not the adjacency matrix of the physical topology of connections but the admittance matrix, defined as the Laplacian matrix for the admittance (a measure of how easily a device will allow a current to flow) between nodes. The stability analysis of the system is performed by assuming small perturbations of the synchronized state, in the same fashion as the  master stability function~\cite{pc98}. This analysis results in a condition for the Lyapunov exponents, connecting topology and dynamics. Other previous works have also addressed the synchronization properties of (synthetic) power-grids~\cite{lbd12,rstw12}. The spontaneous synchronization observed in real power-grids is then described as an outcome of the destabilization of the power-grid.

{\color{black}  Apart from the previous works, the other interesting field where spontaneous synchrony seems to be important is neuroscience. There is an ongoing consensus in biology~\cite{ckw08} about the role of bistability as a natural framework for biological switches. This bistability is many times reported as an hysteresis cycle controlling the response of the system to  external stimuli, with synchronization being a paradigmatic example behind many phenomena. Following these ideas, many abrupt transitions and hypersensitive responses to perturbations  observed in neuroscience have been recently revisited from the perspective of ES. Examples include the conscious-unconscious transition~\cite{kmmvt16,kim17}, epileptic seizures in the brain~\cite{wtdl17}, the sensitive frequency detection of the cochlea~\cite{pgnl16}, the hypersensitivity of Fibromyalgia patients~\cite{lklkckmh18}, and the activity of the choroid plexus controlling mammalian circadian clocks~\cite{Myung18}.}

In~\cite{kmmvt16,kim17} the authors hypothesized that the conditions for explosive synchronization in human brain networks would be present in the anesthetized brain just over the threshold of unconsciousness. In particular, network conditions for explosive synchronization have been found in empirically-derived functional brain networks~\cite{jfehhk13}.  Unconsciousness and resting states~\cite{dpmmrh13} seem to be related to a bifurcation point from which the dynamical system confronts multistability that may lead to spontaneous synchrony.

{\color{black} In~\cite{wtdl17} the authors investigate the onset of epileptic seizures as the onset of ES. This study reveals that ES may be induced due to a rewiring of the underlying network that increases local clustering.} 

Also recently, the authors in~\cite{pgnl16} proposed a model in which acoustical signal transduction in the cochlea (a spiral-shaped cavity forming the auditory portion of the inner ear) presents an abrupt synchronization transition which provides a plausible explanation for the system's ability for frequency selectivity. 

All of these works are only beginning to reveal the horizon, and more work is still necessary 
to accurately relate the phenomenon of ES to observations in real systems. One example linking the observation of ES with the presence of one of its related mechanisms is the connection between Fibromyalgia and ES. By analyzing EEG signals from patients with chronic pain, the authors of~\cite{lklkckmh18} revealed a significant positive correlation between the frequency and degree of cortical areas within the alpha frequency, pointing out that ES (under the CDF framework) could be the underlying mechanism behind the hypersensitivity that characterizes this condition. Another example along this same line~\cite{Myung18} has unveiled that the activity in the choroid plexus of mice 
adjusts the suprachiasmatic nucleus clock, responsible for generating the circadian rhythms. When analyzing the dynamical patterns of gene expression found in the choroid plexus, the authors suggest that the frequency and coupling of individual oscillators in the choroid plexus are positively correlated (CFC framework) and that this mechanism is critical to how synchronization and spatial patterning arise in the choroid plexus.



\subsection{ES in a nutshell}

\begin{figure*}[!t]
\begin{center}
\includegraphics[width=.85\textwidth,angle=0,clip=1]{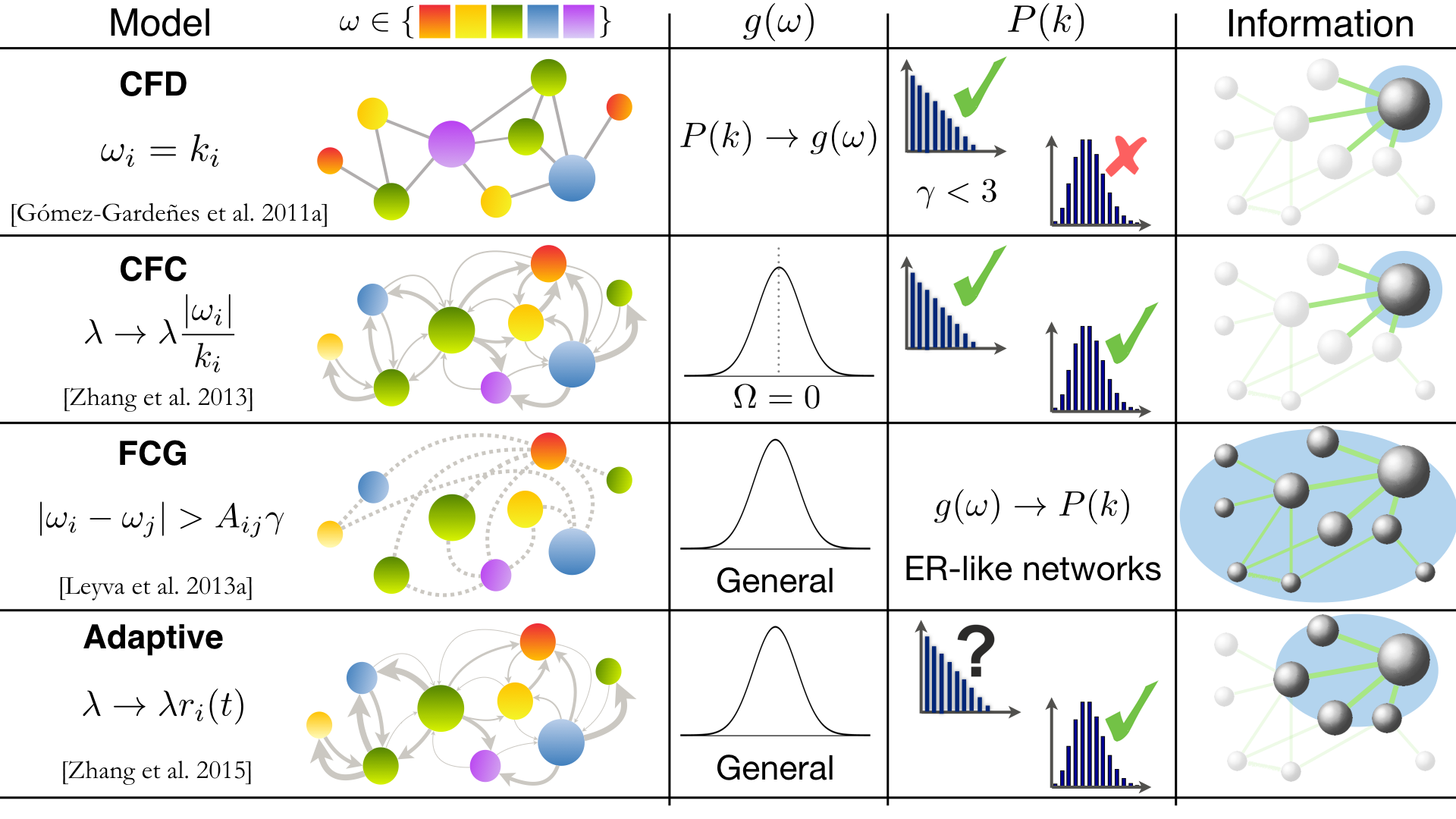}
\end{center}
\caption{Table summarizing the frameworks leading to ES. In the first column we report the mechanisms implemented in each model as applied in the Kuramoto framework together with the references that introduced them. The second column illustrates schematically the implementation of each rule in a toy network where the colors of each node account for their corresponding natural frequency (see legend on the top of the column). The third and fourth column summarize the dependence of the ES on the frequency $g(\omega)$ and degree $P(k)$ distributions. In the fifth column the information horizon needed to implement each rule is illustrated.}
\label{fig:summaryES}
\end{figure*}

To round off this section devoted to ES we summarize and compare the main frameworks leading to this phenomenon. In a similar fashion as for EP (recall Fig.~\ref{fig:EP_classes}) this summary is reported in Fig.~\ref{fig:summaryES} where, in the first column, the basic rules of the CFD, CFC, FCG and Adaptive frameworks are shown as they are implemented in the Kuramoto model together with the references of the works introducing each framework. 

The toy networks drawn in the second column show how the former rules are implemented. The colors of the nodes represent the natural frequencies of nodes, see legend on the top (adjacent colors represent close values for $\omega$). This way, the assignation of $\omega$ in the CFD model is made according to the degree of each node. In the cases of the CFC, FCG and Adaptive frameworks the frequencies have been randomly assigned. The distribution of frequencies affect, in these cases, the interactions among the nodes. Namely, for the FCG and Adaptive  frameworks the backbone network is preserved but  the redefinition made for the interaction coupling of a node $i$ with its neighbors turns the original network into a directed one. In addition, the directional links are weakened or reinforced depending on the attributes of the nodes (CFC model) or the dynamical state of the neighborhoods (Adaptive framework). For the FCG case the frequency distribution biases the skeleton  of interactions since the existence of a link is restricted to those pairs of nodes with a pre-defined frequency gap (dashed lines).

The third and fourth column of Fig.~\ref{fig:summaryES} report the dependence of ES on the frequency and degree distributions. While for the CFD model the degree distribution imposed 
automatically forces the frequency distribution to be the same, the rest of the ES cases reported have been 
for a variety of typically even frequency distributions (such as Gaussian, Lorentizian, homogeneous, etc). Note that in the CFC framework the definition of $g(\omega)$ has to be made keeping in mind that, due to the loss of rotational invariance of the equations introduced by the redefinition of the coupling, the distribution has to be centered around $\Omega=0$. The existence of ES is also influenced by the $P(k)$ of the underlying network. For instance, in the CFD model only SF networks with $\gamma<3$ display ES while for the FCG model the $P(k)$ of the network is imposed by the frequency assignment given by $g(\omega)$, giving rise to homogeneous (ER-like) topologies. For the CFC and Adaptive models there is no restriction for $P(k)$ although, for the Adaptive case, the existence ES in SF  networks is still to be studied.

Finally, the fifth column in Fig.~\ref{fig:summaryES} reports the information horizon needed to implement the different rules leading to ES. For the CFD and CFC this horizon is purely local since only the attributes of single nodes are needed. This horizon expands for the Adaptive framework since the dynamical states of the neighbors of each node $i$ are needed to construct $r_i$. For the FCG case the information horizon expands further to the whole network since the construction of the contact network implies the search to all possible pairs of nodes that fulfill the frequency gap condition.

\section{Connection between EP and ES}\label{sec:EPES}

The phenomenology behind EP and ES share similar ingredients, but 
there are also apparent mechanistic and phenomenological dissimilarities. Both EP and ES are fundamentally based on the implementation of some microscopic rules that avoid the formation of a macroscopic (connected or synchronized) component that leads to the onset of the collective phenomenon at work. 
In EP these rules are explicit. In ES the rules are implicit and result from the combination of the network topology and the natural frequency distribution of the nodes.
Table~\ref{fig:EPESdifferences} summarizes some apparent key distinctions between EP and ES.
Whereas EP is usually defined as a kinetic process with links between nodes 
added or removed,
ES is based on a dynamical process where nodes are either in sync or not, depending on their phase difference
and frequencies. In ES the network dynamics run on a underlying fixed network whereas in EP the underlying structure evolves as nodes across distinct clusters are linked together. Typically, EP is based on a non-local cluster merging process while in ES cluster merging is a local node-to-node based process. 
EP requires a sufficiently homogeneous initial cluster size distribution
whereas in ES the node frequency distribution is usually heterogeneous.
Finally, in EP, at $p_c$, a gap between small-sized and larger-sized clusters is observed, whereas for ES 
a necessary condition can be formulated in terms of frequency gaps.

\begin{table*}[t]
\begin{center}
\begin{tabular}{|l|l|l|} \hline

\ \ Explosive Percolation &  \ \ Explosive Synchronization \\ \hline

\ \ Kinetic process & \ \ Dynamical process\\

\ \ Cluster nodes are linked & \ \ Cluster nodes are in sync   \\

\ \ Typically non-local cluster merging & \ \  Local node-to-node-based cluster merging  \\

\ \ Sufficiently homogeneous initial cluster size distribution \ \ \ & \ \  
Heterogeneous node frequency distribution (sufficient condition) \ \ \   \\

\ \ Gap between small-sized and large-sized clusters (powder keg) \ \ \ & 
\ \ Frequency gap condition (Sec:~\ref{subsec:FGC}) \ \ \ \\

\hline 

\end{tabular}
\end{center}
\caption{Key distinctions between EP and ES.}
\label{fig:EPESdifferences}
\end{table*}%

\begin{figure}[b]
\begin{center}
\includegraphics[width=.55\textwidth,angle=0,clip=1]{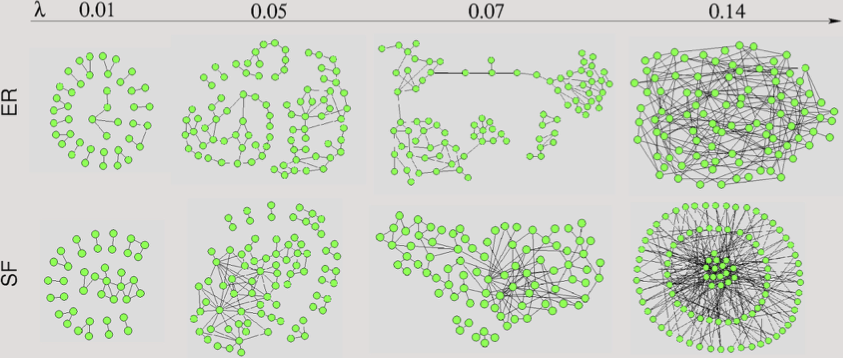}
\end{center}
\caption{Traditional path to synchronization for the Kuramoto model. Schematic illustration of the synchronized clusters for several values of $\lambda$ for an ER and a SF networks of $N=100$ nodes. In the case of the ER graph, the giant synchronized component emerges as a result of the aggregation of small clusters of synchronized nodes, while for the SF network there is always a giant cluster that grows as $\lambda$ increases. [Reprinted from~\cite{gma07a}].}
\label{fig:EPES1}
\end{figure}

While an exact map between EP and ES is yet to be established, in~\cite{zzbl14} the authors monitor the  formation of microscopic synchronized clusters in order to translate the synchronization transition as the coupling strength $\lambda$ varies in terms of a percolation-like process.
In particular, the authors apply the technique introduced in~\cite{gma07a} to monitor the formation of synchronized links by measuring the local degree of synchronization between two nodes $i$ and $j$ as
\begin{equation}
R_{ij}=\left|\lim_{T\rightarrow\infty}\int_{\tau}^{\tau+T}{\mbox{e}}^{{\mbox{i}}\left[\theta_{j}(t)-\theta_{i}(t)\right]}{\mbox{d}}t\;\right|,
\label{local_corr}
\end{equation}
where $T$ denotes the time-interval over which integration is performed. Following this definition, $R_{ij}=1$ when two nodes are fully synchronized, and $R_{ij}=0$ when their dynamics are completely incoherent. 

For the Kuramoto model in complex networks, Eq.~(\ref{ekurageneral}), the computation of $R_{ij}$ as a function of $\lambda$ allows for the study of how synchronized clusters emerge by monitoring the formation of each synchronized link between two nodes, say $i$ and $j$, whenever $R_{ij}>\eta\rightarrow 1$. In~\cite{gma07a} the creation of synchronized links in ER and SF networks was monitored using this technique establishing two qualitatively different paths toward synchronization. Namely, as shown in Fig.~\ref{fig:EPES1}, for ER graphs the transition starts with the formation of separated clusters of 
synchronized nodes that, at $\lambda=\lambda_c$, coalesce forming a macroscopic (in terms of nodes) 
cluster. For $\lambda>\lambda_c$ the clusters grow incorporating links between the already present nodes until full synchronization is reached~\cite{gma11b}. On the other hand, for SF networks the synchronization path is governed by a unique cluster initially formed by the hubs. This cluster grows monotonically by incorporating new nodes as $\lambda$ increases.

\begin{figure*}
\begin{center}
\includegraphics[width=.85\textwidth,angle=0,clip=1]{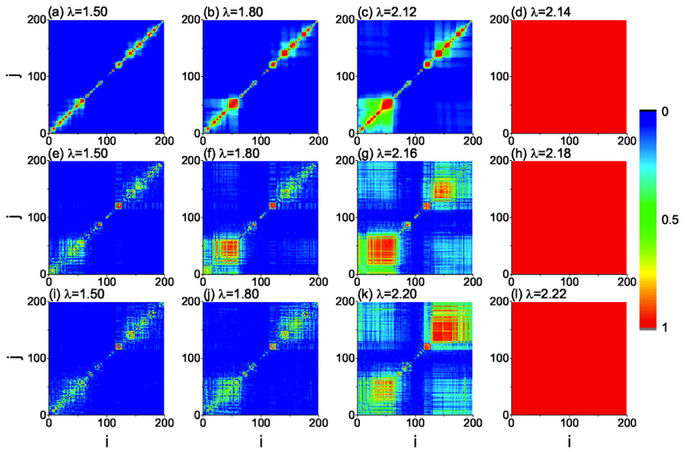}
\end{center}
\caption{Evolution of matrix ${\bf R}$ for four different coupling strengths $\lambda$. Panels in the top correspond to a fully connected network while those in the middle and bottom are for an ER graph and a SF network respectively. In the three cases the coupling strength of the first three panels starting from the left correspond to $\lambda<\lambda_c^{f}$, while the third matrix ${\bf R}$ is the one observed just before the onset of synchronization $\lambda\lesssim\lambda_c^{f}$. The fourth matrix from the left corresponds to $\lambda\gtrsim\lambda_c^{f}$. The rows and columns of the matrix are arranged in ascending order of the natural frequencies. $N=200$ for all the networks and $\langle k\rangle=6$ for the ER and SF graphs. [Reprinted from~\cite{zzbl14}].}
\label{fig:EPES2}
\end{figure*}

In~\cite{zzbl14} the authors follow the same procedure but apply it to the CFC Kuramoto model, Eq.~(\ref{CFCKuramoto}). This way they implement the forward (backward) continuation in $\lambda$ of the CFC model and measure the correlation matrix ${\bf R}$ for $\lambda$ values around $\lambda_c^{f}$ ($\lambda_c^{b}$). The results are shown in Fig.~\ref{fig:EPES2} for three network topologies with $N=200$ nodes: (top row) fully connected network, (middle row) ER graph, and (bottom row) SF network. In all the cases $g(\omega)$ is Gaussian with $\Omega=0$. The first three panels in each row (starting from the left) correspond to $\lambda$ values before the onset of ES in the forward branch, while the last one shows the matrix ${\bf R}$ immediately after $\lambda_c^{f}$. The rows (and columns) are arranged in ascending order of natural frequencies so that nodes with similar $\omega_i$ correspond to adjacent rows (columns) in matrix $R$.

The results of this study point out that when $\lambda<\lambda_c^{f}$ small synchronization clusters appear. However, these clusters are formed along the diagonal, pointing out that they are formed by nodes of similar frequencies, avoiding the creation of synchronized links connecting nodes of disparate frequencies (far from the diagonal). These clusters remain isolated even at the $\lambda\lesssim \lambda_c^{f}$, 
in contrast to the traditional paths to synchronization observed in complex networks (Fig.~\ref{fig:EPES1}). Whereas for $\lambda\gtrsim\lambda_c^{f}$ all clusters 
merge together forming a fully synchronized network. 

To illustrate how small clusters of synchronized nodes avoid merging together even when $\lambda\lesssim\lambda_c^{f}$ the authors in~\cite{zzbl14} derived a {\em suppressive rule} for the formation of synchronized links in the CFC Kuramoto model. Consider two oscillators with $\omega$ and $\omega'$ as natural frequencies, from Eq.~(\ref{effectiveCFC}) in Sec.~\ref{sec:CFCtheory}, the evolution for their phase difference $\Delta\theta=\theta-\theta'$ is
\begin{equation}
\Delta\dot{\theta}=\omega-\omega'+\lambda r\left[|\omega|\sin(\Phi-\theta)-|\omega'|\sin(\Phi-\theta')\right]\;.
\end{equation}
Therefore, these two oscillators will be locked, $\Delta\dot{\theta}=0$, when
\begin{equation}
\frac{|\omega-\omega'|}{|\omega|+|\omega'|}\leq \lambda r\;.
\label{suppressive}
\end{equation}
The suppressive rule points out that those oscillators with close frequencies that, in addition, lie  on the tails of the distribution $g(\omega)$ (centered at $\Omega=0$) will be those establishing synchronized links while for those around $\Omega=0$ the condition of Eq.~(\ref{suppressive}) is difficult to fulfill. This becomes evident from the absence of synchronized links in the central part of the matrices ${\bf R}$ in Fig.~\ref{fig:EPES2} corresponding to $\lambda<\lambda_c^f$, as these entries correspond to pairs of nodes with close and small absolute value frequencies. 

{\color{black}  Recently, another study has shed light on the behavior of synchronized clusters at the ES onset~\cite{kkmhb17}. Focusing on the backward transition, they study how the giant synchronized cluster is destroyed at $\lambda_c^{b}$. Their results reveal that, both for the CFD and CFC frameworks, the synchronized cluster disappears progressively in time, step by step, keeping the probability distribution of the phases of its constituents invariant during the process. Thus, this finding indicates the self-similarity of the synchronized clusters that are observed at $\lambda_c^{b}$.}

A mapping between EP and ES phenomena may be enabled by establishing connections between the explicit rules of EP and the implicit rules of ES. For instance, the suppressive rule of~\cite{zzbl14} is one such attempt.
It is also worth noting that not all of the explicit rules that lead to EP directly suppress the growth of the largest cluster, for instance the Devil's staircase rule operates by preferentially merging clusters of similar size~\cite{naglerPRX}. Likewise, one can initialize a set of clusters to have a ``powder keg" and then any edge addition rule, even random addition, leads to EP~\cite{powderkeg09,ChoClusterAgg}. 
Similarly, an ES transition may be prepared by a sufficiently broad natural frequency
distribution of the nodes, but the sufficiency is dependent on the underlying network topology. 
All-to-all coupling enhances the conditions for a continuous transition as the fraction of out-of-sync nodes
slows the growth of the largest cluster. On the other hand, a network topology with pronounced community structure
or degree disassortativity may enable ES.
Thus, a framework that unifies explosive transitions appears possible for some but not for all explosive phenomena.

\section{Other models with explosive behavior}\label{sec:Other}

\subsection{Explosive congestion in complex networks}

Transportation dynamics on networks can be, in general, interpreted as the flow of elements from an origin node to a destination node. When the network is used by simultaneous transportation processes, many elements necessarily travel through the same node or link. This, in combination with the possible physical constraints of nodes and links, such as the capacity to  
route a packet in finite time, can lead to network congestion, a phenomenon in which the amount of elements in transit on the network grows proportional to time~\cite{gdvca02,gadg02,zyckn05}. Network congestion is always a function of the network topology and the routing protocol. The most widely used transportation protocol 
is shortest path routing, a protocol consisting in minimizing the distance (or time) between two places.

\begin{figure}
\begin{center}
\includegraphics[width=.45\textwidth,angle=0,clip=1]{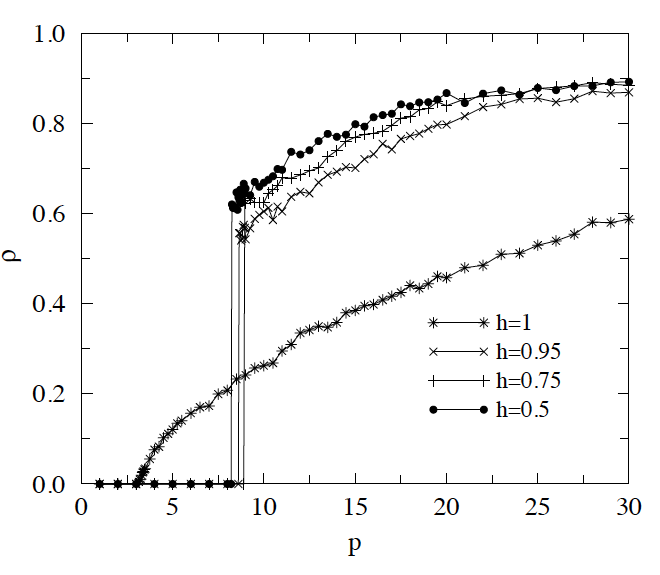}
\end{center}
\caption{
Explosive congestion transitions measure by the density of congested nodes $\rho$ as a function of the Poisson injection rate of packets to the network p. {\color{black} The parameter h weights the deviation from shortest path due to congestion awareness, Eq.(\ref{congawar})}. Note that h = 1 corresponds to the standard strategy in which traffic awareness is absent. As soon as traffic conditions are taken into account, the explosive congestion transition is reminiscent of a first-order phase transition and the critical point shifts rightward [Reprinted from~\cite{eggm05}].}
\label{fig:Echenique2004}
\end{figure}

The pioneering work by~\cite{eggm05} explored how the average network performance depends on the ability of the routing protocol to divert traffic across paths other than the shortest ones. To this end, the authors numerically explored a model in which a tunable parameter accounts for the degree of traffic awareness incorporated into a packet's delivery. Suppose that a node $l$ is holding a packet whose destination is node $j$.  Any node $i$ that is a neighbor to $l$, has an
effective distance between $i$ and $j$ defined as
\begin{equation}
\delta_i =h d_i +(1-h)c_i , 
\label{congawar}
\end{equation}
where $d_i$ is the minimum number of hops one has to pass by to reach $j$, i.e., the shortest path between $i$ and $j$, and $c_i$ is the length of the queue of packets to be delivered by node $i$. Additionally, $h$ is a tunable parameter that accounts for the degree of traffic awareness incorporated in the delivery algorithm. The main finding is that the onset of the congested phase is reminiscent of a first or a second order phase transition depending on whether or not the routing combines a shortest path delivery strategy with traffic aggregation at a local scale. This abrupt transition can be understood as a shifting of the critical point of a second order phase transition, that when attained reveals the large distribution of congestion across many points in the network, see Fig.~\ref{fig:Echenique2004}.
The explosive emergence of congestion in scale-free networks was further analyzed by~\cite{hwrqy07} proposing a model where the node queue length is proportional to the node degree and its delivering ability proportional to the queue length. They found a cycle of hysteresis, indicating that the system is bistable in a certain range of packet density. For different packet densities, the system can self-organize to four different phases: free-flow, saturated, bistable, and jammed. The bistability is related to the existence of a first order phase transition in the flow.

Later on, in~\cite{ddbm09} the authors proposed a minimal model to reveal the mechanisms triggering the emergence of congestion. These mechanisms are reminiscent of a bootstrap percolation process, where a node is active (here, congested) if the number of active neighbors exceeds a given threshold (in this case the length of the queue). 
A similar finding of the explosive transition to congestion, was exposed in~\cite{wzrgy09} for finite buffer nodes. The authors propose also control strategies, based on the idea of random packet dropping, to prevent complete congestion. Such control strategies can modify the order of the phase transition from first to second order. 

In a more general framework of dynamic information routing~\cite{ktb16}, the authors explore a generic mechanism to route information on top of collective dynamical reference states in complex networks. A dynamical reference state is simply a global solution of the dynamical process (e.g. synchronization). They prove specifically for a Kuramoto-like oscillatory dynamics, how individual unit properties, the network topology and external inputs  
act together in concert
to organize information routing. Interestingly, they found that changing for example the unit properties (in the case of the Kuramoto model, the intrinsic frequency of oscillators), the transition in information routing has a switch-like dependency on the control parameter reminiscent of the explosive transitions above mentioned.


\subsection{Explosive epidemics in networks}\label{subsec:explosiveEpidemics}

\begin{figure}[t]
\begin{center}
\includegraphics[width=.45\textwidth,angle=0,clip=1]{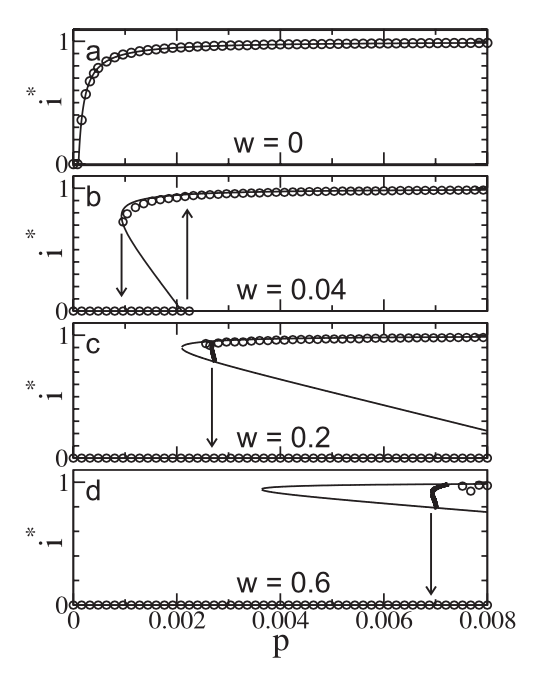}
\end{center}
\caption{
Bifurcation diagram of the {\color{black} density of the infected individuals $i^{\star}$ as a function of the infection probability $p$ for different values of the rewiring rate $w$.} 
Solid lines corresponds to the theoretical predictions, and symbols to explicit simulation. Without rewiring only a single continuous transition occurs. (a). By contrast, with rewiring a number of discontinuous transitions, bistability, and hysteresis loops (indicated by arrows) are observed (b), (c), (d). Fast rewiring (c), (d) leads to the emergence of limit cycles (thick lines indicate the lower turning point of the cycles), which have been computed numerically [Reprinted from~\cite{gdb06}].}
\label{fig:Gross2006}
\end{figure}

Epidemic spreading has a long tradition in the physics literature~\cite{pscvmv15} with  
a wide interest in the determination of the epidemic threshold: a value of the infectivity that separates two scenarios, non-invasive and invasive regimes, respectively. In particular, the epidemic threshold concept is analogous to the concept of critical point for a phase transition in non-equilibrium systems. This epidemic threshold phase transition has been proven to be continuous and second-order for the vast majority of epidemic dynamical models. However, from the pioneering work by~\cite{gdb06} on adaptive networks, which includes the possibility for individuals to protect themselves by avoiding contacts with infected people, the authors found the presence of a hysteresis loop and first order transitions in the fraction of infected individuals as a function of the rewiring (defined as the probability of redirecting connections to non-infected individuals, $w$), see Fig.~\ref{fig:Gross2006}. 

Usually, epidemic models assume that the transition to endemic states in a population can be fully explained in terms of  contagions between pairs of individuals. However, 
several epidemic models are also used to understand social information contagion, and the dynamics of social transmission depend not only 
on the characteristics of the transmitting and receiving individuals (e.g. on attitude or persuasiveness) but also  
on the context of the transmission event. In~\cite{gltp17}, the authors propose a model that accounts for explosive contagion in complex networks. Essentially, the model relies on a definition of the transmission from a transmitter $j$ to a receiver $i$ with synergistic rate given by
\begin{equation}
\lambda_{j\rightarrow i} = \alpha f (n^{h}(i)),
\end{equation}
\noindent where $\alpha$ is the intrinsic value of the spreading phenomenon in the absence of the context. The function $f(x)$ encodes the synergy of $x$ on the result of the transmissibility, and the argument $n^{h}(i)$ is the number of ignorant/healthy individuals connected with the receiver $i$. For two different choices of $f(x)$, linear and exponential, the authors explore the effect of a multiplicative parameter $\beta$ on $n^{h}(i)$, such that if $\beta>1$ the synergetic effect is positive, while for $\beta<1$ the synergy is destructive. Using a microscopic Markov Chain approach (MMCA)~\cite{gabmm10,ggma11}, it is possible to analytically describe the evolution of $p_i(t)$, which is the probability that a node $i$ becomes  
infected at time $t$ in a susceptible-infected-susceptible (SIS) epidemic model as follows
\begin{equation}
p_{i}(t+1) = p_{i}(t) (1-\mu) + (1- p_{i}(t)) q_{i}(t).
\end{equation}
\noindent Here $\mu$ is the recovery rate of infected individuals, and $q_{i}(t)$ the probability of being infected by a neighbors, which reads
\begin{equation}
q_{i}(t) = 1- \Pi_{j=1}^{N} (1- \lambda_{j\rightarrow i} A_{ij} p_{j}(t)),
\end{equation}
\noindent where $A_{ij}$ is the network adjacency matrix. The function $\lambda_{j\rightarrow i}$ is thus:  
\begin{equation}
 \lambda_{j\rightarrow i} = \alpha \exp[\beta\sum_{l=1}^{N}A_{il} (1- p_{l}(t))].
\end{equation}

\begin{figure}
\begin{center}
\includegraphics[width=.45\textwidth,angle=0,clip=1]{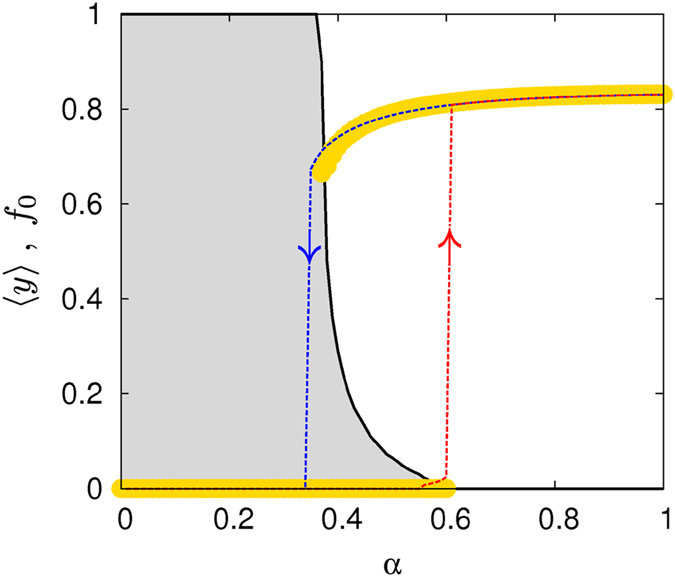}
\end{center}
\caption{
Average of infected individuals, $\langle y\rangle$ as a function of {\color{black} the intrinsic value of the spreading phenomenon in the absence of the context, $\alpha$} for the SIS process in a homogeneous random network {\color{black}with average degree $\langle k \rangle=6$ and for a value of the synergistic effect $\beta =-0.5$.}
The dashed curves indicate the solution obtained by solving the Markovian evolution equations whereas the solid amber circles correspond to the results obtained by using MC simulations ($10^3$ realizations for each value of $\alpha$). The hysteresis effect points out the existence of a bi-stability region. The solid curve shows the fraction $f_0$ of realizations (in the MC simulations) that end up in the fully ignorant solution, $\langle y\rangle=0$ . The recovery rate is $\mu=0.2$ [Reprinted from~\cite{gltp17}].}
\label{fig:Ggardenes2016}
\end{figure}

The phase diagram of the system is depicted in Fig.~\ref{fig:Ggardenes2016}. Both the Monte Carlo simulations and the MMCA theory predict the coexistence of endemic infected and infection-free states and the corresponding hysteresis effect with discontinuous transitions between these regimes. Similar results have been reported in the dynamics of the susceptible-infected-recovered (SIR) model with reinfections~\cite{gbpa15}.

\begin{figure}
\begin{center}
\includegraphics[width=.45\textwidth,angle=0,clip=1]{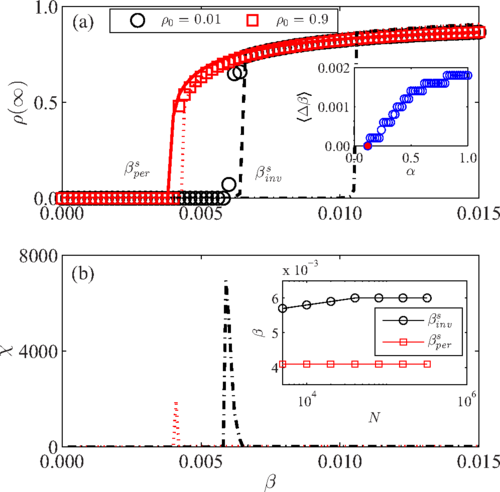}
\end{center}
\caption{
Steady state infected density $\rho(\infty)$ and susceptibility measure $\chi$ for random regular networks. (a) The density
$\rho(\infty)$ versus {\color{black} the infectivity parameter $\beta$ rescaled with $\alpha =0.9$, see Eq.(\ref{rescbeta})}, where the red squares and black circles are simulation results with initial infected density $\rho_0=0.9$ and
$\rho_0=0.01$, respectively. The red solid and black dashed lines are the
theoretical results with the same respective initial seed fractions. The red dotted and black dotted dashed lines are
results from a mean-field approximation. The quantities $\beta_{inv}$ and $\beta_{per}$ are, respectively, the simulated invasion and persistence thresholds determined via the susceptibility measure. (b) Susceptibility measure $\chi$ versus $\beta$ with the same parameters as in (a). To discern the extremely small value of $\chi$ for $\rho_0=0.9$, the dotted line in (b) is plotted one thousand times larger than the original values. The inset in (b) shows the width of the hysteresis loop versus $\alpha$. Other parameters are  $\mu= 0.1$ and $k = 10$. [Reprinted from~\cite{qwmzl17}].}
\label{fig:Liu2017}
\end{figure}

A prescription that also shows the phenomenology of explosive epidemics is found in~\cite{qwmzl17}, 
with a focus on local synergy, 
quantified by the interaction of elements that when combined produce a total effect greater than the sum of the individual elements. In the case of epidemic contagion, the authors study a model in which the usual infectivity parameter $\beta$ is rescaled as
\begin{equation}
\beta ' = 1- (1-\beta)^{1+\alpha m}
\label{rescbeta}
\end{equation}
\noindent where $\alpha$ is the strength of the synergistic effect, and $m$ the number of infected neighbors. The effect of synergy is an explosive epidemic outbreak characterized by a first order phase transition, see Fig.~\ref{fig:Liu2017} for details.

A very interesting effect was reported in~\cite{bwahh15} when the recovery of sick individuals depends on the availability of healing resources that are generated by the healthy population. In this case, the limitation of resources triggers the emergence of an abrupt transition in the epidemic outbreak.  
Abrupt percolation transitions in complex networks can also be produced by the spreading of cooperative contagions~\cite{ccgg15}. Let us imagine a set up in which the agents can be infected by a single disease A with probability $p$, and that nodes infected with A can be infected with a different disease B with a raised 
probability $q$. The net effect of this dynamical process reveals that cooperation can lead to phase transitions of different orders, causing either a mild or an abrupt massive outbreak  
at the threshold. An abrupt outbreak emerges only if the network structure supports a bottleneck for cascading mutual infections. Other specific situations where epidemics can 
emerge in an explosive transition are, for example, when the infection spreads and in addition we require that the recovery of infected nodes can only occur 
when they remain connected to a predefined central node via  
a path that contains only healthy nodes~\cite{bwghh16}. This particular situation induces the possibility that healthy nodes become encapsulated by infected nodes and thus are absorbed into an infected cluster all at once, thereby promoting abrupt transitions at the epidemic outbreak.


\subsection{Explosive phenomena on multilayer complex networks}

\begin{figure}
\begin{center}
\includegraphics[width=.35\textwidth,angle=0,clip=1]{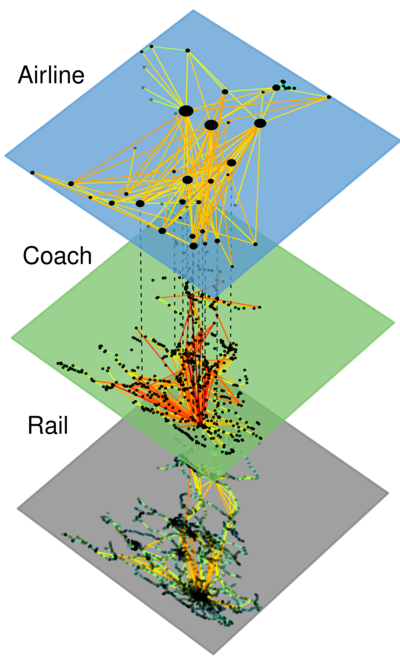}
\end{center}
\caption{
Example of multilayer complex network. Each node represents a geographical area, different layers account for different transportation networks. [Reprinted from~\cite{dpa15}].}
\label{fig:Muxviz2015}
\end{figure}

The explosive phenomena surveyed in the previous sections where grounded in ``traditional'' complex networks, in that the focus was on individual networks. However,  
there is an increasing interest in so-called ``networks of networks", accounting for 
a diversity of distinct network types, node and link types, and interconnection patterns.
A useful paradigm for organizing the complexity of many such systems is the framework of multilayer networks~\cite{kabgmp14,bbcggr14}, see Fig.~\ref{fig:Muxviz2015} for an example. 
Here we briefly review the findings of abrupt (explosive) transitions that emerge from the  
structural aspects of a multilayer network's topology, as well as from the interplay between structure and certain dynamical processes on multilayer networks. 

Interdependent and multilayered networks 
are used to define classes
of complex systems where networks are not considered as isolated entities but instead as interacting with one another. More specifically, we can consider
a multiplex network, which is 
a multilayer network in which each distinct network layer implements a different categorical relationship between the entities in that layer, and the same exact set of nodes exists in each network layer as replicas of each other.  
Note, the notion of
interdependent networks is a more general framework 
where nodes can be different in each network, and 
the functioning of nodes in one network can depend upon interconnections to nodes in  
another network, e.g. the electric grid and the cyber-infrastructure used for its control. 
Many, if not all, real networks are ``coupled'' with other real networks. Examples can be found in several domains: social networks (e.g., Facebook, Twitter, etc.) are coupled because they share the same actors~\cite{slt10}; multimodal transportation networks are composed of different layers (e.g., bus, subway, etc.) that share the same locations~\cite{b11}; the functioning of communication and power grid systems might depend one on the other~\cite{bppsh10}. 
The majority of phenomena that have been studied on interdependent networks, including percolation~\cite{bppsh10, sgcgp12}, epidemics~\cite{smsb12}, evolutionary games~\cite{ggraf12}  and linear dynamical systems~\cite{gdgpma13}, have provided results that differ significantly from their counterparts obtained
in the case of isolated complex networks~\cite{dgpa16}. Sometimes the difference is striking:  
for example, while the large-scale connectivity of isolated scale-free networks is robust against random failures of nodes or edges~\cite{rjb00}, scale-free interdependent networks can instead be 
 very fragile with respect to the breakdown of large-scale connectivity~\cite{bppsh10, sgcgp12}.

\subsubsection{Abrupt structural transitions}

The merging of
independent networks into an interconnected network-of-networks 
undergoes a structurally sharp transition as the 
interconnectivity between the networks increases. 
Depending on the relative importance of  inter- and intra- layer connections, it has been found~\cite{ra13,ra16} that the 
system can exist in one of  
two regimes: in one regime, the various layers are structurally decoupled and they act as independent entities; in the other regime, network layers are indistinguishable and the whole system behaves as a single-level network, with an abrupt transition between the two regimes. 

Focusing on the analysis of the spectrum of the supra-Laplacian matrix, defined as
\begin{equation}
\mathcal{L} = 
 \left(
\begin{array}{cc}
\mathcal{L}_{A} + p \mathbbm{1} & - p \mathbbm{1}\\
- p \mathbbm{1} & \mathcal{L}_{B} + p \mathbbm{1}
\end{array} 
\right) \, ,
\label{eq:lap}
\end{equation}

\noindent where the blocks present in $\mathcal{L}$ are square symmetric matrices of dimensions $N \times N$.
In particular, $\mathcal{L}_{A}$ and $\mathcal{L}_{B}$ are the Laplacians of the networks $A$ and $B$, respectively, and $p$ the inter-connectivity weight.
 The spectrum of the Laplacian of a graph is a fundamental mathematical object for the study of the structural properties of the graph itself. The authors in~\cite{ra13} unveiled the origin of the structural changes of the merging of networks in an interconnected system. Note that, for any graph, all eigenvalues of its Laplacian are non-negative numbers. The smallest eigenvalue is always equal to zero and the eigenvector associated to it is trivially a vector whose entries are all identical.
The second smallest eigenvalue $\lambda_2$, also called the {\em algebraic connectivity}, is one of
the most significant eigenvalues of the Laplacian. It is strictly larger than zero only if the graph is connected. More importantly, the eigenvector associated to $\lambda_2$, which is called the {\em characteristic valuation} of a graph, provides even deeper information about its structure. For example, the components of this vector associated to the various nodes of the network are used in spectral clustering algorithms for the bisection of graphs. By studying the behavior of the second smallest eigenvalue of the supra-Laplacian matrix $\mathcal{L}$ and its characteristic valuation as a function of $p$, given the single-layer network Laplacians $\mathcal{L}_A$
and $\mathcal{L}_B$, they proved that
\begin{equation}
\lambda_2\left(\mathcal{L}\right) =
\left\{
\begin{array}{ll}
2p & \textrm{  if } p\leq p^*\\
\leq \frac{1}{2} \lambda_2\left(\mathcal{L}_A + \mathcal{L}_B\right)  & 
\textrm{  if } p\geq p^*
\end{array}
\right. \; .
\label{eq:final}
\end{equation}
Thus indicating that the algebraic connectivity of the interconnected system follows two distinct
regimes, one in which its value is independent of the structure of the two layers, and the other
in which its upper bound is limited by the algebraic connectivity of the weighted superposition of the two layers
whose Laplacian is given by $ \frac{1}{2} \left( \mathcal{L}_A +  \mathcal{L}_B\right)$.
More importantly, the discontinuity in the first derivative of $\lambda_2$  
indicates
a radical change
of the structural properties of the system  
at $p^*$. Such dramatic change
is visible in the coordinates of characteristic valuation of the nodes of the two network layers. In the regime $p\leq p^*$, the components of the
vector corresponding to network $A$ are 
\begin{equation}
\left|v_A\right> = 
- \left|v_B\right> \qquad \textrm{ where } \left|v_A\right> = \pm \frac{1}{\sqrt{2N}} \left|1\right> \; .
\label{eq:sign1}
\end{equation}
This means that the two network layers are structurally disconnected and independent. For $p\geq p^*$, we have
\begin{equation}
 \left<v_A|1\right>=\left<v_B|1\right>=0  \; ,
\label{eq:sign2a}
\end{equation}
which means that the components of the vector corresponding to interdependent
nodes of network $A$ and $B$ have the same sign, while nodes in the same layer have
alternating signs. Thus in this second regime, the system connectivity is dominated by inter-layer connections, and
the two network layers are structurally indistinguishable.

Further studies~\cite{r14} on the spectral properties of the normalized Laplacian of random interconnected networks show richer behavior than those of  isolated networks. Depending on the combination of basic structural properties as e.g. degree distribution, degree correlation, and strength difference between intra- and interlayer connections, topological and dynamical phase transitions associated with the spectrum of the normalized Laplacian can be either discontinuous or continuous, and different regimes can disappear or even coexist.
The rigorous generalization to the previous findings to m-layers and its application in diffusion processes can be found in~\cite{ssvm15,vm16,cm16}.


\subsubsection{Abrupt percolation transitions}

In 2009-2010 the analysis of the percolation properties of interdependent networks started~\cite{leicht2009,bppsh10}, with the influential paper by~\cite{bppsh10} addressing dependencies between networks.  Focusing on the specific case of two interconnected networks A and B, they proposed a simple model of cascading failure propagation, and they discover a sharp transition in the probability of existence of a mutual giant component. The mutual giant component is defined as the set of nodes in network A which belong to an infinite-sized 
cluster in network A and also have their corresponding nodes in network B belong to  
an infinite-sized cluster in network B (or vice-versa). Their main  
focus in on the probability of the existence of a large cascading failure in the mutual giant component.
The proposed model was analytically studied in terms of avalanches in~\cite{bdgm2012} pointing out the existence of a hybrid phase transition. Several studies have focused on avoiding catastrophic failures via optimal coupling between networks based on their connectivity patterns 
e.g.~\cite{bsl12,rhbacsm14,gosiaPRE2019}, showing that interconnectivity can be an asset.  

Later the problem in~\cite{bppsh10} was revisited by~\cite{sgcgp12} demonstrating that the original study 
can be simplified conceptually by deleting all references to cascades of failures. The model is formulated in terms an order parameter $S$ which accounts for the probability that a randomly chosen node is a member of 
the infinite percolating cluster, thus they can apply classical percolation techniques.
Nevertheless, the authors point out that for interdependent networks that are correlated with each other, or that are spatially embedded, the transition is in general not first order, and that interdependence can make the transition even less sharp than found in ordinary percolation.

The apparent mismatch between the results of~\cite{bppsh10} and~\cite{sgcgp12}  
has been recently solved in a unifying framework proposed by~\cite{lcskk2016}. The idea is to formalize the problem in terms of a hybrid phase transition, akin to that first reported in~\cite{bdgm2012}, for the order parameter $m(z)$ which is the size of the giant mutually connected cluster per node ($z$ refers to the average degree of the network). As defined in Sec.~\ref{sec:type5} the order parameter for a hybrid transition is characterized as:
\[
m(z) =
\left\{
\!
\begin{aligned}
&0 & \text{ if } z<z_c\\
&m_0 + r (z-z_c)^{\beta_m} &\text{ if } z\ge z_c
\end{aligned}
\right.
\]
\noindent where $m_0$ and $r$ are constants and $\beta_m$ is the critical exponent of the order parameter. Recall that the exponent must obey $0<\beta<1$ for the hybrid transition to hold.
The first distinction offered by the authors in the cascade failure model, is between clusters and avalanches. Clusters are mutual connected components. Avalanches consist of mutual connected components separated from the giant component as a consequence of a removed node that triggers  
subsequent cascade. While the avalanche dynamics exhibits a critical pattern for a critical average degree $z_c$, the finite mutual connected components at $z_c$ mostly consist of one or two nodes, which is in discord with the power-law behavior of the cluster size distribution at a transition point characteristic of the conventional second-order percolation transition. From the proposed hybrid phase transition scheme, the authors classify avalanches in the critical region as finite and infinite avalanches. When an infinite avalanche occurs, the mutual giant component collapses, and the system falls into an absorbing state. 

Apart from the previous abrupt transitions found, several authors have observed a sharp contrast to ordinary percolation between multilayer networks and traditional (monolayer) networks. The principal difference is rooted in the definition of the mutual component whereby  
a node is a member of the mutual component only if all the nodes that are interdependent with this node also belong to the mutual component. In monolayer networks, nodes of high degree belong to the giant connected component (percolation cluster) with higher probability, however, in multilayer networks the inclusion 
of nodes of high ``superdegree'' (that indicates the number of interdependent nodes in other layers of each individual node) into the mutual component is more difficult. The nontrivial point is that 
when the ``superdegree" of the nodes in the layers are distributed heterogeneously among the layers, 
each given degree can enter the mutual component not smoothly but through a discontinuous transition~\cite{bd14}. 
It has been shown that for a large class of multilayer networks, the mutually connected component does not depend on the global topology. The mutual component problem for these networks can be solved by reduction to that for multiplex networks~\cite{bdm15}. In multiplex networks,~\cite{clzgb13} presented a general framework for studying the emergence of a mutual giant component in multiplex networks with overlap. They show the presence of a critical value of edge overlap in a two-layer multiplex network that can change the order of the phase transition from hybrid first order to second order. The model is a directed version of percolation of interdependent multiplex 
networks with overlap which admits an epidemic spreading interpretation. This generalized percolation model reduces the to the model of interdependent multiplex networks in absence of overlap but in presence of overlap it is distinct. In~\cite{cdb16} the authors provide a message passing theory for characterizing the percolation transition in multiplex networks with link overlap and an arbitrary number of layers, showing that the percolation transition has a discontinuous phase transition, with the exception of the trivial case in which all the layers completely overlap.

Recently, the authors in~\cite{rb17} propose a percolation model where the condition that makes a node functional is that the node is functioning in at least two of the layers of the network, redefining then the concept of mutual giant component in an M-layer multiplex setting.  According to this model, interdependencies make the system more fragile than it would be by considering each layer independently. This fact is consistent with the traditional model used to study percolation in multiplex networks, and it is apparent from the fact that the percolation transition is abrupt for any number of network layers considered in the inter-dependent model.

The nuanced contrasts between the specific results reviewed here on interdependent networks show that
the landscape is broad and that
interconnectivity between networks can be a liability or an asset depending on the context.

\subsubsection{Abrupt synchronization transitions}

Another example of explosive phenomena in multilayer networks is that of explosive synchronization described in Sec.\ref{sec:ES}. Several works have found abrupt synchronization transitions in this framework. We review them here.

In~\cite{zbgl15} the authors address the adaptive scheme described in Sec.\ref{ssec:adaptive} in scale-free networks as layers of a multiplex network. In particular, the authors study a multiplex system composed of two networks (A and B) in which each of the $N$ nodes of a layer is associated (coupled) to another node of the other network. The evolution equations for the two layers read:
\begin{eqnarray}
\dot{\theta}^{A}_i=\omega^{A}_i+\lambda\alpha^{A}_i(t)\sum_{j=1}^{N}A_{ij}\sin(\theta^{A}_j-\theta^{A}_i)\;,\label{eq:adaptivemultiplex1}\\
\dot{\theta}^{B}_i=\omega^{B}_i+\lambda\alpha^{B}_i(t)\sum_{j=1}^{N}B_{ij}\sin(\theta^{B}_j-\theta^{B}_i)\;,
\label{eq:adaptivemultiplex2}
\end{eqnarray}
where we denote with the same index $i$ each pair 
of nodes, one belonging to A and the other to B, that are coupled one-to-one. Matrices ${\bf A}$ and ${\bf B}$ correspond to the adjacency matrices of layer A and B respectively.

\begin{figure}[!t]
\begin{center}
\includegraphics[width=.55\textwidth,angle=0,clip=1]{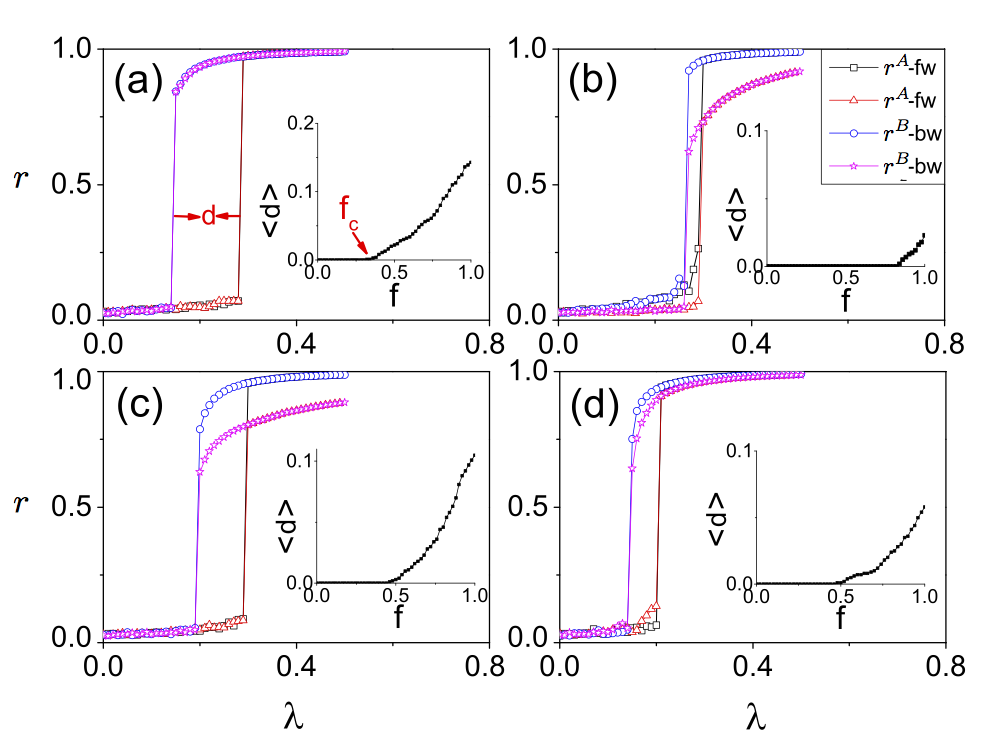}
\end{center}
\caption{Forward and backward synchronization diagrams for different multiplex networks of $N=1000$ nodes and $f=1$. {\color{black} Synchronization is represented by the Kuramoto order parameter $r$ as a function of the coupling constant between oscillators $\lambda$.}
Layer A is an ER network with $\langle k\rangle=12$  and an homogeneous frequency distribution $g(\omega)$ in the interval $[-1,1]$. The second layer is (a) another ER network with the same characteristics, (b) an ER network with $\langle k\rangle=6$, (c) the same ER as in A but with $g(\omega)$ being a Lorentzian distribution ($\gamma=0.5$), and (d) a SF network with $\langle k\rangle=12$.  The inset of each panel shows the evolution of the length of the hysteresis region as a function of $f$. [Reprinted and adapted from~\cite{zbgl15}].}
\label{fig:Adaptive3}
\end{figure}

Given Eqs.~(\ref{eq:adaptivemultiplex1})-(\ref{eq:adaptivemultiplex2}), the nodes are split into the same two groups I and II as in the case of a single network. In this case, if a node $i$ in layer A is of type I (II) the corresponding node $i$ in layer B will be also of type I (II). Those $(1-f)N$ pairs 
of nodes of type II are not dynamically coupled, $\alpha_i^{A}=\alpha_i^{B}=1$, and they are only affected by the dynamics of their neighbors in their corresponding layer. On the other hand, if the pair 
of nodes are of type I they interact directly 
in the following way: $\alpha_i^{A}=r_i^{B}(t)$ and $\alpha_i^{B}=r_i^{A}(t)$. Therefore, the $fN$ nodes of type I in one layer are affected by the local degree of synchronization among the neighbors of node $i$ in the other network layer.

Figure~\ref{fig:Adaptive3} reports the synchronization diagrams for $f=1$ in three different ER-ER multiplex networks [panels (a), (b) and (c)], while panel (d) shows the transition obtained when the multiplex is composed by an ER and a SF layer with the same average degree $\langle k\rangle=12$. Interestingly, in all the four cases the explosive transition happens almost simultaneously for both layers. 

A similar approach is developed in~\cite{hjmqymz16}, but they also impose a new phase to the Eqs.~(\ref{eq:adaptivemultiplex1})-(\ref{eq:adaptivemultiplex2}) in the multiplex defined as $\theta=p\theta^{(A)}+(1-p)\theta^{(B)}$. In the particular case of a power-law degree distributed network (A), and a Watts-Strogatz network (B), the weighted version of the phase $\theta$, undergoes a first order phase transition as a function of $p$. 
Yet, recently, Khanra {\it et al.}~\cite{kkhp18} also investigated a coupled phase-frustrated Sakaguchi-Kuramoto dynamics on a bilayer multiplex network. Essentially, the authors investigate Eqs.~(\ref{eq:adaptivemultiplex1})-(\ref{eq:adaptivemultiplex2}) in a framework where $\theta^{X}_j-\theta^{X}_i \rightarrow \theta^{X}_j-\theta^{X}_i -\alpha$, for every layer $X$. The results show that the frustrated system can undergo a sharp transition when coupled in a multilayer network for a wide region of the parameter space, moreover the critical points of the forward and backward transition can be computed analytically.

Very recently Nicolaou et al.~\cite{Nicolau}, inspired by the Kuramoto dynamics, introduced a new nonlinear model, called Janus oscillators, in which each unit $i$ splits into two oscillators. These oscillators have independent phases, $\theta_{i}^{A}$ and $\theta_{i}^{B}$, and different natural frequencies, $\omega_i^{A}$ and $\omega_{i}^{B}$, that are related in the following way: $\omega_i^{A}=\Omega+\omega/2$ and $\omega_{i}^{B}=\Omega-\omega/2$ (where as usual $\Omega$ is the average frequency of oscillators). These two faces of the same unit $i$ can be viewed as the two representations of each node $i$ in the layers of a duplex.  However, the duplex induced by the coexistence of two oscillators in each dynamical unit $i$ takes a particular form, as revealed from the dynamical equations at work:
\begin{eqnarray}
\dot{\theta}^{A}_i=\omega^{A}_i+\lambda^{\prime}\sin(\theta^{B}_i-\theta^{A}_i)+\lambda\sum_{j=1}^{N}A_{ij}\sin(\theta^{B}_j-\theta^{A}_i)\;,\\
\dot{\theta}^{B}_i=\omega^{B}_i+\lambda^{\prime}\sin(\theta^{A}_i-\theta^{B}_i)+\lambda\sum_{j=1}^{N}A_{ji}\sin(\theta^{A}_j-\theta^{B}_i)\;.
\label{eq:Janus}
\end{eqnarray}
From these equations we observe that the two oscillators in dynamical unit $i$ interact with each other through coupling $\lambda^{\prime}$ while oscillator $i$ in layer $A$ ($B$) interacts with those neighbors in layer $B$ ($A$) as dictated by the adjacency matrix ${\bf A}$.  Thus, this duplex takes a bipartite form in the sense that the interactions of oscillators in layer $A$ ($B$) are always with those in layer $B$ ($A$). This structure automatically induces a frequency gap $\omega$ between each pair of connected oscillators. The frequency gap induced by the Janus model produces a variety of interesting behaviors, even in the case of simple network structure. Among these behaviors, the authors found ES by varying $\lambda$, thereby tuning from the asynchronous state to one of multiple synchronized states that can range from phase-locking to a variety of partially synchronized states.

Another interesting phenomena arises when coupling two different dynamics (one on each layer) interacting via a duplex (two-layer multiplex). Critical phenomena associated with this kind of interaction where first analyzed by~\cite{cga13,cga14} in the context of epidemic spreading. However, the observation of an abrupt transition for interacting dynamical processes was found again in the context of synchronization. In~\cite{nsal17} the authors couple two dynamical processes, one for each layer of a duplex with their respective topologies. Assuming that the
dynamics of the entire system are governed by the following equations:
\begin{equation}
  \left\{ \begin{array}{c}
   \dot{x}_i = F_{\omega_i} ( {\bf x} , A )
   \\
   \dot{y}_i = G_{\chi_i} ( {\bf y} , B )
   \\
\end{array}
\right.
\qquad 
i=1,2,\ldots N
\label{eq:gen}
\end{equation}
where ${\bf x}$ and ${\bf
  y}$ denote the vectors of states of
the two dynamical processes, while the topologies of the two layers are
encoded in the adjacency matrices $A$ and
$B$ respectively.

\begin{figure*}[t]
\begin{center}
\includegraphics[width=.85\textwidth,angle=0,clip=1]{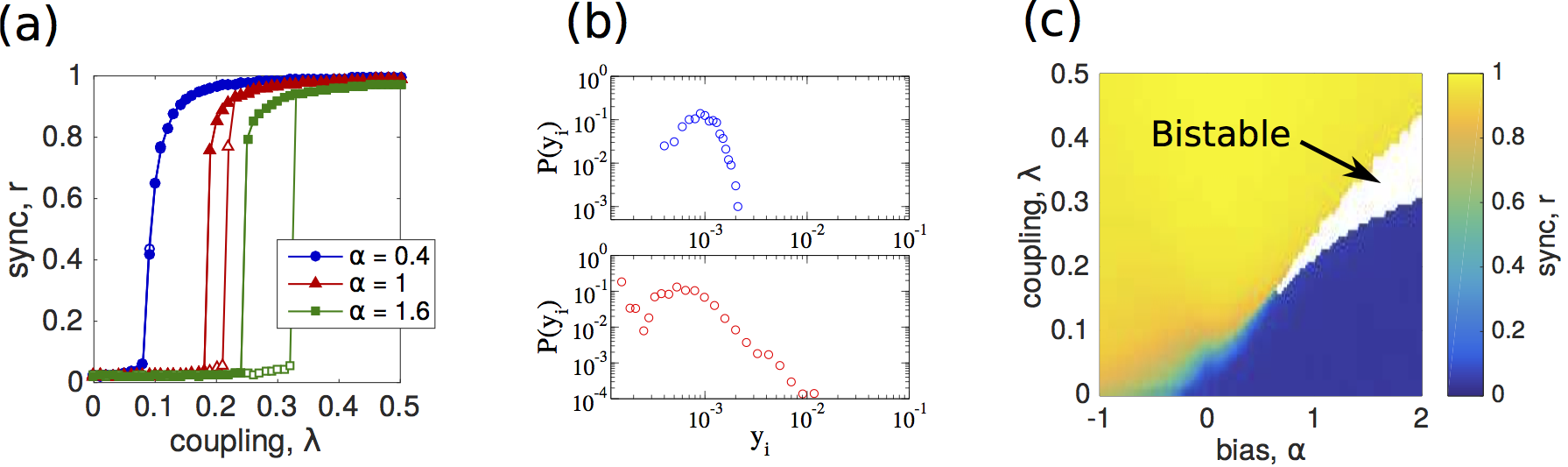}
\end{center}
  \centering
  \caption{Spontaneous explosive synchronization induced by
      the multiplex coupling of the two processes: {\color{black}synchronization and biased random walkers}.  (a) Level of
    synchronization $r$ vs the coupling intensity $\lambda$ at layer A for bias exponents of the random walks $\alpha=0.4$, $1.0$, and $1.6$ (blue, red, and green,
    respectively).  (b) Distribution $P(y_i)$ of steady-state random
    walker fractions $y_i$ at layer B for $\alpha=1.0$, when the
    oscillators at layer A are incoherent ($\lambda=0.1$, top, blue)
    and synchronized ($\lambda=0.4$ bottom, red).  (c) Synchronization
    phase diagram showing $r$ as a function of coupling $\lambda$ and
    bias exponent $\alpha$. The bistable region is colored in
    white and corresponds to the region where the transition towards synchronization is abrupt. 
    Networks are of size $N=1000$ power-law distributed with exponent $\gamma=3$ and $\langle
    k^{[A]}\rangle=\langle k^{[B]}\rangle = 10$.  [Reprinted from~\cite{nsal17}].
  }
\label{fig:ESmulti}
\end{figure*}

The dynamical evolution of the two network
processes are governed  
respectively by the functions $F_{\omega}$ and
$G_{\chi}$, which depend on the sets of parameters $\omega$ and
$\chi$, so that the state $x_i$ ($y_i$) of node $i$ at layer $A$ ($B$) is a function of the state ${\bf x}$ (${\bf y}$) and of
the topology $A$ ($B$) of the first (second) layer. The
key ingredient that connects the two dynamical processes is provided
by the nature of the correspondence between layers. In fact, the
parameter $\omega_i$ in function $F_{\omega_i}$ at layer A is itself a
function of time which depends on the dynamical state $y_i$ of node
$i$ at layer B, while the parameter $\chi_i$ at layer B depends on the
state $x_i$ of node $i$ at layer A.  Namely, we have:
\begin{equation}
  \left\{ \begin{array}{c}
   \dot{\omega}_i = f ( \omega_i, y_i) 
   \\
   \dot{\chi}_i = g (\chi_i, x_i)
   \\
\end{array}
\right.
\qquad 
i=1,2,\ldots N
\label{eq:gen2}
\end{equation}
where $f$ and $g$ are two assigned functions. The specific problem analyzed consists of the coupling between a Kuramoto model in one layer, and a biased random walk on the network in the other layer:
\begin{align}
  \dot{x}_i = \omega_i+ \lambda \sum_{j=1}^N a_{ij}
  \sin(x_j-x_i),
  \label{eq:Kuramoto}
\end{align}
where $\omega_i$ corresponds to the natural frequency of the
oscillator $i$ and $\lambda$ is the {\em coupling strength} of the Kuramoto model, and 
\begin{align}
  \dot{y}_i =\frac{1}{\tau_y} \sum_{j=1}^N \left(
  \pi_{ij}-\delta_{ij}\right) y_j = \frac{1}{\tau_y} \sum_{j=1}^N
  \left( \frac{b_{ji} \chi_i^\alpha}{\sum_l b_{jl}
    \chi_l^\alpha} - \delta_{ij} \right) y_j
 \label{eq:RandomWalk} 
\end{align}
where $\pi_{ij}$ is the transition probability from node $j$ to node
$i$, $\tau_y$ is the time scale of the random walker dynamics, and we
have assumed that the random walk is biased on a node property
$\chi_i$, with a tuneable {\em bias exponent}
$\alpha$. The results of the model are depicted in Fig.~\ref{fig:ESmulti}, where the explosive synchronization phenomenon is revealed for the Kuramoto order parameter $r$.
Very recently, the general formalism introduced in Eq.~(\ref{eq:gen}), has been used to study a multiplex in which a SIS epidemic model and the Kuramoto dynamics coexist~\cite{Soriano}. The authors have found that both epidemic and synchronization transitions become discontinuous when the two dynamics interact in a cooperative way. This happens when synchronization in the first layer is promoted by the spreading activity in the second one and, simultaneously, the contagion rate in the second layer is fostered by the dynamical coherence in the first.

Finally, during the revision of this manuscript, a relevant result concerning the emergence of different types of synchronization transitions in multilayer networks has appeared~\cite{dbbh19}. In this work, the authors analyze interdependent 
and competitive interactions between dynamical processes on multilayer networks, including 
synchronization dynamics. They define a function that accounts for the interaction between layers of the multiplex, 
in particular 
the influence that each node $i$ in layer A has on its replica node $i$ in layer B for a two-layered multiplex. The function relies on a local order parameter, $z_i$, on one layer, specified in~\cite{dbbh19} to be 
the local level of synchronization on that layer.  This local order parameter is the input signal 
for the corresponding replica 
node in a different layer, for example: $F_i^{A\rightarrow B}(t) = | z_i^A |(t)$ specifies how much a node $i$ in layer A affects a node $i$ in layer B. For competitive interactions
(i.e., those that discourage synchronization),
the proposed form of the function is $ F_i^{A\rightarrow B}(t) = | 1- z_i^A |(t)$. Using this last specification, the global synchronization level undergoes a discontinuous backward transition as the coupling strength is decreased, reminiscent of an explosive process resulting from the frustration imposed on the replica nodes by that choice for the interaction function between layers.

\section{Conclusions and Open Challenges}\label{sec:Conclusions}

The paradigm of explosive phenomena in complex networks offers many novel advances, from the basic theoretical understanding of phase transitions to the ability to model a broad array of new phenomena on networks. 
Despite the large body of work on explosive phenomena and efforts to develop a deeper understanding, many abrupt transitions are still not well understood.  This review offers an attempt to organize the extensive existing literature into thematic categories and identify fruitful directions for future research. 

Since the start of vigorous activity in the field in 2009, there has been a chain of fascinating discoveries about the different exotic behaviors and universality classes of explosive phenomena. More specifically, we now understand a myriad of mechanisms that lead to EP and ES 
and have discovered many processes and models that implement these mechanisms. 
Through EP in particular we see the massive impact that a repeated, small intervention can achieve.  This also provides new opportunities for modeling network failure modes in technological and economic systems, as well as interventions and consequences, as discussed briefly in~\cite{helbing2013globally}. 

The phenomenology behind EP and ES share similar ingredients. In particular, both EP and ES are based on the implementation of some microscopic rules that avoid the formation of a macroscopic (connected or synchronized) component. 
In EP the rules are typically explicit and lead to the growth of a powder-keg distribution for the cluster-sizes at the percolation transition point. 
In ES the rules are implicit and result from the interplay of the natural frequency distribution and the network connectivity patterns. More understanding of the sub-critical synchronized cluster evolution for ES on different network topologies may provide an explicit framework to connect EP and ES. 

Beyond the tour de force needed for establishing a mathematically grounded connection between EP and ES, there are still several challenges that remain unexplored for each phenomenon independently. For example, in EP, there is the need to quantify the tradeoff between the extent of global versus local information required. 
In ES, there is the question about how local order develops, e.g. do chimera states exist? 
EP has opened new opportunities for modeling networked systems, such as modular networks, optimal immunization strategies, network dismantling, and information spreading. Major outstanding challenges in the subject of EP are developing tailored control interventions for such systems.  Major challenges in the subject of ES are that of relating the spectral properties of the graph with the outcome of the synchronization transition, or that of finding real-world examples where ES could be a plausible mechanism for switching on and off synchronization. ES is proving to be a useful paradigm for understanding neural disorders, which is extremely promising.

Many models of EP and ES have been designed to produce anomalous features via a built-in cluster growth suppression mechanism. 
Identifying additional natural, physical and socio-technological systems, where the suppression of large cluster mergers is 
an 
intrinsic feature of the dynamics is a compelling research avenue. The existence of such features in real-world systems may help to better understand and control these systems. 

The array of behaviors, underlying mechanisms, and processes that implement explosive transitions on complex networks are still being discovered.  And the explosive transitions surveyed in this review lead to many new research questions. 
We refer the reader to a number of recent reviews with extensive additional details and perspectives~\cite{bastas2014review, AraujoPercReview2014, SaberiPhysReports2015, boccaletti2016}.

In summary, abrupt transitions may underlie many important phenomena and functions of systems found in nature, and it is of great importance to be able to describe and understand the mechanisms through which they occur. The explosive phenomena reported so far are opening up new perspectives on the types of transitions we can observe in networked systems and we envisage that they will catalyze increased future research activity on dynamical processes unfolding on top of complex networks.

\acknowledgements{We gratefully acknowledge support from funding agencies and the reviewers of this manuscript. In particular RMD acknowledges support from the U.S. Army Research Office MURI Award No. W911NF-13-1-0340 and Cooperative Agreement No. W911NF-09-2-0053, and DARPA Award No. W911NF-17-1-0077.  JGG acknowledges support from MINECO through project No. FIS2017-87519-P, and from the Departamento de Industria e Innovaci\'on del Gobierno de Arag\'on y Fondo Social Europeo (FENOL group E-19). 
AA acknowledges support by Ministerio de Econom\'{\i}a y Competitividad (grant FIS2015-71582-C2-1, PGC2018-094754-B-C21), Generalitat de Catalunya (grant 2017SGR-896), and Universitat Rovira i Virgili (grant 2017PFR-URV-B2-41), ICREA Academia and the James S. McDonnell Foundation (grant \# 220020325).
We benefited from critical feedback from G. Bianconi, B. Ziff, B. Khang, S. G\'omez, C. Granell, M. Timme, Y. Fender, and L. Fender.}




\end{document}